\documentclass[aps, amsmath, amssymb, twocolumn, 10pt, 
longbibliography, superscriptaddress, nofootinbib]{revtex4-2}

\usepackage{textcomp}
\usepackage{amsmath}
\usepackage{amsfonts}
\usepackage{calc}
\usepackage{mathrsfs}
\usepackage{array}
\usepackage{xcolor}
\usepackage{bm}
\usepackage{graphicx}
\usepackage{hyperref}
\usepackage{tensor}
\usepackage{mathtools}
\usepackage[caption=false]{subfig}
\usepackage{float}
\usepackage{orcidlink}
\usepackage[normalem]{ulem}
\usepackage{placeins}
\usepackage{siunitx}
\usepackage{upgreek}

\graphicspath{{Figures/}}

\newcommand{\scri}{{\mathscr I}}

\newcommand{\diff}[2]  {\frac{d #1}{d #2}}
\newcommand{\sdiff}[2]  {\frac{d^2 #1}{d #2^2}}
\newcommand{\pdiff}[2]  {\frac{\partial #1}{\partial #2}}
\newcommand{\psdiff}[2]  {\frac{\partial^2 #1}{\partial #2^2}}
\newcommand{\eqn}[1] {Eq.~\eqref{eq:#1}}
\newcommand{\fig}[1] {Fig.~\ref{fig:#1}}

\newcommand{\secref}[1] {Sec.~\ref{sec:#1}}
\newcommand{\tblref}[1] {Table.~\ref{tbl:#1}}

\newcommand{\nn}{\nonumber}

\newcommand{\lm}  {\ell m}

\newcommand{\rstar} {r_{*}}

\newcommand{\barh} {\bar{h}}
\newcommand{\rp} {r_{p}}
\newcommand{\ellmax} {\ell_{\rm max}}
\newcommand{\bmat}[1]{\boldsymbol{\mathcal{#1}}}
\DeclarePairedDelimiterX\braket[1]{\langle}{\rangle}{#1}

\newcommand{\beq}{\begin{equation}}
\newcommand{\eeq}{\end{equation}}
\newcommand{\bec}{\begin{cases}}
\newcommand{\eec}{\end{cases}}

\definecolor{colour1}{HTML}{0571b0} 
\definecolor{colour2}{HTML}{92c5de} 
\definecolor{colour3}{HTML}{f4a582} 
\definecolor{colour4}{HTML}{ca0020} 
\definecolor{colour5}{HTML}{fe4a49} 

\hypersetup{colorlinks=true, linkcolor=colour1, citecolor=colour1,
filecolor=colour1, urlcolor=colour1}

\begin{document}
\title{Gravitational self-force with hyperboloidal slicing and spectral methods}

\author{Benjamin Leather\,\orcidlink{0000-0001-6186-7271}} 
\affiliation{Max Planck Institute for Gravitational Physics (Albert Einstein
Institute), Am M\"{u}hlenberg 1, 14476 Potsdam, Germany}
\date{\today}


\begin{abstract}
We present a novel approach for calculating the gravitational self-force (GSF) in the Lorenz gauge, 
employing hyperboloidal slicing and spectral methods. 
Our method builds on the previous work that applied hyperboloidal surfaces and spectral approaches to a scalar-field 
toy model~[Phys. Rev. D 105, 104033 (2022)],
extending them to handle gravitational perturbations. 
Focusing on first-order metric perturbations, we address the construction of the hyperboloidal foliation, 
detailing the minimal gauge choice. 
The Lorenz gauge is adopted to facilitate well-understood regularisation procedures, 
which are essential for obtaining physically meaningful GSF results. 
We calculate the Lorenz gauge metric perturbation for a secondary on a quasicircular orbit in a Schwarzschild background 
via a (known) gauge transformation from the Regge-Wheeler gauge.
Our approach yields a robust framework for obtaining the metric perturbation components needed to calculate key physical quantities, 
such as radiative fluxes, the Detweiler redshift, and self-force corrections.  Furthermore, the compactified hyperboloidal
approach allows us to efficiently calculate the metric perturbation throughout the entire spacetime.
This work thus establishes a foundational methodology for future second-order GSF calculations within this gauge, offering 
computational efficiencies through spectral methods.
\end{abstract}

\maketitle

\section{Introduction}
\label{sec:introduction}
Observations of compact binary coalescences from black holes and black holes and neutron stars are now a routine
occurence within the field of gravitational wave astronomy.  To date, the LIGO-Virgo-Kagra (LVK) collaboration 
\cite{LIGOScientific:2007fwp, LIGOScientific:2014pky, LIGOScientific:2014qfs, PhysRevD.102.062003, 
PhysRevLett.123.231107, VIRGO:2012dcp, VIRGO:2014yos, PhysRevLett.123.231108, 10.1093/ptep/ptaa125, 
PhysRevD.88.043007, Somiya:2011np}
has detected nearly 100 compact binary coalescences, with the majority of these detections being binary black 
hole mergers.  
As our ground based detectors have improved over the course of the LVK observing runs, we have seen glimpses of 
compact binary coalescences with greater mass asymmetry \cite{LIGOScientific:2020stg, LIGOScientific:2020zkf} 
between the two consitutents.  This trend is expected to continue with the current (O4) and planned observing 
runs as the number of gravitational wave (GW) detections increase.  Beyond our current range of ground based detectors,
we shall soon be able to detect a lower range of frequencies of GW with new observatories—including 
next-generation (XG) ground-based detectors like the 
Einstein Telescope \cite{Punturo:2010zz} and Cosmic Explorer \cite{Reitze:2019iox, Evans:2021gyd}, 
and the space-based LISA mission \cite{LISA:2017pwj}.  
We expect this millihertz frequency regime will be populated by the most asymmetric compact binaries in nature:
intermediate-mass-ratio inspirals (IMRIs) and extreme-mass-ratio inspirals (EMRIs).  EMRIs in particular, are
systems characterised by a mass-ratio $\epsilon := \mu / M \lesssim 10^{-4}$, where $M$ and $\mu$ are the masses
of primary and secondary bodies respectively.  The task of modelling such systems is most naturally done
through gravitational self-force (GSF) theory.

GSF theory is based upon black hole perturbation theory (BHPT), where the smaller of the two objects is 
treated as a perturber within the stationary, background spacetime of the primary black hole.
In such binaries, the disparate masses lead to asymmetric time-scales of the differing aspects of the
evolution: the orbital timescale $T_{\rm orb} \sim M$ and the radiation-reaction timescale 
$T_{\rm rad} \sim M / \epsilon$.  For EMRIs, since $\epsilon \ll 1$, $T_{\rm orb} \ll T_{\rm rad}$,
and hence such binaries have a clear separation of scales.  On the orbital timescale, the motion of the
secondary closely resembles that of a geodesic in a background spacetime.  Yet, over the course of the
evolution, as we approach the radiation-reaction timescale, the secondary gradually deviates away from
such geodesic motion due to the backreaction between the secondary and the gravitational peturbation 
it sources. This interaction can be thought of as gravitational ``self-force'' which accelerates the
secondary away from geodesic motion in the background spacetime.

The gravitational self-force up to second perturbative order will lead to a cumulative correction of
${\cal O}(\epsilon^{0})$ in the motion of the secondary on the radiation reaction timescale.
More pertinently, it has been shown that corrections up to such perturbative order will contribute
${\cal O}(\epsilon^{0})$ \cite{Hinderer-Flanagan:08,Burke:2023lno} to the accumulated phase of the 
gravitational waveform.  Therefore, if one is to accurately model EMRIs over the duration of an inspiral, 
one must consider an expansion of the metric up-to-and-including second-order in the mass ratio.

To date, second-order calculations have addressed the binding energy \cite{Pound:2019lzj}, 
energy flux \cite{Warburton:2021kwk}, and gravitational waveform \cite{Wardell:2021fyy} for a smaller 
spinning secondary \cite{Mathews:2021rod, MathewsCapra} orbiting a Schwarzschild black hole, 
as well as for cases involving a slowly rotating black hole \cite{MathewsCapra}.
Such calculations leverage the disparity between the orbital and radiation-reaction timescales to
construct the waveforms from two distinct components: (i) an offline calculation of the Fourier
domain field equations on a grid of slow-evolving parameters and (ii) a quick (online)
step that solves the simple ODEs on this grid.

A recent development within this offline step is the use of \emph{hyperboloidal slicing} to foliate the
spacetime.  Hyperboloidal foliation penetrates the black hole horizon and future null infinity directly,
as opposed to the usual spacetime time coordinate, which will only intersect the bifurcation sphere $({\cal B})$
and spatial infinity $(i^{0})$.  Such a foliation is extremely advantageous when considering Fourier domain
self-force calculations.  For example, in Refs.~\cite{Pound:2019lzj, Wardell:2021fyy}, the use of
hyperboloidal slicing is an important element to tame the nature of the source terms that appear in the
multiscale expansion \cite{Miller:2020bft, Miller:2023ers}.  

More recently, there has been a concerted
effort to realise the full potential of hyperboloidal methods in BHPT 
\cite{Zenginoglu:2007jw, Zenginoglu:2011jz, PanossoMacedo:2018hab, PanossoMacedo:2019npm, PanossoMacedo:2024nkw} 
and GSF calculations \cite{PanossoMacedo:2022fdi, PanossoMacedo:2024pox}.
This has culminated in Ref.~\cite{PanossoMacedo:2022fdi}, hereafter known as Paper I, which presented
a novel approach to self-force calculations that compactified the hyperboloidal surfaces, and solved the 
resulting Fourier domain problem on a full $(\ell, m)$ spherical decomposition with a spectral method.  
Ref.~\cite{PanossoMacedo:2024pox}, then built upon this method by introducing an alternative $m$-mode 
decomposition for the problem in attempt to overcome a major bottleneck in exisiting 
second-order calculations.

Paper I and Ref.~\cite{PanossoMacedo:2024pox}, however, only applied this novel approach to a scalar 
field toy model for different scenarios that mimic the problems that the gravitational case faces.  
In this work, we seek to extend this approach from a scalar toy model to gravitational perturbations.
We present the \emph{first complete gravitational self-force calculation through hyperboloidal slicing 
and spectral methods}.

We shall conduct a calculation at first-order in the mass-ratio in the \emph{Lorenz gauge}.
The Lorenz gauge is particularly advantageous when considering GSF calculations for numerous reasons.
Firstly, the GSF is most well-understood in this gauge, particularly
the procedure of regularisation that is required to obtain physically meaningful results.  
In self-force calculations, the secondary body is “skeletonised” into a point-particle singularity, 
enabling the subtraction of the dominant singular component of the particle’s gravitational field. 
This approach isolates the regular residual part of the field, which physically contributes to the self-force.
In the Lorenz gauge, this singularity is \emph{isotropic}, and the procedures for regularising
this singularity are well established.  In other gauges, such as the radiation gauge, one
encounters extended singularities that are more difficult to reconcile\footnote{An alternative scheme
is avaliable to conduct self-force calculations within the radiation gauge that is free from such 
gauge singularites is the Green-Hollands-Zimmerman metric reconstruction scheme 
\cite{Toomani:2021jlo, Bourg:2024vre}.}.
Furthermore, the asymptotic behaviour of the Lorenz gauge metric perturbation has been studied 
extensively, and is not divergent towards either null infinity and horizon.

Another motivation for conducting our calculation within the Lorenz gauge is that it currently
forms the basis of all GSF calculations at second perturbative order.  It is therefore natural
to conduct our calculation, albeit at first-order, within the same gauge.  Eventually one envisions
future second-order calculations being completely based along the methodology presented in this work,
Paper I, and Ref.~\cite{PanossoMacedo:2024pox}.  As such, the calculation here could serve as a 
key ingredient for future second-order calculations.

To provide a clear overview of the methodology presented in this work, we breifly outline
the main steps involved in our calculation:
\begin{enumerate}
	\item Formulate the Regge-Wheeler(-Zerilli) equations on hyperboloidal slices.
	\item Formulate Berndtson's equations for the gauge transformation 
	from Regge–Wheeler gauge to Lorenz gauge on the same slicing.
	\item Solve both systems numerically using a collocation-point spectral method
	(with analytical mesh refinement).
	\item Construct the Regge-Wheeler gauge metric perturbation from the 
	Regge-Wheeler(-Zerilli) fields, and transform it to the Lorenz gauge using
	Berndtson's equations.
	\item Compute the flux, Lorenz-gauge self-force and gauge-invariant
	quantities via mode-sum regularization.
\end{enumerate}

This paper is organized as follows. We begin in \secref{hyperboloidal_slicing} with a review of 
the construction of our hyperboloidal foliation and the specific choice of \emph{minimal gauge} 
slicing. While this follows the same construction presented in Paper I, we provide a more 
comprehensive description here. In \secref{lorenz_gauge} and \secref{regge_wheeler_gauge}, 
we outline the Lorenz and Regge-Wheeler gauges within the context of GSF calculations. 
The gauge transformation between them is reviewed in \secref{gauge_transformations}. 
Our hyperboloidal method for solving the necessary Regge-Wheeler(-Zerilli) fields to facilitate 
the transformation from the Regge-Wheeler gauge to the Lorenz gauge is presented in 
\secref{hyperboloidal_rwz_equation}, and the gauge fields involved in this calculation are 
emphasised in \secref{gauge_fields}. We then outline the techniques for full metric 
reconstruction (\secref{metric_reconstruction}), the calculation of fluxes 
(\secref{gravitational_wave_fluxes}), the gravitational self-force (\secref{self_force}), 
and the Detweiler redshift (\secref{detweiler_redshift}). Finally, \secref{conclusion} 
provides a summary of our results and a discussion of future research directions. 
Additional details are presented in the Appendices.

\subsection*{Notation}
We begin by establishing the conventions used throughout this work. 
Geometrised units are employed, setting  $G = c = 1$, and the metric signature 
is taken to be $(- + + +)$.
We employ Boyer-Lindquist coordinates 
$(t, r, \theta, \varphi)$ with the usual metric function
$f(r) = 1 - 2M / r$, where $M$ is the mass of the primary.  
The standard line element for the Schwarzschild solution in these coordinates then
reads
\beq
	ds^{2} = -f(r)^{2}dt^{2} + f(r)^{-1}dr^{2} + r^{2}(\sin^{2}\theta\, d\theta^{2} + d\varphi^{2}).
	\label{eq:schwarzschild_line_element}
\eeq
We shall denote the metric of the unit 2-sphere by the alternative notation, 
$dw^{2} := \sin^{2}\theta\, d\theta^{2} + d\varphi^{2}$, to avoid a conflict of notation
with the conformal factor we shall introduce.

\section{Hyperboloidal slicing and radial compactification}
\label{sec:hyperboloidal_slicing}
\subsection{Hyperboloidal foliation}
We shall introduce compact, horizon-penetrating hyperboloidal coordinates $(\tau,
\sigma, \theta, \varphi)$ through the use of scri-fixing and the \emph{height-function technique} \cite{Zenginoglu:2007jw}.  
Following the approach of Refs.~\cite{PanossoMacedo:2018hab, PanossoMacedo:2019npm},
we begin with the horizon-penertrating ingoing Eddington-Finkelstein coordinates 
$(v, r, \theta, \varphi)$, where $v = t + \rstar(r)$, that then can be suitably augmented 
to that reach both the horizon and null infinity.  Here we have first introduced the tortoise
coordinate, which is defined by the relation $d\rstar/dr = f(r)^{-1}$.

In these coordinates, the line element from \eqn{schwarzschild_line_element} takes the form
\beq
	ds^{2} = -f(r)dv^{2} + 2 dv dr + r^{2}dw^{2}.
\eeq
Hypersurfaces along $v = \rm{const}$, extend from the future horizon ${\cal H}^{+}$
as $\rstar \rightarrow -\infty$ to past null infinity $\scri^{-}$ as $\rstar \rightarrow \infty$.
This is exemplified, as in previous hyperboloidal
treatments \cite{PanossoMacedo:2018hab, PanossoMacedo:2019npm}, by considering
the ingoing and outgoing null vectors $k^{\alpha}$ and
$l^{\alpha}$ in the spacetime:
\begin{align}
	l^{\alpha} &= \frac{1}{\nu} \left( \delta^{\alpha}_{v}
	+ \frac{f(r)}{2} \delta^{\alpha}_{r} \right),
	\label{eq:outgoing_null_vector_eddington_finklestein}
	\\
	k^{\alpha} &= -\nu \delta^{\alpha}_{r},
	\label{eq:ingoing_null_vector_eddington_finklestein}
\end{align}
where $\nu$ is a free boost-parameter under a Lorentz transformation.  One can fix this 
freedom when one switches to our hyperboloidal foliation.

To construct a foliation that reaches both the horizon and future null infinity, $\scri^{+}$,
we define a transformation to hyperboloidal coordinates $(\tau, \sigma, \theta, \varphi)$ via
the transformation from the horizon-penertrating ingoing Eddington-Finkelstein coordinates:
\beq
	v = \lambda \big( \tau - h(\sigma) \big),\quad\
	r = \lambda\frac{\rho(\sigma)}{\sigma}.
	\label{eq:hyperboloidal_transformation_ingoing_form}
\eeq
Here $\lambda$ is an associated
length scale of the spacetime, unspecified here, whereas in
Paper I it was set to $\lambda = 4M$.
Since our spacetime is spherically symmetric, the radial function,
$\rho(\sigma)$, is the areal radius on the conformal representation of the
spacetime whilst the conformal factor, identified in terms $\sigma$, is given
by
\beq
	\Omega = \frac{\sigma}{\lambda}.
	\label{eq:conformal_factor}
\eeq
The conformal spacetime metric can be given in terms of the line element
$d\tilde{s}^{2} = \Omega^{2}ds^{2}$ such that
\begin{multline}
	d\tilde{s}^{2} = 
	-\sigma^{2} F(\sigma) d\tau^{2}
	+ h^{\prime}(\sigma) ( 2\beta(\sigma)
	- \sigma^{2} F(\sigma) h^{\prime}(\sigma) ) d\sigma^{2} \\
	- 2 ( \beta(\sigma)
	- \sigma^{2} F(\sigma) h^{\prime}(\sigma) ) d\tau d\sigma
	+ \rho(\sigma)^{2} dw^{2},
	\label{eq:conformal_line_element}
\end{multline}
where ${}^{\prime}$ denotes differentiation with respect to $\sigma$,
$F(\sigma) = f(r(\sigma))$, and the radial component of the shift is given by
\beq
	\beta(\sigma) = \rho(\sigma) - \sigma\rho^{\prime}(\sigma).
	\label{eq:radial_shift_beta}
\eeq

In starting from the ingoing Eddington-Finkelstein coordinates, one ensures our
foliation is horizon penetrating.
The function, $h(\sigma)$, is then specified to
ensure the hyperboloidal slices foliates future null infinity $\scri^{+}$. 
To find the appropriate form for $h(\sigma)$, one can consider the
the ingoing and outgoing null vectors $\tilde{k}^{\alpha}$ and
$\tilde{l}^{\alpha}$ in the conformal spacetime, as in 
Refs.~\cite{PanossoMacedo:2018hab, PanossoMacedo:2019npm}.  
The null vectors are
normalised as $\tilde{g}_{\alpha\beta}\tilde{l}^{\alpha}\tilde{k}^{\beta} =
-1$, and are related to the null vectors in Schwarzschild spacetime
by $\tilde{l}^{\alpha} = \Omega^{-1} l^{\alpha}$ and 
$\tilde{k}^{\alpha} = \Omega^{-1} k^{\alpha}$.
In contrast to the null vectors appearing in 
Eqs.~(\ref{eq:outgoing_null_vector_eddington_finklestein})
-(\ref{eq:ingoing_null_vector_eddington_finklestein}),
the hyperboloidal surfaces parametrised with a
height function as in \eqn{hyperboloidal_transformation_ingoing_form} do reach
$\scri^{+}$ at $\sigma = 0$ if $\tau$ satisfies
\beq
	\tilde{k}^{\alpha}\partial_{\alpha}\tau = 1.
	\label{eq:good_parameter}
\eeq
This condition thereby fixes the boost parameter, $\nu$, and leads to the
following null vectors in the conformal spacetime,
\begin{align}
	\tilde{l}^{\alpha} &= \frac{h^{\prime}(\sigma)}{2\beta^{2}} 
	\left( 2\beta - \sigma^{2} F(\sigma) h^{\prime}(\sigma) \right)
	\delta^{\alpha}_{\tau} - \frac{\sigma^{2} F(\sigma)}{2\beta^{2}}
	h^{\prime}(\sigma)\delta^{\alpha}_{\sigma}, \\
	\tilde{k}^{\alpha} &= \delta^{\alpha}_{\tau} + \frac{1}{h^{\prime}(\sigma)}
	\delta^{\alpha}_{\sigma}.
	\label{eq:null_vector_hyperboloidal}
\end{align}

In conjuction with \eqn{good_parameter}, one sets the hypersurface where
$\sigma = 0$ to be future null infinity,
\beq
	\lim_{\sigma \rightarrow 0} \tilde{k}^{\alpha} = \delta^{\alpha}_{\tau},
\eeq
which leads to the following condition for $h(\sigma)$,
\beq
	\lim_{\sigma \rightarrow 0} \frac{1}{h^{\prime}(\sigma)} =
	0.
	\label{eq:height_function_condition}
\eeq
At this point one must proceed with caution to ensure this condition does not
threaten the regularity of the outgoing null vector, $\tilde{l}^{\alpha}$, at
future null infinity.  As in Ref.~\cite{PanossoMacedo:2018hab}, one can assume that
since $\rho(\sigma)$ is the areal radius of the conformal spacetime it is a
regular function on its domain that yields non-vanishing, positive values. 
This is important when considering the shift, $\beta(\sigma)$, that
appears in the outgoing null vector in \eqn{null_vector_hyperboloidal}.  If you
assume this phase shift has the generic form of an expansion of $\sigma$,
\beq
	\beta(\sigma) = \beta_{0} + \beta_{1}\sigma + {\cal O}(\sigma^{2}),
	\label{eq:beta_expansion}
\eeq
then this will yield
\beq
	\rho(\sigma) = \beta_{0} + \rho_{1}\sigma - \beta_{1}\sigma \ln\sigma.
	\label{eq:phase_solution_generic}
\eeq
Therefore, to avoid a logarithmic divergence, one must have $\beta_{1} = 0$.
Identifying $\beta_{0} = \rho_{0}$ for notational simplicity and 
inserting the result for $\rho(\sigma)$ into
\eqn{hyperboloidal_transformation_ingoing_form}, we find
\beq
	F(\sigma) = 1 - \frac{2M}{\lambda\rho_{0}}\sigma + {\cal O}(\sigma^{2}),
\eeq
and
\beq
	h^{\prime}(\sigma) = \frac{2\rho_{0}}{\sigma^{2}}
	\left[ 1 + \frac{2M}{\lambda\rho_{0}}\sigma \right] 
	+ {\cal O}(\sigma^{0}).
	\label{eq:height_function_derivative}
\eeq
This obeys the condition given in \eqn{height_function_condition} and thus
$\tilde{l}^{\alpha}$ is indeed regular.  

Now we have appropiately behaved
(conformal) null vectors, we are free to integrate
\eqn{height_function_derivative} to give the general form, 
\beq
	h(\sigma) = h_{0}(\sigma) + A(\sigma),
	\label{eq:generic_height_function}
\eeq
where 
\beq
	h_{0}(\sigma) := -2\rho_{0} \left[
	\frac{1}{\sigma} - \frac{2M}{\lambda\rho_{0}}
	\ln \sigma \right].
	\label{eq:h0_height_function}
\eeq
and $A(\sigma)$, is a gauge function. Together with $\beta(\sigma)$, $A(\sigma)$
captures the full gauge freedom\footnote{The only condition on these gauge degrees 
of freedom is that the hyperboloidal surfaces are spacelike everywhere such that
$\tilde{\nabla}_{\alpha}\tau\tilde{\nabla}^{\alpha}\tau < 0$.  In turn, this
imposes the following restriction on the derivative of the height function
\cite{PanossoMacedo:2018hab}:
\beq
	0 < \sigma^{2} h^{\prime}(\sigma) < \frac{2\beta(\sigma)}{F(\sigma)}.
\eeq} of the hyperboloidal foliation.  Specifically,
$\beta(\sigma)$ represents the freedom to choose the conformal
representation of the spacetime, while $A(\sigma)$ characterises the foliation.
Notably, there is significant freedom in the choice of $A(\sigma)$.  For example,
letting the function have an angular dependence, i.e. $A = A(\sigma, \theta)$,
will allow one to construct a framework applicable for a background Kerr
geometry \cite{PanossoMacedo:2019npm}.

\subsection{Minimal Gauge}
\label{sec:minimal_gauge}
\begin{figure}[t!]
	\centering
	\includegraphics[width=\columnwidth]{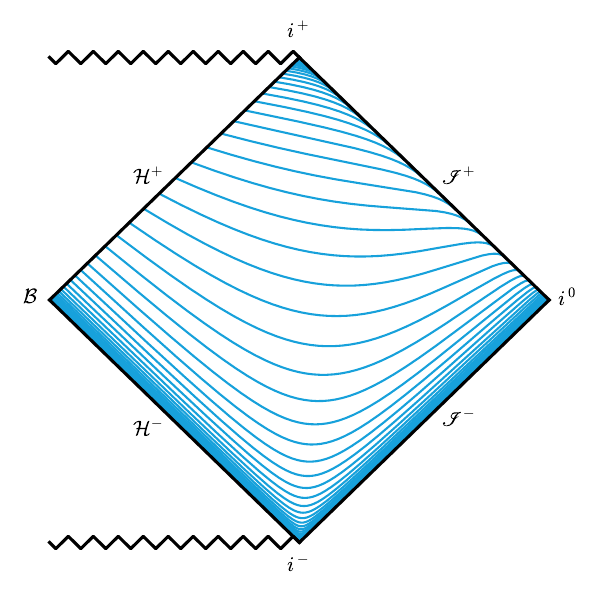}
	\caption{Penrose diagram for the Schwarzschild exterior spacetime region
	with minimal gauge hyperboloidal slices.
	The solid blue curves depict hyperboloidal time surfaces 
	$\tau-$constant extending between the black-hole horizon ${\cal H}^+$ at $\sigma=1$ 
	and future null infinity ${\scri^+}$ as $\sigma=0$.
	Here, we set $\lambda = 4M$ and $\rho_{0} = 1/2$  to align with the exact minimal
	gauge slicing adopted in this work.
	These coordinates provide a smooth foliation on the full exterior spacetime, which means that 
	both the horizon and null infinity can be included in the computational domain.
	One should note our hyperboloidal slices are spacelike everywhere, and therefore
	identical to the null vectors exactly at ${\scri^+}$.  
	The hypersufaces might appear to be tangential to this surface due to the 
	compactification of our spacetime by the mapping of the Penrose diagram.}
\label{fig:minimal_gauge_penrose}
\end{figure}

Since we now have completely identified the gauge degrees of freedom, we shall
now discuss the main choice of gauge when applying hyperboloidal formulations
in this work.  The \emph{minimal gauge} \cite{PanossoMacedo:2018hab,
PanossoMacedo:2019npm}, assumes the gauge functions $\beta(\sigma)$
and $A(\sigma)$ take their simplest form, whereby 
\beq
	\beta(\sigma) = \rho_{0},\quad\quad A(\sigma) = 0.
\eeq
Fixing $\beta$ to be a constant implies $\rho(\sigma)$ assumes the form
$\rho(\sigma) = \rho_{0} + \sigma \rho_{1}$ and the entirety of the degrees of
freedom collapses to a choice of the length scale $\lambda$ and the
parameters $\rho_{0}$ and $\rho_{1}$.  In fact, one can go further and demand
that the location of the event horizon within the compactified regime is
at $\sigma_{H} = 1$.  This choice is useful for numerical purposes and also
contrains the parameters such that
\beq
	\rho_{0} = \frac{2M}{\lambda} - \rho_{1}.
	\label{eq:minimal_gauge_rho_constraint}
\eeq
Hence the entirety of the degrees of freedom are now simply a choice of
$\lambda$ and $\rho_{1}$.  For simplicity we choose $\rho_{1} = 0$.
This minimal structure for the hyperboloidal hypersufaces are demonstrated in the 
Penrose Diagram shown in \fig{minimal_gauge_penrose}.

In our treatment of the frequency domain problem, it will be useful to introduce the mapping
from Boyer-Linequist coordinates $(t, r, \theta, \varphi)$ to hyperboloidal coordinates 
\beq
	t = \lambda \left( \tau - H(\sigma) \right),\quad
	r = \frac{2M}{\sigma},
	\label{eq:hyperboloidal_transformation_general_form}
\eeq
where it can be seen through comparison with \eqn{hyperboloidal_transformation_ingoing_form}
that the function $H(\sigma) := h(\sigma) + x(\sigma)$, where $x(\sigma) = \rstar(\sigma) / \lambda$ 
is the rescaled tortoise coordinate.
As in Paper I, we shall choose $\lambda = 4M$\footnote{The choice of $\lambda = 4M$ is motivated by the 
fact it leads to siginifcant simplifications in the case of Schwarzschild spacetime.}, which
leaves us with $\rho_{0} = 1/2$ and a height function given by
\beq
	H(\sigma) = \frac{1}{2} \left[ \ln(1 - \sigma) - \frac{1}{\sigma} + \ln\sigma \right].
	\label{eq:height_function}
\eeq

\section{Lorenz Gauge}
\label{sec:lorenz_gauge}
In this section, we provide a brief overview of the Lorenz gauge, focusing on
the Lorenz-gauge field equations for a metric-perturbation to first-order
in the mass-ratio and their subsequent decomposition onto a tensorial harmonic basis.
A more complete description of the Lorenz gauge formulation up-to-second-order in the
mass ratio can be found in \cite{Miller:2023ers}, which has been utilised in the
calculations \cite{Pound:2019lzj, Warburton:2021kwk, Wardell:2021fyy}.

\subsection{Field Equations}
We shall consider a linear metric perturbation, denoted $h_{\mu\nu}$, to a Schwarzschild background, 
$g_{\mu\nu}$ such that the full spacetime metric is given by their sum:
$\mathbf{g_{\mu\nu}} = g_{\mu\nu} + h_{\mu\nu}$.  In this context, the problem of solving the Einstein
field equations reduces to solving the linearised Einstein field equations for the metric perturbation.
The resulting perturbation satisfies a linear system of partial differential equations given by:
\begin{multline}
	\Box \barh_{\mu\nu} + 2 \tensor{R}{^{\alpha}_{\mu}^{\beta}_{\nu}}\barh_{\alpha\beta} \\
	+ g_{\mu \nu} \nabla_{\sigma} Z^{\sigma} - 2 \nabla_{(\mu} Z_{\nu)} = -16\pi T_{\mu\nu},
	\label{eq:linearised_einstein_eqn}
\end{multline}
where $\Box := g^{\mu\nu}\nabla_{\mu}\nabla_{\nu}$ is the d'Alembertian operator, 
$\nabla_{\nu}$ is the covariant derivative, and $R$ is the Riemann tensor defined on the
the background spacetime metric $g_{\mu\nu}$.  Additionally, we have introduced the notation, 
$Z_{\mu} := \nabla^{\nu}\barh_{\mu\nu}$, and the trace-reversed metric perturbation\footnote{The
``trace-reversal'' refers to the fact that ${\rm Tr}(\barh) = -{\rm Tr}(h)$.}, $\barh_{\mu\nu}$,
defined as
\beq
	\barh_{\mu\nu} := h_{\mu\nu} - \frac{1}{2} g_{\mu\nu} h.
	\label{eq:trace_reversed_metric_perturbation}
\eeq
Here, $h := g^{\mu\nu}h_{\mu\nu}$ is the trace of the metric perturbation.
The gauge freedom residing in \eqn{linearised_einstein_eqn} can be constrained by imposing
the \emph{Lorenz-gauge} condition, $Z_{\mu} = 0$, which leads to a manifestly hyperbolic wave equation,
\beq
	\Box \barh_{\mu\nu} + 2 \tensor{R}{^{\alpha}_{\mu}^{\beta}_{\nu}}\barh_{\alpha\beta}
	= -16\pi T_{\mu\nu}.
	\label{eq:lorenz_gauge_eqn}
\eeq
At first order in the mass ratio, which is the regime considered in this work, our wave equation 
can be treated as being sourced by a point particle. 
Consequently, the stress-energy tensor in \eqn{lorenz_gauge_eqn} takes on a distributional 
form, given by:
\beq
	T_{\mu\nu}(x^{\alpha}) = \mu \int^{\infty}_{-\infty} \frac{u_{\mu} u_{\nu}}{\sqrt{-{\rm det}(g)}} 
	\delta^{(4)}[x^{\alpha} - x^{\alpha}_{p}(\hat{\tau})]d\hat{\tau},
	\label{eq:point_particle_stress_energy_tensor}
\eeq
where $\mu$ is the mass of the point particle, $u_{\mu}$ is its four-velocity, $\hat{\tau}$ is the proper time
measured along the particle's worldline, and ${\rm det}(g)$ is the determinant of the metric of the background 
spacetime.  In the case of a Schwarzschild background, this is simply ${\rm det}(g) = -r^{4}\sin^{2}\theta$.
The point particle distributional form is only valid if the particle's worldline, $x^{\alpha}_{p}(\hat{\tau})$,
follows a geodesic in the background spacetime since then the gauge condition, $Z_{\mu} = 0$, is consistent
with \eqn{lorenz_gauge_eqn}; otherwise the stress-energy tensor would not be conserved.
Moreover, the breakdown of the point-particle assumption means that an altogether different treatment is 
required beyond linear-order.  One such treatment is a generalisation of the point particle, where
the source of a small body is reduced to a singularity endowed with the multipole moments of the extended body.
We shall refer the reader to Refs.~\cite{Miller:2023ers,Pound:2009sm,Pound:2014xva,Pound:2015wva} for further
details.

In this work we shall restrict ourselves to the case of the secondary body being on a quasicircular trajectory
in a Schwarzschild background.  Given the spherical symmetry of the background spacetime, we can also restrict
the secondary's motion to be confined to the equatorial plane such that 
$x_{p}^{\mu} = (t_{p}, r_{p}, \pi/2, \varphi_{p})$, without any loss in generality.  
Note hereafter we shall denote any quantity evaluated at the particle's position by the subscript $p$.
The azimuthal motion of a circular geodesic can be parameterised in terms of the coordinate time, $t$, as a
monotonically increasing function where
\beq
	\varphi_{p}(t) = \Omega_{\varphi} t,\quad\quad \Omega_{\varphi} = \sqrt{\frac{M}{\rp^{3}}}. 
	\label{eq:phi_p_t}
\eeq
The non-zero components of the four-velocity of the particle are in turn given by the straighforward analytic 
expressions,
\beq
	u^{t} = \frac{1}{\sqrt{1 - 3M/r_{p}}},\quad\quad 
	u^{\varphi} = \sqrt{\frac{M/r_{p}^{3}}{1 - 3M/r_{p}}}.
	\label{eq:four_velocity_components}
\eeq
The geodesic equations for the motion of a timelike test body lead to the first integrals for the constants of
motion for specific energy and angular momentum, ${\cal E} = -u_{t}$ and ${\cal L} = u_{\varphi}$, respectively,
which can be solved to find
\beq
	{\cal E} = \sqrt{\frac{f^{2}_{p}}{1 - 3M/r_{p}}},\quad\quad
	{\cal L} = \sqrt{\frac{M r_{p}}{1 - 3M/r_{p}}}.
	\label{eq:constants_of_motion}
\eeq

The point-particle nature of the source renders the metric perturbation in \eqn{lorenz_gauge_eqn} singular at the
worldline.  This precludes direct numerical treatment of the field equations in the vicinity of the particle using
physical retarded boundary conditions.  One way to circumvent this issue is to decompose the metric perturbation
into (tensor) spherical harmonic modes, rendering the individual modes of the metric perturbation finite and
continuous throughout the spacetime.  In Schwarzschild spacetime, the inherent spherical symmetry of the problem 
allows one to define an appropiate, second-rank tensorial harmonic basis on the 2-sphere, where $t$
and $r$ are held fixed.  The precise minutiae of the tensorial harmonic basis varies from work to work, but here 
we follow the conventions of Barack, Lousto and Sago and use their eponymous Barack-Lousto-Sago (BLS) basis
as defined in Ref.~\cite{Barack:2007tm}.  An alternative formulation by Martel and 
Poisson can be found in Ref.~\cite{Martel:2005ir}.

The BLS harmonics form a complete, orthornormal, ten-dimensional basis for any rank-2 symmetric tensor 
within Schwarzschild spacetime.  One can then expand the trace-reversed metric perturbation concurrently with
the stress energy tensor given in \eqn{point_particle_stress_energy_tensor}, and substitute the results into
the linearised Einstein field equations in \eqn{lorenz_gauge_eqn}, to obtain a set of ten coupled, hyperbolic,
partial differential equations for the harmonics of the metric perturbation.  The spherical harmonic modes
of the ten independent components, $i \in \{ 1, \dots, 10 \}$, of the metric perturbation are given by
\beq
	\barh^{(i)}_{\lm}(t, r) = \frac{r}{\mu a^{(i)}_{\ell}} \int^{2\pi}_{0} \int^{\pi}_{0} 
	\barh^{(i)}_{\alpha\beta} \eta^{\alpha\mu} \eta^{\beta\nu} Y^{(i) \lm *}_{\mu\nu} dw,
	\label{eq:metric_perturbation_harmonic_decomposition}
\eeq
where $dw = \sin\theta d\theta d\varphi$ is the volume element on the 2-sphere, 
$\eta^{\alpha\mu} = {\rm diag}(1, f^{2}, r^{-2}, r^{-2}\sin^{-2}\theta)$, and the asterisk denotes complex
conjugation.  The basis, $Y^{(i) \lm *}$, forms an orthornormal set whereby
\beq
	\int^{2\pi}_{0} \int^{\pi}_{0} \eta^{\alpha\mu} \eta^{\beta\nu} 
	Y^{(i) \lm}_{\mu\nu} Y^{(j) \ell^{\prime} m^{\prime} *}_{\alpha\beta} dw
	= \delta_{ij}\delta_{\ell \ell^{\prime}}\delta_{m m^{\prime}},
	\label{eq:bls_orthogonality}
\eeq
except for the $i = 3$ harmonic, which is also orthogonal, but the norm is $f^{2}$ as opposed to 1.
The coefficients, $a^{(i)}_{\ell}$, that appear in \eqn{metric_perturbation_harmonic_decomposition}, are defined
as
\beq
a^{(i)}_{\ell} = \frac{1}{\sqrt{2}}
	\begin{cases}
		1, & i = 1, 2, 3, 6, \\
		(\ell(\ell + 1))^{-1/2}, & i = 4, 5, 8, 9, \\
		(\ell(\ell - 1)(\ell + 1)(\ell + 2))^{-1/2}, & i = 7, 10.
	\end{cases}
\eeq
We have also followed the convention of BLS in the decomposition seen in \eqn{metric_perturbation_harmonic_decomposition},
by pulling out a factor of $r^{-1}$ to further simplify the resulting field equations.

With the multiscale expansion (see Ref.~\cite{Miller:2023ers}) in mind, one can further decompose the multipole modes 
into the Fourier domain, where one can completely decouple the radial and time dependence through a Fourier 
transform,
\beq
	\barh^{(i)}_{\lm}(t, r) = \frac{1}{2\pi}\int^{\infty}_{-\infty} \barh^{(i)}_{\lm}(\omega, r) 
	e^{-i\omega t}  d\omega.
	\label{eq:Fourier_transform}
\eeq
However, since the motion of the secondary is a geodesic, the periodicity reduces this integral over all frequencies to 
a discrete sum of Fourier harmonics, where the mode frequency is simply an overtone of the azimuthal frequency defined 
in \eqn{phi_p_t}, i.e. $\omega_{m} = m\Omega_{\varphi}$.  Hence, for circular orbits, the Fourier transform in \eqn{Fourier_transform}
becomes trivial and we are left with
\beq
	\barh^{(i)}_{\lm}(t, r) = \barh^{(i)}_{\lm}(r) e^{-i\omega_{m} t}.
	\label{eq:Fourier_transform_circular_orbit}
\eeq
Substituting the full spherical harmonic and Fourier decomposition into the linearised Einstein field equations given
in \eqn{lorenz_gauge_eqn}, whilst at the same time expanding the stress-energy tensor in \eqn{point_particle_stress_energy_tensor}
in the same manner, one can obtain a system of 10 coupled, 2nd-order, ordinary differential equations for the radial modes,
\beq
	\Box_{\lm} \barh^{(i)}_{\lm}(r) - 4 f^{-2} {\cal M}^{(i)}_{(j)} \barh^{(j)}_{\lm}(r) = {\cal J}^{(i)}_{\lm}(r).
	\label{eq:lorenz_radial_field_equations}
\eeq
Here, $\Box_{\lm}$ is the scalar wave operator,
\beq
	\Box_{\lm} := \sdiff{}{r} + \frac{f^{\prime}}{f} \diff{}{r} - \frac{1}{f^{2}} \left[ V_{\ell}(r) - \omega_{m}^{2} \right],
	\label{eq:scalar_wave_operator}
\eeq
with the prime indicating differentiation with respect to $r$, and the potential, $V_{\ell}(r)$, is given by
\beq
	V_{\ell}(r) = f(r) \left( \frac{\ell(\ell + 1)}{r^{2}} + \frac{2M}{r^{3}} \right).
	\label{eq:lorenz_potential}
\eeq
${\cal M}^{(i)}_{(j)}$ is a matrix operator that couples the different modes of the metric perturbation, and ${\cal J}^{(i)}_{\lm}(r)$
is the decomposition of the source term into the Fourier domain which schematically reads ${\cal J}^{(i)}_{\lm}(r) \propto \delta(r - r_{p})$.
This result is by no means new, and has been given in numerous references including 
\cite{Barack:2005nr, Barack:2007tm, Akcay:2010dx, Akcay:2013wfa, Wardell:2015ada, Durkan:2022fvm}.  
The radial field equations presented in \eqn{lorenz_radial_field_equations} are much more numerically tractable, and 
form the starting point for the majority of pre-existing self-force calculations in the Fourier domain.  
It is important to note that the BLS harmonics of the metric perturbation 
are not themselves coupled together, but form two disjoint sets that correspond to even- and 
odd-parity\footnote{Parity is refering to the fact 
that under the parity operation $(\theta, \varphi) \rightarrow (\pi - \theta, \varphi + \pi)$, 
the \emph{even-parity} basis elements are invariant, 
whilst the \emph{odd-parity} modes change sign.} perturbations.
This can be most easily identified from the source terms that appear on the right-hand side of 
\eqn{lorenz_radial_field_equations}, which are given in Appendix B. of Ref.~\cite{Wardell:2015ada}.  
Since, ${\cal J}^{(i = 1, \dots, 7)} \propto [Y^{\lm}(\pi/2, \varphi_{p})]^{*}$,
then the source terms and corresponding modes, $\barh^{(i = 1, \dots, 7)}$, are only \emph{non-zero} for $\ell + m = {\rm even}$. 
Similarly for ${\cal J}^{(i = 8, \dots, 10)} \propto [Y^{\lm}_{,\theta}(\pi/2, \varphi_{p})]^{*}$, 
hence the source terms, and by extension, $\barh^{(i = 8, \dots, 10)}$ are only \emph{non-zero} for $\ell + m = {\rm odd}$.

In fact, the Lorenz gauge condition, $Z_{\mu} = 0$, can be used to simplify the problem even further by reducing 
the number of components that need to be solved for independently for each $(\ell, m)$ multipole mode.  
Decomposing the gauge condition in the same manner as the field equation results in three even-parity equations for the gauge condition
\cite{Akcay:2010dx, Wardell:2015ada},
\begin{align}
	i\omega_{m}\barh^{(1)}_{\lm} =& -f\left(i\omega_{m}\barh^{(3)}_{\lm} + \barh^{(2)}_{\lm,r} + \frac{\barh^{(2)}_{\lm} - \barh^{(4)}_{\lm}}{r}\right), \label{eq:gauge_even1}\\
	i\omega_{m}\barh^{(2)}_{\lm} =& -f\barh^{(1)}_{\lm,r} + f^2\barh^{(3)}_{\lm,r} \nn \\
	&-\frac{f}{r}\left(\barh^{(1)}_{\lm}-\barh^{(5)}_{\lm}-f\barh^{(3)}_{\lm}-2f\barh^{(6)}_{\lm}\right), \label{eq:gauge_even2} \\
	i\omega_m\barh^{(4)}_{\lm} =& - \frac{f}{r}\left(r\barh^{(5)}_{\lm,r} + 2\barh^{(5)}_{\lm} + \ell(\ell + 1)\barh^{(6)}_{\lm} - \barh^{(7)}_{\lm}\right), \label{eq:gauge_even3}
\end{align}
and one odd-parity equation for the gauge condition,
\beq
	i\omega_{m}\barh^{(8)}_{\lm} = -\frac{f}{r}\left(r\barh^{(9)}_{\lm, r} + 2\barh^{(9)}_{\lm} - \barh^{(10)}_{\lm}\right).
	\label{eq:gauge_odd}
\eeq
The full hierarchical structure, where the complete set of BLS components is computed using a combination of the field equations 
and the gauge conditions, is outlined in Table I of Ref.~\cite{Akcay:2010dx, Wardell:2015ada}. 
Mode computation can be further simplified by exploiting the symmetry of the field equations under complex conjugation. 
Specifically, the modes for $m < 0$ can be derived 
from the modes for $m > 0$ using the relation $\barh^{(i)}_{\ell, -m} = (-1)^{m}\barh^{(i)*}_{\ell, m}$.

Thus far, we have outlined an appropiate framework for solving the Lorenz-gauge field equations in the Fourier domain, which has been
implemented in Refs.~\cite{Akcay:2010dx, Akcay:2013wfa, Wardell:2015ada, Miller:2023ers}.  This implementation has involved the \emph{traditional approach}
of solving the individual ODEs through the method of \emph{variation of parameters}.  In this method, one first solves for a basis of
homogeneous solutions, comprised of ``inner'' and ``outer'' solutions, that are regular at the horizon and at infinity, respectively.
This basis of solutions is then used to construct appropiate weighting coefficients by convolving the homogeneous solutions against the 
source.  It is these weighting coefficients, together with the homogeneous solutions, that are then used to construct the full, 
inhomogeneous solutions.  

This approach has been immensely successful, leading to the first second-order computations in the Lorenz gauge \cite{Pound:2020, Warburton:2021kwk, Wardell:2021fyy}. 
However, it is not without its drawbacks. Fundamentally, this methodology is not directly applicable when the background spacetime is a Kerr black hole. 
In Kerr spacetime, there is no known ansatz to reduce the partial differential equations to a set of decoupled ordinary differential equations.
Therefore one must consider an alternative formulation if one would like to solve for the Lorenz gauge metric peturbation in a Kerr spacetime.
One such approach, outlined in \cite{Dolan:2023enf, Wardell:2024yoi}, is to construct a full, Lorenz gauge metric perturbation from
a linear combination of differential operators acting on solutions to the Teukolsky equation and related scalars, which are all solutions
found through ODEs in the Fourier domain.
The analogue of this approach in the case of Schwarzschild background is to construct the metric perturbation from the Regge-Wheeler-Zerilli 
functions through an appropiate gauge transformation of the metric perturbation from Regge-Wheeler gauge to Lorenz gauge.  
This approach was outlined independently by Berndtson in \cite{Berndtson:2007gsc} and Hopper and Evans\footnote{The transformation outlined
in \cite{Hopper:2012ty} is only given for the odd-parity sector, but the transformation is equivalent to the one presented by Berndtson in 
\cite{Berndtson:2007gsc}.} in \cite{Hopper:2012ty}.  This approach has seperately been implemented by Durkan and Warburton in \cite{Durkan:2022fvm}
in the context of solving for the slow-evolution of the metric perturbation in Lorenz gauge, a crucial component of the multiscale 
expansion for second-order calculations.

From a computational perspective, the method of variation of parameters has significant limitations. 
Firstly, the homogeneous solutions are usually computed on a finite numerical grid using either asymptotic or Frobenius expansions. 
These boundary conditions, which are inconvenient to derive, can suffer from convergence problems if evaluated outside the “wave-zone”. 
A heuristic guide is that the wave-zone scales as $\sim1/\omega$ away from the source, and the boundary conditions should be 
evaluated at least this far to achieve convergence. For low-frequency radiative modes, this occurs at a very large radius, 
increasing the computational burden since the integration of the homogeneous solutions must be carried out over this entire domain, 
accumulating substantial numerical error.  Additionally, in the Lorenz gauge, the homogeneous solutions with retarded boundary 
conditions are not regular near the opposite boundary from where they are defined 
\cite{Akcay:2013wfa}. 
This requires integrating over the large numerical domain from the large outer radius, where the homogeneous solutions can 
span many orders of magnitude, leading to precision loss. Furthermore, from the gauge conditions in Eqs.~(\ref{eq:gauge_even1})-(\ref{eq:gauge_odd}), 
it is evident that in the limit $\omega \rightarrow 0$, some of the Fourier modes of the metric perturbation become linearly dependent \cite{Akcay:2010dx}.
These issues are problematic in a variation of parameters implementation in the Lorenz gauge, as the inhomogeneous solutions 
are typically constructed through (numerical) matrix inversion. Degeneracies and other such issues lead to ill-conditioned matrices 
with large condition numbers, resulting in significant numerical error.

In Paper I, we introduced a novel approach for solving the scalar self-force using hyperboloidal slices and spectral methods. 
This method demonstrated high efficiency and accuracy across a range of source types, including distributional sources, 
worldtube sources, and sources with unbounded support. It performed exceptionally well for large-radius circular orbits 
$(\rp \geq 1000M)$, and high spherical harmonic modes $(\ell \geq 100)$. To extend this framework to Lorenz-gauge gravitational perturbations, 
we adopt Berndtson’s method, which maps the Regge-Wheeler gauge metric perturbation to Lorenz gauge. 
To lay the foundation for this calculation, we will first introduce the Regge-Wheeler gauge.
\section{Regge-Wheeler Gauge}
\label{sec:regge_wheeler_gauge}
To introduce the Regge-Wheeler gauge that aligns with the notation we shall use in this work, we first consider
a decomposition of the metric perturbation into even- and odd-parity pieces, owing to the spherical symmetry of
the Schwarzschild background, $h^{\lm}_{\mu\nu} = h_{\mu\nu}^{e, \lm} + h_{\mu\nu}^{o, \lm}$, where 
the superscrips $e$ and $o$ are the even and odd parity pieces of the metric perturbation, respectively.
The perturbations can be decomposed onto a complete basis of tensorial spherical harmonics, where the even
perturbation is given by
\begin{widetext}
	\beq
		h_{\mu\nu}^{e,lm} = 
		\begin{pmatrix}
			\vspace{0.2cm} f(r) {\mathbb H}_{0}^{\lm} Y_{\lm} 
			& {\mathbb H}_{1}^{\lm} Y_{\lm} 
			& {\rm h}_{0}^{\lm} \dfrac{\partial Y_{\lm}}{\partial \theta} 
			& {\rm h}_{0}^{\lm} \dfrac{\partial Y_{\lm}}{\partial \varphi} \\
			\vspace{0.2cm} * & \dfrac{{\mathbb H}_{2}^m Y_{\lm}}{f(r)} 
			& {\rm h}_{1}^{\lm} \dfrac{\partial Y_{\lm}}{\partial \theta} 
			& {\rm h}_{1}^{\lm} X_{\lm} \\
			\vspace{0.2cm} * & * 
			& r^2 \left( {\mathbb K}^{\lm} Y_{\lm} + {\mathbb G}^{\lm} W_{\lm} \right) 
			& r^2 \sin\theta\, {\mathbb G}^{\lm} X_{\lm} \\
			\vspace{0.2cm} * & * & * 
			& r^2 \sin^2\theta \left( {\mathbb K}^{\lm} Y_{\lm} - {\mathbb G}^{\lm} W_{\lm} \right)
		\end{pmatrix}
	\eeq
\end{widetext}
where asterisks denote symmetric matrix components.  
Following Berndtson's work \cite{Berndtson:2007gsc}, the basis of spherical harmonics are given by,
\begin{align}
	W_{\lm}(\theta, \varphi) &= \psdiff{Y_{\lm}}{\theta} - \cot\theta \pdiff{Y_{\lm}}{\theta}
	- \frac{1}{\sin^{2}\theta} \psdiff{Y_{\lm}}{\varphi},\\
	X_{\lm}(\theta, \varphi) &= \frac{2}{\sin\theta} \pdiff{}{\varphi} \left(\frac{\partial Y_{\lm}}{\partial \theta}
	- \cot\theta\, Y_{\lm}\right).
\end{align}
The odd-parity perturbation is given by
\begin{widetext}
	\beq
		h_{\mu\nu}^{o,lm} = 
		\begin{pmatrix}
			\vspace{0.2cm} 0 & \hspace{1.0cm} 0 
			& \hspace{1.0cm} {\rm h}_{0}^{\lm} \csc \theta \dfrac{\partial Y_{\lm}}{\partial \varphi} 
			&  \hspace{1.0cm} -{\rm h}_{0}^{\lm} \sin \theta \dfrac{\partial Y_{\lm}}{\partial \theta} \\
			\vspace{0.2cm} * & \hspace{1.0cm} 0 
			&  \hspace{1.0cm} {\rm h}_{1}^{\lm} \csc \theta \dfrac{\partial Y_{\lm}}{\partial \varphi} 
			&  \hspace{1.0cm} -{\rm h}_{1}^{\lm} \sin \theta \dfrac{\partial Y_{\lm}}{\partial \theta} \\
			\vspace{0.2cm} * & \hspace{1.0cm} * 
			& \hspace{1.0cm} -{\rm h}_{2}^{\lm} X_{\lm} 
			& \hspace{1.0cm} {\rm h}_{2}^{\lm} \sin \theta\, W_{\lm} \\
			\vspace{0.2cm} * & \hspace{1.0cm} * 
			& \hspace{1.0cm} * 
			& \hspace{1.0cm} {\rm h}_{2}^{\lm} \sin^2 \theta\, X_{\lm}
		\end{pmatrix},
	\eeq
\end{widetext}
One can see that the even-parity perturbation, within Berndtson's conventions, has 7 degrees of freedom, manifest
through the functions ${\mathbb H}_{0}^{\lm}$, ${\mathbb H}_{1}^{\lm}$, ${\mathbb H}_{2}^{\lm}$, ${\mathbb K}^{\lm}$,
${\mathbb G}^{\lm}$, ${\rm h}_{0}^{\lm}$, ${\rm h}_{1}^{\lm}$, whilst the odd-parity perturbation has only 3 degrees of freedom
which are given by the functions ${\rm h}_{0}^{\lm}$, ${\rm h}_{1}^{\lm}$, ${\rm h}_{2}^{\lm}$.  
The Regge-Wheeler gauge is defined by setting the odd-parity function, ${\rm h}_{2}^{\lm} = 0$, and the even-parity
functions, ${\rm h}_{0}^{\lm} = {\rm h}_{1}^{\lm} = {\mathbb G}^{\lm} = 0$ 
\cite{PhysRev.108.1063, Durkan:2022fvm, Berndtson:2007gsc}.

In our context, the stress-energy tensor can also be decomposed into a tensorial spherical harmonic basis and
decomposed into Fourier modes.  This decomposition is given in Appendix.~\ref{sec:stress_energy_tensor}.
The problem of solving for the metric perturbation in Schwarzschild spacetime is therefore condensed into solving
for certain master functions, representative of the remaining degrees of freedom in the odd- and even-parity 
perturbations.  This is the approach followed by Regge and Wheeler \cite{PhysRev.108.1063}, and 
Zerilli \cite{PhysRevD.2.2141} in their seminal works.  We shall follow the same approach here, where our master
functions shall be the solutions to the generalised Regge-Wheeler-Zerilli master equation 
\cite{Leaver:1985ax,Berndtson:2007gsc,Durkan:2022fvm,Hughes:2000pf}:
\beq
	{\cal L}_{s} \psi_{s}(r) = S_{s}(r),\quad s \in \{0, 1, 2\}.
	\label{eq:generic_rwz_eqn}
\eeq
Here, ${\cal L}_{s}$ is a differential operator given by
\beq
	{\cal L}_{s} := \sdiff{}{\rstar} - P_{\ell}(r) + \omega^{2},
\eeq
and $s$ is the spin-weight.  
The potential, $P_{\ell}(r)$, has the structural form of
\beq
	P_{\ell}(r) = \frac{f(r)}{r^{2}} V_{\ell}(r),
\eeq
where the notation has been written in the suggestive form to prempt our forray into
the hyperboloidal domain.  The form of $V_{\ell}(r)$ is dependent on whether one
is solving the Regge-Wheeler master equation, in which case it is given by:
\beq
	V_{\ell}(r) = \ell(\ell + 1) + \frac{2M(1 - s^{2})}{r},
	\label{eq:rw_potential}
\eeq
or the (Regge-Wheeler-)Zerilli equation, where it is:
\begin{multline}
	V_{\ell}(r) = \frac{1}{\Lambda^{2}} \bigg[ 2\mu_{\ell}^{2} \left( \mu_{\ell} 
	+ 1 + \frac{3M}{r} \right) \\
	+ \frac{18 M^{2}}{r^{2}} \left( \mu_{\ell} + \frac{M}{r} \right) \bigg].
	\label{eq:rwz_potential}
\end{multline}
The Zerilli potential is only defined for $s = 2$, and $\Lambda = \mu_{\ell} + 3M/r$ with 
$\mu_{\ell} = (\ell + 1)(\ell + 2)/2$.

After a short derivation, one can express the radial functions, in terms of the master
solutions of \eqn{generic_rwz_eqn}.  Here, we shall only present solutions in the odd-parity 
sector, since the even-sector solutions are rather lengthy and hence we shall direct the
reader to Refs.~\cite{Zerilli:1970wzz, Nakano:2003he, Berndtson:2007gsc}.  
Recall, ${\rm h}^{\rm RW}_{2} = 0$ in RW gauge, and thus the 
non-zero odd-parity functions (for $\ell \geq 2$ and $\omega \neq 0$) are given by
\cite{Berndtson:2007gsc, Durkan:2022fvm},
\begin{align}
	{\rm h}^{\rm RW}_{0} &= \frac{f}{i\omega} \left( r \diff{\psi_{2}}{r} - \psi_{2} 
	- 16\pi f T^{o}_{22} \right) \\
	{\rm h}^{\rm RW}_{1} &= \frac{r}{f}\psi_{2}.
\end{align}
Note the inclusion of the source term, $T^{o}_{22}$, which is from the decomposition of the stress-energy tensor
given in Appendix.~\ref{sec:stress_energy_tensor}.

The main component of reconstruction of the Schwarzschild Regge-Wheeler metric perturbation, and quantities derived
from this such as the radiative fluxes, and associated self-force, are the solutions to the inhomogeneous
RWZ equations, namely \eqn{generic_rwz_eqn}.  These solutions can be computed in the same manner as the Lorenz-gauge
BLS modes in \secref{lorenz_gauge} through the now traditional method in self-force computations, the method of
variation of parameters.  We shall depart from this methodology, and instead compute the RWZ solutions through the
use of hyperboloidal slicing and spectral methods, as outlined in the frequency domain in Paper I.
Before moving onto the intricacies of this, we shall first outline the gauge transformation that allows one to
transform from the RW gauge to Lorenz gauge, thereby expressing the metric perturbation in Lorenz gauge in terms of
the solutions to the RWZ equations.
\section{Gauge Transformations}
\label{sec:gauge_transformations}
Perturbation theory inherently possesses \emph{gauge freedom} due to the diffeomorphism invariance of Einstein's field equations.
This diffeomorphism allows one to make an infinitesimal coordinate tranformation of the form
\beq
	x^{\mu} \rightarrow x^{\prime \mu} = x^{\mu} + \epsilon\, \xi^{\mu},
	\label{eq:infinitesimal_coordinate_transformation}
\eeq
where $\xi^{\mu}$ is the vector field that generates the diffeomorphism.  If we consider how this transformation
affects the metric perturbation, one recalls under an infinitesimal trasnformation of this nature a geometrical quantity
like a tensor field is perturbatively expanded as $Q = Q + \epsilon\, \delta Q$.  The perturbation, $\delta Q$,
under a gauge transformation given in \eqn{infinitesimal_coordinate_transformation} transforms as
\beq
	Q \rightarrow Q^{\prime} = Q + \epsilon (\delta Q - {\pounds}_{\xi} Q) + {\cal O}(\epsilon^{2}),
	\label{eq:tensor_gauge_transformation}
\eeq
where ${\pounds}_{\xi}$ is the Lie derivative along the vector field $\xi^{\mu}$.  Therefore the metric perturbation,
$h_{\mu\nu}$, transforms under a gauge transformation as
\begin{align}
	h_{\mu\nu} \rightarrow h^{\prime}_{\mu\nu} &= h_{\mu\nu} - {\pounds}_{\xi} g_{\mu\nu}, \\
	& = h_{\mu\nu} - 2\xi_{(\mu;\nu)}.
	\label{eq:metric_gauge_transformation}
\end{align}
If we now consider a transformation from Regge-Wheeler gauge to Lorenz gauge,
\beq
	h^{\rm L}_{\mu\nu} = h^{\rm RW}_{\mu\nu} - 2\xi_{(\mu;\nu)},
	\label{eq:regge_wheeler_to_lorenz_gauge}
\eeq
where the superscripts ``L" and ``RW" denote Lorenz gauge and Regge-Wheeler gauge respectively.

If we apply the Lorenz gauge condition, $Z_{\mu} = 0$, a short calculation leads to the following condition for the
gauge vector in order to transform to the Lorenz gauge,
\beq
	\Box\tensor{\xi}{_{\mu}} = h^{\rm RW}_{\mu\nu}.
	\label{eq:gauge_vector_condition}
\eeq
The gauge vector can then be decomposed into the frequency domain, utilising the tensorial spherical harmonics, as
\beq
	\xi_{\mu}(t, r, \theta, \varphi) = \sum_{\ell = 0}^{\infty} \sum_{m = -\ell}^{\ell} \xi^{\lm}_{\mu}(r, \theta, \varphi) e^{-i\omega t},
	\label{eq:gauge_vector_decomposition}
\eeq
which itself is split into an \emph{even-} and \emph{odd-parity} part, $\xi^{\lm}_{\mu} := \xi^{\lm, e}_{\mu} + \xi^{\lm, o}_{\mu}$,
where the even-parity part is given by
\begin{align}
    \xi^{\lm, e}_{t}(r, \theta, \varphi) &= {\mathbb M}^{\lm}_{0} Y_{\lm} 
	 + \delta_{\ell 0}\frac{{\mathbb C}_{0}f\,t}{\sqrt{4\pi}}\\
	\xi^{\lm, e}_{r}(r, \theta, \varphi) &= {\mathbb M}^{\lm}_{1} Y_{\lm},\\
    \xi^{\lm, e}_{\theta}(r, \theta, \varphi) &= {\mathbb M}^{\lm}_{2} \pdiff{Y_{\lm}}{\varphi},\\
	\xi^{\lm, e}_{\varphi}(r, \theta, \varphi) &= {\mathbb M}^{\lm}_{3} \pdiff{Y_{\lm}}{\theta},
\end{align}
and the odd-parity part is given by
\begin{align}
    \xi^{\lm, o}_{t}(r, \theta, \varphi) &= \xi^{\lm, o}_{r}(r, \theta, \varphi) = 0,\\
    \xi^{\lm, o}_{\theta}(r, \theta, \varphi) &=  
    \begin{alignedat}[t]{2}
        {\mathbb{Z}}^{\lm} \csc&\theta \pdiff{Y_{\lm}}{\varphi} \\
        &{} + \delta_{\ell 1}{\mathbb{C}_{1}} t\, r^{2} \csc\theta \pdiff{Y_{1m}}{\varphi},
    \end{alignedat} \\
    \xi^{\lm, o}_{\varphi}(r, \theta, \varphi) &= 
    \begin{alignedat}[t]{2}
        -\bigg( {\mathbb{Z}}^{\lm} &\sin\theta \pdiff{Y_{\lm}}{\theta} \\
        &{} + \delta_{\ell 1}{\mathbb{C}_{1}} t\, r^{2} \sin\theta \pdiff{Y_{1m}}{\theta} \bigg).
    \end{alignedat}
\end{align}
Here the gauge functions, ${\mathbb M}^{\lm}_{i}$, and ${\mathbb Z}^{\lm}$ are functions of $r$ only, and the 
$\mathbb{C}_{i}$ are constants of integration which, when determined, remove the residual gauge freedom and fully determine
the gauge.  One can find explicit expressions for the gauge functions in Ref.~\cite{Berndtson:2007gsc}.
Note, we have followed Berndtson's convention and choosen a covariant gauge vector, $\xi_{\mu}$, but a contravariant
form, $\xi^{\mu}$, is used by Detweiler and Poisson in \cite{PhysRevD.69.084019}.

The gauge transformation, once decomposed, can then be determined and used to push the RW metric perturbation into Lorenz gauge
for each multipole mode.  For the even-sector, these transformations are given for an individual multipole mode, $(\ell, m)$, by
\cite{Berndtson:2007gsc, Durkan:2022fvm}
\begin{align}
	{\mathbb H}_{0}^{\rm L} &= {\mathbb H}_{0}^{\rm RW} + \frac{2 i \omega}{f} {\mathbb M}_{0} 
	+ \frac{2 M}{r^{2}} {\mathbb M }_{1},\\
	{\mathbb H}_{1}^{\rm L} &= {\mathbb H}_{1}^{\rm RW} - \frac{2M}{r^{2} f} {\mathbb M}_{0} 
	+ i \omega {\mathbb M}_{1} - \diff{{\mathbb M}_{1}}{r},\\
	{\mathbb H}_{2}^{\rm L} &= {\mathbb H}_{2}^{\rm RW} - \frac{2M}{r^{2}} {\mathbb M}_{1} 
	- 2f\diff{{\mathbb M}_{1}}{r},\\
	{\mathbb K}^{\rm L} &= {\mathbb K}^{\rm RW} + \frac{2 f}{r} {\mathbb M}_{1} 
	+ \frac{2(\mu_{\ell} + 1)}{r^{2}} {\mathbb M}_{2},\\
	{\rm h}_{0}^{\rm L} &= {\rm h}_{0}^{\rm RW} - {\mathbb M}_{0} + i\omega {\mathbb M}_{2},\\
	{\rm h}_{1}^{\rm L} &= {\rm h}_{1}^{\rm RW} - {\mathbb M}_{1} + \frac{2}{r}{\mathbb M}_{2} 
	+ \diff{{\mathbb M}_{2}}{r},\\
	{\mathbb G}^{\rm L} &= {\mathbb G}^{\rm RW} - \frac{1}{r^{2}}{\mathbb M}_{2}.
	\label{eq:even_gauge_transformation_rw_to_lorenz}
\end{align}
The odd-parity sector is given by
\begin{align}
	{\rm h}_{0}^{\rm L} &= {\rm h}_{0}^{\rm RW} + i\omega {\mathbb Z},\label{eq:h0L_odd} \\
	{\rm h}_{1}^{\rm L} &= {\rm h}_{1}^{\rm RW} + \frac{2}{r}{\mathbb Z} 
	+ \diff{{\mathbb Z}}{r},\label{eq:h1L_odd} \\
	{\rm h}_{2}^{\rm L} &= {\rm h}_{2}^{\rm RW} + {\mathbb Z}.\label{eq:h2L_odd}
\end{align}

With this gauge transformation at hand, one can then directly express the desired Lorenz gauge metric perturbation in
terms of radial fields from solving the generalised Regge-Wheeler-Zerilli equations in the Fourier domain.
The expressions are cumbersome, particularly in the even-parity sector, so we present the odd-sector expressions here 
and refer the reader to the even-parity expressions in Ref.~\cite{Berndtson:2007gsc}.  
The odd-parity solutions can be written, for radiative modes ($\omega \neq 0$), and $\ell \geq 2$, as
\begin{align}
	{\rm h}_{0}^{\rm L}(r) &= \frac{1}{i\omega}\left(\psi_{1} + \frac{2\mu_{\ell}}{3}\psi_{2}\right)
	\label{eq:h0L_RW_tranformation},\\
	{\rm h}_1^{\rm L}(r) &=
	\begin{alignedat}[t]{2}
		\frac{1}{(i \omega)^2} &\bigg(-\frac{2\mu_{\ell}}{3} \diff{\psi_{2}}{r} + 
		\frac{2\psi_{1}}{r} - \frac{2 \mu_{\ell} \psi_{2}}{3r}\\
		&{}- \diff{\psi_{0}}{r} 
		+ 16 \pi f T^{o}_{12} \bigg),
		\label{eq:h1L_RW_tranformation}
	\end{alignedat} \\
		{\rm h}_2^{\rm L}(r) &= 
	\begin{alignedat}[t]{2}		
		\frac{1}{(i \omega)^2} &\bigg( r f \diff{\psi_{2}}{r} 
		+ \psi_{1} + \frac{(2\mu_{\ell} + 3)r - 6M}{3r} \psi_{2}\\
		&{} + 16 \pi f T^{o}_{22} \bigg).
		\label{eq:h2L_RW_tranformation}
	\end{alignedat}
\end{align}
Here $T^{o}_{12}$ and $T^{o}_{22}$ are source terms from the decomposition of the stress-energy tensor, see 
Appendix.~\ref{sec:stress_energy_tensor}.
Note the $(\ell, m)$ modes have been dropped here for brevity, and to align the expressions with those of 
Eqs.~(\ref{eq:h0L_odd})-(\ref{eq:h2L_odd}).  
The expressions here are interesting in and of themselves, as the Lorenz gauge metric perturbation is smooth and 
$C^{0}$-differentiable on the worldline of the particle.  If the expressions for the Lorenz gauge metric perturbation were
simply functions of the RWZ solutions, which for some spin-weights are discontinuous on the worldline, then the Lorenz
gauge metric perturbation would be not have the expected smoothness.  Hence within our expressions the differentiability is
restored by inclusion of suitable distributional terms, like the last terms in \eqn{h1L_odd} and \eqn{h2L_odd}, preserving
the expected differentiability of the metric perturbation.  This will also serve as a stringent test of our numerical
calculation, since preserving this differentiability will involve very accurate numerical solutions of the RWZ equations.

The required RWZ fields for construction of the Lorenz gauge in the odd- and even-parity sectors in summarised in 
Table.~\ref{tbl:rwz_fields}.
{\renewcommand{\arraystretch}{1.4}
\begin{table}[htb]
\begin{tabular*}{\columnwidth}{c @{\extracolsep{\fill}} c}
\hline
\hline
Even Sector: $(\ell + m) =$ even &
Odd Sector: $(\ell + m) =$ odd \\
\hline
$\psi_{0}$, $\psi_{1}$, $\psi_{2}$,
&$\psi_{1}$, $\psi_{2}$\\
$\psi_{0b}$, $M_{2af}$ $(\chi := \psi_{0b} + M_{2af})$,
&\\
\hline
\hline
\end{tabular*}
\caption{A table summarising the RWZ fields required for the even- and odd-parity reconstruction of the Lorenz gauge fields
for the radiative components $(\omega \neq 0$ and $\ell \geq 2)$.  In our treatment, unlike in
Refs.~\cite{Berndtson:2007gsc, Durkan:2022fvm}, we shall combine the purely gauge contrubutions, $\psi_{0b}$ and $M_{2af}$,
into a single gauge field.  This is detailed further in \secref{gauge_fields}. Note in the even-parity sector, $\psi_{2}$,
will be the Zerilli field with the RWZ-potential defined in \eqn{rwz_potential} as opposed to the RW potential, \eqn{rw_potential},
for the odd-parity sector.}
\label{tbl:rwz_fields}
\end{table}
}
In the odd-sector, we require the RW fields $\psi_{1}$, $\psi_{2}$ and their radial derivatives.  
The even-sector, however, is slightly more complicated.  We require the generalised RWZ fields 
$\psi_{0}$, $\psi_{1}$, $\psi_{2}$, $\psi_{0b}$, as well as purely gauge piece $M_{2af}$.  
Firstly, $\psi_{0}$, $\psi_{1}$, are the RW fields analogous to those in the odd-sector, while $\psi_{2}$ corresponds
to the RWZ field.
Specifically, $\psi_{2}$ satisfies the generalised RWZ equation with the Zerilli potential, \eqn{rwz_potential}.

The new field, $\psi_{0b}$, is defined within Berndtson's work as a RW field, with the standard potential for a 
spin weight $s = 0$, but with a substantially different source from the original scalar field, $\psi_{0}$.
For example, the source for $\psi_{0b}$ contains a derivative of a Dirac delta function, which is not present in
the source term for $\psi_{0}$.  If one considers homogeneous, source free perturbations, then the field $\psi_{0b}$ 
is equivalent to $\psi_{0}$.  

Now, let us consider the field with the peculiar moniker, $M_{2af}$.  If one takes the deriviation of 
As elucidated by Durkan and Warburton in Ref.~\cite{Durkan:2022fvm}, who eruditely reported that $M_{2af}$ generates
the scalar, tracefree perturbation of the metric, and is a purely gauge contribution.
From a more practical standpoint, the purely gauge field, $M_{2af}$, can be related to the RW master function, 
$\psi_{0}$, through the field equation \cite{Berndtson:2007gsc,Durkan:2022fvm},
\beq
	{\cal L}_{0} M_{2af}(r) = f(r)\psi_{0}(r).
	\label{eq:m2af_field_eqn}
\eeq
This is essentially the RW equation for the spin-$0$ master function, but with the added complexity of an unbounded 
source emanating from the presence of the $\psi_{0}$ source term. In Ref.~\cite{Durkan:2022fvm}, this field equation 
was solved using the technique of \emph{partial annihilators} \cite{Hopper:2012ty}, which involves applying 
an additional differential operator to reduce the field equation to a fourth-order differential equation with a compact, 
distributional source term. 
This approach allows for the construction of an inhomogeneous solution by convolving the distributional 
source term with a basis of four homogeneous solutions that span the space of the ``annihilated" fourth-order 
homogeneous equation. The technique of partial annihilators has also been demonstrated in the context of gauge 
transformations from Regge-Wheeler to Lorenz gauge in \cite{Hopper:2012ty}. 
However, as a numerical method, the technique of partial annihilators encounters the same difficulties we 
highlighted in the context of variation of parameters. In fact, the situation is even more challenging, as 
one not only needs to find an appropriate operator to ``annihilate" the source term but also an appropriate 
basis that spans the fourth-order ODE, which is by no means trivial.  We shall show that our method, as in 
Paper I is far more suited to the complexities of the unbounded right-hand side of
\eqn{m2af_field_eqn}, and provides a more efficient and accurate solution to the problem.

In the even sector, similarly to the odd-sector, the expressions construct a metric perturbation that 
is $C^{0}$-differentiable on the worldline through suitable additions of distributional terms.  
Once again, this will form a rigorous test of our numerics.

Thus far, we have outlined how one can construct a Lorenz gauge metric perturbation from the RWZ fields where such
a perturbation is decomposed onto Berndtson's spherical harmonic basis.  In \secref{lorenz_gauge}, we outlined
how for practical calculations that have been done previously, such as Refs.~\cite{Akcay:2013wfa, Wardell:2015ada},
the Lorenz gauge metric perturbation has been decomposed onto a different basis known as the BLS basis.
In this work we shall be no different.  We shall construct the Lorenz gauge metric perturbation in the BLS basis,
and as such we shall require an appropiate transformation from Berndtson's variables to the BLS basis.
This can be straightforwardly achieved by comparing the expressions appearing in Ref.~\cite{Berndtson:2007gsc} 
and Ref.~\cite{Wardell:2015ada}, as was done by in Ref.~\cite{Durkan:2022fvm}.  
The even-sector BLS components in terms of Berndtson's variables are given by
\begin{align}
	\barh^{(1)}_{\lm} &= r f \left({\mathbb H}_{0}^{\lm} + {\mathbb H}_{2}^{\lm}\right)\\
	\barh^{(2)}_{\lm} &= 2 r f {\mathbb H}_{1}^{\lm}\\
	\barh^{(3)}_{\lm} &= 2 r {\mathbb K}^{\lm}\\
	\barh^{(4)}_{\lm} &= 2 \ell(\ell + 1) {\rm h}_{0}^{\lm}\\
	\barh^{(5)}_{\lm} &= 2 f \ell(\ell + 1) {\rm h}_{1}^{\lm}\\
	\barh^{(6)}_{\lm} &= r \left({\mathbb H}_{0}^{\lm} - {\mathbb H}_{2}^{\lm}\right)\\
	\barh^{(7)}_{\lm} &= 2 r \ell(\ell + 1)(\ell + 2)(\ell - 1) {\mathbb G}^{\lm}.
	\label{eq:even_bls_berndtson}
\end{align}
The odd-sector BLS components are given by
\begin{align}
	\barh^{(8)}_{\lm} &= 2 \ell(\ell + 1) {\rm h}_{0}^{\lm}\\
	\barh^{(9)}_{\lm} &= 2 f \ell(\ell + 1) {\rm h}_{1}^{\lm}\\
	\barh^{(10)}_{\lm} &= \frac{2\ell(\ell + 1)(\ell + 2)(\ell - 1)}{r} {\rm h}_{2}^{\lm}.
	\label{eq:odd_bls_berndtson}
\end{align}
Having now provided a detailed review of the Lorenz gauge, Regge-Wheeler gauge, and the gauge transformation 
between them, we will shift our focus to the hyperboloidal method for solving the RWZ equations and, 
by extension, the Lorenz gauge metric perturbation.

\section{Hyperboloidal method to solving the RWZ equation}
\label{sec:hyperboloidal_rwz_equation}
We now specialise to the minimal gauge discussed in \secref{minimal_gauge}.  
In this coordinate system, the black hole horizon is located at $\sigma = 1$, whilst future
null-infinity lies at $\sigma = 0$.
The change to hyperboloidal time slicing leads to the following transformation for the
RWZ master functions $\psi_{s}$\footnote{Note that this choice of transformation differs from that of 
Paper I as that chooses to include the conformal factor within the rescaling factor $Z$.},
\beq
	\psi_{s} = Z \tilde{\psi}_{s},\quad Z = e^{\zeta H},\quad \zeta = -i\omega\lambda.
	\label{eq:hyperboloidal_transformation_field}
\eeq
From hereafter, we will denote the rescaled (conformal) quantities with a tilde, e.g.,
$\tilde{\psi}_{s}$.

The conformal RWZ master functions will in turn satsify equations of the form
\beq
	\tilde{\cal L}_{s} \tilde{\psi}_{s} = \tilde{S}_{s},
	\label{eq:hyperboloidal_rwz_eqn}
\eeq
where the operator and source are related to the original RWZ quantities by
\beq
	{\cal L}_{s} \psi_{s} = {\cal F} \tilde{\cal L}_{s} \tilde{\psi}_{s},
\eeq
and
\beq
	\tilde{S}_{s} = {\cal F}^{-1} S_{s}.
\eeq
Here, in line with Paper I, we have introduced the re-scaling factor 
${\cal F}$, which for this particular set-up is given simply by
\beq
	{\cal F} = \frac{Z f}{r^{2}}.
\eeq
The operator, $\tilde{\cal L}_{s}$, can be written in terms of the new hyperboloidal radial
coordinate as
\beq
	\tilde{\cal L}_{s} = \alpha_{2} \sdiff{}{\sigma} + \alpha_{1} \diff{}{\sigma} + \alpha_{0}
	\label{eq:hyperboloidal_operator},
\eeq
with $\alpha_{i}$ for $i \in \{0, 1, 2\}$ being polynomials in $\sigma$.  These coefficients are
given by
\begin{align}
	\alpha_{2} &= \sigma^{2}(1 - \sigma), \label{eq:alpha_2_rwz}\\
	\alpha_{1} &= 2\sigma(1 - 3\sigma^{2}) - \zeta(1 - 2\sigma^{2}),\\
	\alpha_{0} &= -\left(\zeta^{2}(1 + \sigma) + \zeta\sigma^{2} + V_{\ell}\right).
\end{align}

The main difference between the distributional source terms, $\tilde{S}_{s}$, and the distributional
source in Paper I, is that some of the source terms have additional distributional content for certain
spin-weights. In general, the original RWZ source terms have the form,
\beq
	S_{s}(r) = \kappa_{s}\delta(r - \rp) + \eta_{s}\delta^{\prime}(r - \rp).
	\label{eq:generic_source_term_rwz}
\eeq
The source coefficients, $\kappa_{s}$ and $\eta_{s}$, are given for circular equatorial orbits in
Appendix~\ref{sec:RWZ_source_terms}.
To obtain source terms for our conformal fields, $\tilde{\psi}_{s}$, one must consider the change of coordinates
of \eqn{generic_source_term_rwz}, in particular the change of coordinates of the Dirac delta functions.
The presence of the derivative of the Dirac delta function means one must extend the concept of a derivative to
the distributional sense.  This is done through the concept of a distributional derivative, which is defined through
integrating these distributions against a test function to yield
\beq
	f(x)\delta^{\prime}(x - a) = f(x)\delta^{\prime}(x - a) - f^{\prime}(a)\delta(x - a),
	\label{eq:distributional_derivatives}
\eeq
Another important result is the composition of a function with a smooth, continuously 
differentiable function, $g(x)$, and a Dirac delta distribution, $\delta(x - x_{i})$, is given by
\beq
	\delta[g(x)] = \sum_{i} \frac{\delta(x - x_{i})}{|g^{\prime}(x_{i})|},
	\label{eq:delta_function_composition}
\eeq
where $x_{i}$ are the roots of the function $g(x)$, which are assumed to be simple roots and are such that
$g(x_{i}) \neq 0$.  Since the source terms involve derivatives of the delta function, it is necessary to 
consider how the derivative of an appropriately smooth function interacts with the derivative of the delta function.  
Differentiating \eqn{delta_function_composition}, and utilising the 
relation in \eqn{distributional_derivatives}, one can show that
\beq
	\delta^{\prime}[g(x)] = \sum_{i} 
	\left(\frac{\delta^{\prime}(x - x_{i})}{|g^{\prime}(x_{i})|g^{\prime}(x_{i})}
	+ \frac{\delta(x - x_{i}) g^{\prime\prime}(x_{i})}{|g^{\prime}(x_{i})|g^{\prime}(x_{i})^{2}}\right).
	\label{eq:delta_prime_function_composition}
\eeq
The relations in \eqn{delta_function_composition} and \eqn{delta_prime_function_composition} can then be
used to transform the source in \eqn{generic_source_term_rwz} to our hyperboloidal coordinates.
The new source, denoted $\tilde{S}_{s}$, will then have the form
\beq
	\tilde{S}_{s}(\sigma) = \tilde{\kappa}_{s}\delta(\sigma - \sigma_{p}) 
	+ \tilde{\eta}_{s}\delta^{\prime}(\sigma - \sigma_{p}),
	\label{eq:generic_source_term_hyperboloidal}
\eeq
where the coefficients $\tilde{\kappa}_{s}$ and $\tilde{\eta}_{s}$ are related to the original coefficients by
\begin{align}
	\tilde{\kappa}_{s} &= \frac{\sigma_{p}^{2}}{2M {\cal F}}\kappa_{s}, \label{eq:kappa_tilde}\\
	\tilde{\eta}_{s} &= \frac{\sigma_{p}^{4}}{4M {\cal F}}\eta_{s}. \label{eq:eta_tilde}
\end{align}
Note, the source terms for the conformal fields that correspond to $M_{2af}$ and $\psi_{0b}$,
will involve different rescalings, as we shall outline in \secref{gauge_fields}.

We now have a full, conformal equation for the hyperboloidal fields that correspond to the RWZ fields.
In effect, we have transformed the mixed boundary problem to a regular boundary value problem.
To find the conformal solution that corresponds to the physical solution in the original coordinates, one
need only a find a regular solution to the conformal equation.  The correct boundary behaviour is enforced
by the nature of our hyperboloidal transformation.  Specifically, the $Z$ factor in the transformation in
\eqn{hyperboloidal_transformation_field} automatically includes the ingoing and outgoing behaviour of the
original retarded behaviour of the RWZ fields due to the geometrical interpretation of the height function
in Paper I.  

As in Paper I, the vanishing principal part of the differential operator at $\sigma = 0, 1$ in \eqn{alpha_2_rwz}, 
means that a regular solution satisfies the regularity condition at the boundaries of the compactified domain,
\beq
	\bigg( \alpha_{1}\tilde{\psi}_{s} + \alpha_{0}\tilde{\psi}_{s} \bigg)
	\bigg|_{\substack{\sigma = 0, \\ \sigma = 1}} = 0.
	\label{eq:regularity_conditions_rw_fields}
\eeq
As such, one requires no external boundary conditions since the source is distributional in nature, and
therefore finite at the boundaries of our compactified domain.  For the gauge pieces required to reconstruct
the Lorenz gauge metric perturbation, the source to $M_{2af}$ is unbounded, but still finite and so one 
will have a different regularity condition.  We shall outline this in the following section.
Once again, however, we shall find that geometric construction ensures there is no incoming characteristics
at our domain boundaries and thus the treatment of boundary conditions is purely behavioural rather than numerical.

The distributional source of the form of \eqn{generic_source_term_hyperboloidal} will lead to jumps in the field
and its derivative at $\sigma_{p}$, which will uniquely determine the retarded field (along with regularity condition).
Defining,
\begin{align}
	\left( \tilde{\psi}_{s, \pm} \right) \bigg|_{\sigma = \sigma_{p}}
	&= \lim_{\epsilon \rightarrow 0} \tilde{\psi}_{s, \pm}(\sigma_{p} \pm \epsilon),
	\label{eq:jump_definition} \\
	\left( \diff{\tilde{\psi}_{s, \pm}}{\sigma} \right) \Bigg|_{\sigma = \sigma_{p}}
	&= \lim_{\epsilon \rightarrow 0} \diff{\tilde{\psi}_{s, \pm}}{\sigma}(\sigma_{p} \pm \epsilon),
	\label{eq:jump_definition_derivative}
\end{align}
we find the junction conditions at the particle are given by
\begin{align}
	\left( \tilde{\psi}_{s, +} - \tilde{\psi}_{s, -} \right) \bigg|_{\sigma = \sigma_{p}}
	&= \frac{\tilde{\eta}_{s}}{\alpha_{2}}\bigg|_{\sigma = \sigma_{p}} := {\cal J}_{s}, 
	\label{eq:jump_condition}\\
	\left( \diff{\tilde{\psi}_{s, +}}{\sigma}  - \diff{\tilde{\psi}_{s, -}}{\sigma} 
	\right) \Bigg|_{\sigma = \sigma_{p}} 
	&= \frac{\tilde{\kappa}_{s}}{\alpha_{2}}
	- \left( \frac{\alpha_{1}}{\alpha_{2}}{\cal J}_{s} + \diff{{\cal J}_{s}}{\sigma} \right)
	\bigg|_{\sigma = \sigma_{p}}.
	\label{eq:jump_condition_derivative}
\end{align}
These jump conditions will need to be appropiately handled with our multi-domain spectral method to ensure the
non-differentiability of the field can be captured in the solution.
\section{Gauge Fields}
\label{sec:gauge_fields}
In Refs.~\cite{Berndtson:2007gsc,Durkan:2022fvm}, the gauge piece that enters into the metric perturbation 
is split into an inhomogeneous solution with an unbounded source, namely $M_{2af}$, and an inhomogeneous 
solution $\psi_{0b}$ with a distributional source.
This is a sensible choice in terms of solving for these fields in terms of the traditional method
of variation of parameters.
However, from the perspective of our spectral method, it is pertinent to combine these fields into a
single gauge piece, $\chi$, that is equivalent to the sum of the two pieces such that $\chi := M_{2af} + \psi_{0b}$.
This function, in turn, satisfies the similar field equation to \eqn{m2af_field_eqn},
\beq
	{\cal L}_{0}\chi(r) = f(r)\psi_{0}(r) + S_{0b}(r) := {\cal S}
	\label{eq:chi_field_eqn}
\eeq

Let us consider a similar minimal gauge hyperboloidal transformation as previously, whereby 
\beq
	\chi = M \Omega^{-1} Z \tilde{\chi}.
	\label{eq:regularised_chi}
\eeq
$\chi$ includes a factor of $\Omega^{-1} \sim r$ to take into account the lack of fall-off of the original $M_{2af}$
since this scales as $\sim r$ for large $r$.  If we perform a similar transformation as before,
we find
\beq
	{\cal L}_{0} \chi(r) = {\cal R} \tilde{\cal D}_{0} \tilde{\chi},
\eeq
and
\beq
	\tilde{\cal S} = {\cal R}^{-1} f(\sigma)\psi_{0}(\sigma).
	\label{eq:hyperboloidal_chi_field_source}
\eeq
Here the form of the differential operator is similar to previously,
\beq
	\tilde{\cal D}_{0} = \gamma_{2} \sdiff{}{\sigma} + \gamma_{1} \diff{}{\sigma} + \gamma_{0},
\eeq
whilst the new re-scaling factor is given by
\beq
	{\cal R} = Z f,
\eeq
and
\begin{align}
	\gamma_{2} &= \sigma^{3}(1 - \sigma), \label{eq:gamma_1} \\
	\gamma_{1} &= \sigma\left(\zeta - (1 + 2\zeta)\sigma^{2}\right), \label{eq:gamma_2} \\
	\gamma_{0} &= -\zeta + \sigma^{2} - \zeta^{2}\sigma(1 + \sigma) - \sigma V_{\ell}. \label{eq:gamma_3}
\end{align}
Note, we do not universally use this form of the differential equation for all our RWZ fields, since we require
precise cancellation for the source term in \eqn{hyperboloidal_chi_field_source}.

The distributional source terms, originating from the $\psi_{0b}$ field in the original nomeclature from Berndtson, have the same form
as Eqs.~(\ref{eq:kappa_tilde}) and (\ref{eq:eta_tilde}).  The difference between the source terms is the replacement of the rescaling
factor ${\cal F} \rightarrow {\cal R}$ in the expressions.  Furthermore, the unbounded nature of the source will mean that it is no longer
zero at the domain boundaries, hence the regularity condition at the boundaries will become
\beq
	\bigg( \alpha_{1}\tilde{\chi} + \alpha_{0}\tilde{\chi} \bigg)
	\bigg|_{\substack{\sigma = 0, \\ \sigma = 1}} 
	= \tilde{\cal S} \big|_{\substack{\sigma = 0, \\ \sigma = 1}}.
	\label{eq:regularity_conditions_chi}
\eeq

Similarly to \secref{hyperboloidal_rwz_equation}, the distributional nature of the field will lead to 
limited differentiability at the particle.  These junction conditions will be of the same form to previously,
\begin{align}
	\bigg( \tilde{\chi}_{+} - \tilde{\chi}_{-} \bigg) \bigg|_{\sigma = \sigma_{p}}
	&= {\cal J}_{0b}, 
	\label{eq:jump_condition_chi}\\
	\left( \diff{\tilde{\chi}_{+}}{\sigma}  - \diff{\tilde{\chi}_{-}}{\sigma} 
	\right) \Bigg|_{\sigma = \sigma_{p}} 
	&= \frac{\tilde{\kappa}_{0b}}{\alpha_{2}}
	- \left( \frac{\alpha_{1}}{\alpha_{2}}{\cal J}_{0b} + \diff{{\cal J}_{0b}}{\sigma} \right)
	\bigg|_{\sigma = \sigma_{p}}.
	\label{eq:jump_condition_chi_derivative}
\end{align}
${\cal J}_{0b}$ is defined as
\beq
	{\cal J}_{0b} := \frac{\tilde{\eta}_{0b}}{\alpha_{2}}\bigg|_{\sigma = \sigma_{p}},
	\label{eq:jump_definition_0b}
\eeq
where $\tilde{\eta}_{0b}$ is defined in the same way as Eqs.~(\ref{eq:kappa_tilde})-(\ref{eq:eta_tilde}), 
with $\eta_{0b}$ given in Appendix~\ref{sec:RWZ_source_terms}.

We now have an appropiate formalism to solve for the radiative modes ($\omega \neq 0$ and $\ell \geq 1$) of the Lorenz gauge metric
perturbation through the generalised RWZ equation and associated gauge fields.  
The static modes and monopole mode require a different treatment, which we handle analytically following 
Refs.~\cite{Akcay:2013wfa,Berndtson:2007gsc,Wardell:2015ada}.  This shall be outlined in Appendix~\ref{sec:static_modes}.
\section{Metric Reconstruction}
\label{sec:metric_reconstruction}
Having outlined the transformation of our generalised RWZ fields and gauge fields to hyperboloidal coordinates,
we now outline how to reconstruct the metric perturbation from these fields in order to then calculate the self-force
and Detweiler redshift invariant in the proceeding sections.  One can replace the RWZ field appearing in 
Berndtson's expressions from Ref.~\cite{Berndtson:2007gsc} with the hyperboloidal fields through the transformation in
\eqn{hyperboloidal_transformation_field}.  The gauge fields, $M_{2af}$ and $\psi_{0b}$, always appear in these expressions
in the combination $M_{2af} + \psi_{0b} = \chi$, and so one can replace these fields the hyperboloidal equivalent, $\tilde{\chi}$.

Simply replacing the fields with their transformed counterparts will not yield a suitably regular metric peturbation in 
the conformal space.  One can utilise the same transformation of the form of \eqn{hyperboloidal_transformation_field} for
the BLS components of the Lorenz gauge metric perturbation such that
\beq
	\barh^{(i)}_{\lm} = Z \tilde{h}^{(i)}_{\lm},
	\quad\quad i \in \{1, 2, \ldots, 10\}.
	\label{eq:hyperboloidal_bls_transformation}
\eeq
Since the expressions for the BLS modes of the metric perturbation in terms of the conformal variables also contain
up-to-second-order derivatives of the fields, one must also transform the derivatives of the fields.
The derivative of the rescaling function will involve a derivative of the height function, which can be simplified
to
\beq
	\partial_{\sigma} H = \frac{1 - 2\sigma^{2}}{2\sigma^{2}(1 - \sigma)}.
\eeq
One should note that the expressions for the BLS components of the metric perturbation in terms of the conformal field
will involve certain cancellations that are delicate, in particular those involving the gauge field, $\chi$.
Let us consider the even-parity sector fields, which after performing the transformations we have mentioned, will 
have the following form when considering the series expansion around $\sigma = 0$,
\beq
	\tilde{h}^{(1)}_{\lm} \sim -\frac{\zeta (\zeta \tilde{\chi} + \tilde{\psi}_{0})\big|_{\sigma = 0}}{\sigma} 
	+ {\cal O}(\sigma^{0}).
	\label{eq:h1_series_null_{i}nfinity}
\eeq
The presence of the $\sigma^{-1}$ term seems problematic for regularity towards null infinity.  However, considering
the field equations of the gauge field, $\chi$, in Eqs.~(\ref{eq:gamma_1})-(\ref{eq:gamma_3}) around the same limit 
one finds the trivial linear relation between the fields, $\tilde{\chi}$ and $\tilde{\psi}_{0}$,
\beq
	\tilde{\chi} \big|_{\sigma = 0} = -\frac{\tilde{\psi}_{0}}{\zeta}\bigg|_{\sigma = 0}.
	\label{eq:chi_psi0_relation}
\eeq
This linear boundary condition at null infinity will ensure the regularity of the metric perturbation at null infinity
by exactly cancelling the initially divergent term in \eqn{h1_series_null_{i}nfinity}.  In our numerical implementation,
to ensure the regularity of the metric perturbation at null infinity, we enforce this boundary condition at $\sigma = 0$
by fixing the value of the BLS component to be equal to the expansion of the field up to ${\cal O}(\sigma^{0})$ and
applying this relation in \eqn{chi_psi0_relation}.  The expression for the conformal BLS metric perturbation remains
unchanged elsewhere in the domain.  The full expressions for the BLS components of the metric perturbation in terms of the
hyperboloidal fields are given in full in Appendix \ref{sec:metric_perturbation_hyperboloidal}.
\section{Gravitational Wave Fluxes}
\label{sec:gravitational_wave_fluxes}
Before considering any metric reconstruction, or the gravitational self-force, one set of gauge invariant quantities 
that can be computed are gravitational wave fluxes.  Fluxes can be computed directly from the generalised 
hyperboloidal RWZ fields exactly at null infinity and the black hole horizon, without the need for any approximations
or expansions around the asymptotic regions.  A relatively short derivation, based on the scalar field case derived in
Paper I, yields the following expressions for the energy flux through the null infinity and the horizon in terms of our
hyperboloidal fields for the generic spin-weight $s$,
\begin{align}
	\dot{E}^{\scri^{+}}_{s} = \sum^{\infty}_{\ell = 2} \sum^{\ell}_{m = -\ell}
	\frac{g_{\lm}(\zeta)}{64\pi\mu}\frac{(\ell + 2)!}{(\ell - 2)!} \big|\zeta \psi_{s}^{\lm}\big|^{2}_{\sigma = 0}, 
	\label{eq:energy_flux_null_{i}nfinity} \\
	\dot{E}^{{\cal H}}_{s} = \sum^{\infty}_{\ell = 2} \sum^{\ell}_{m = -\ell}
	\frac{g_{\lm}(\zeta)}{64\pi\mu}\frac{(\ell + 2)!}{(\ell - 2)!} \big|\zeta \psi_{s}^{\lm}\big|^{2}_{\sigma = 1}. 
	\label{eq:energy_flux_horizon}
\end{align}
Here we have introduced the function $g_{\lm}(\zeta)$, which has the functional form 
\cite{Barack:2010tm,Hopper:2010uv,Durkan:2022fvm}:
\beq
	g_{\lm}(\zeta) = 
	\begin{cases}
		|\zeta\lambda^{-1}|^{2}, & \ell + m \text{ is even}, \\
		\lambda, & \ell + m \text{ is odd}.
	\end{cases}
\eeq
One can also form similar expressions for the angular momentum fluxes for generic spin-weights,
\begin{align}
	\dot{J}^{\scri^{+}}_{s} = \sum^{\infty}_{\ell = 2} \sum^{\ell}_{m = -\ell}
	\frac{i m\, j_{\lm}(\zeta)}{64\pi\mu}\frac{(\ell + 2)!}{(\ell - 2)!}
	\big|\psi_{s}^{\lm}\big|^{2}_{\sigma = 0}, 
	\label{eq:angular_flux_null_{i}nfinity} \\
	\dot{J}^{{\cal H}}_{s} = \sum^{\infty}_{\ell = 2} \sum^{\ell}_{m = -\ell}
	\frac{i m\, j_{\lm}(\zeta)}{64\pi\mu}\frac{(\ell + 2)!}{(\ell - 2)!} 
	\big|\psi_{s}^{\lm}\big|^{2}_{\sigma = 1},
	\label{eq:angular_flux_horizon}
\end{align}
where the function $j_{\lm}(\zeta)$ is given by
\beq
	j_{\lm}(\zeta) = 
	\begin{cases}
		\zeta\lambda^{-1}, & \ell + m \text{ is even}, \\
		-\zeta^{-1}\lambda^{2}, & \ell + m \text{ is odd}.
	\end{cases}
\eeq
Note the expressions for the energy and angular momentum fluxes do not include any contributions for the
multipole modes below $\ell = 2$, or any contributions from the static modes where $m = 0$.  If we are considering
the gravitational wave fluxes then we shall only consider the subset of energy and angular momentum fluxes for $s = 2$.
Thus, from hereafter we shall drop the spin-weight index $s$ from the expressions for the fluxes, e.g. 
$\dot{E}^{\scri^{+}}_{s = 2} = \dot{E}^{\scri^{+}}$.  For completeness, we have also checked the fluxes against 
resources such as the Black Hole Perturbation Toolkit (BHPToolkit) \cite{BHPToolkit} to ensure the correctness of our 
methods.
\section{Self-force}
\label{sec:self_force}
Once we have the components of the metric and their associated derivatives, we have the ability to compute the gravitational self-force.
The crucial element of this computation is the divergence of the retarded self-force at the worldline that necessitates delicate
regularisation to yield the physical self-force.   More precisely, Detweiler and Whiting demonstrated the
physical self-force can be computed from the derivative of a \emph{regular} metric perturbation \cite{Detweiler:2002mi}, 
$\barh^{\rm R}_{\mu\nu}$,
\beq
	F^{\mu}_{\rm self}(x_{p}) = \mu k^{\mu\nu\gamma\delta}\nabla_{\delta}\barh^{\rm R}_{\nu\gamma}(x_{p}),
	\label{eq:self_force_definition}
\eeq
where $k^{\mu\nu\gamma\delta}$ is defined as
\begin{align}
	k^{\mu\nu\gamma\delta} &= \frac{1}{2}g^{\alpha\delta}u^{\beta}u^{\gamma} - g^{\alpha\beta}u^{\gamma}u^{\delta}
	- \frac{1}{2}u^{\alpha}u^{\beta}u^{\gamma}u^{\delta} \nn \\
	&+ \frac{1}{4}u^{\alpha}g^{\beta\gamma}u^{\delta} + \frac{1}{4}g^{\mu\delta}g^{\nu\gamma}.
	\label{eq:projection_operator}
\end{align}
As outlined in \cite{Barack:2001bw},  $k^{\mu\nu\gamma\delta}$ is a \emph{kinematic} tensor constructed such that the self-force
is orthogonal\footnote{This orthogonality ensures the mass is constant along the worldline.} to the four-velocity of the particle, 
e.g., $F^{\alpha}_{\rm self}u_{\alpha} = 0$.

A suitable regular metric perturbation can be found in the neighbourhood of the particle through the subtraction of a singular
perturbation, $\barh^{\rm S}_{\mu\nu}$ from the retarded metric peturbation,
\beq
	\barh^{\rm R}_{\mu\nu}(x_{p}) := \lim_{x \rightarrow x_{p}} 
	\left[ \barh^{\rm ret}_{\mu\nu}(x) - \barh^{\rm S}_{\mu\nu}(x) \right].
	\label{eq:regular_field_definition}
\eeq
This component of the metric perturbation is a solution to the sourced Lorenz-gauge field equations as described in 
\eqn{lorenz_gauge_eqn}. However, it must also ensure that the regular metric perturbation, 
$\barh^{\rm R}_{\mu\nu}(x_{p})$, remains a smooth vacuum solution to the same field equations.
In this way, the retarded and singular metric perturbations must diverge at the worldline in precisely the same manner,
and therefore, the self-force can be written as the difference between the retarded and singular metric perturbations
\begin{align}
	F^{\mu}_{\rm self}(x_{p}) &= \mu \lim_{x \rightarrow x_{p}}
	\left[ k^{\mu\nu\gamma\delta}\nabla_{\delta} \left( \barh^{\rm ret}_{\nu\gamma}(x) 
	- \barh^{\rm S}_{\nu\gamma}(x) \right) \right] \nn \\
	&= \lim_{x \rightarrow x_{p}}
	\left[ F^{\mu}_{\rm ret}(x) - F^{\mu}_{\rm S}(x) \right],
	\label{eq:self_force_retarded_singular_split}
\end{align}
where we have defined
\beq
	F^{\mu}_{\rm ret/S}(x) := \mu k^{\mu\nu\gamma\delta}\nabla_{\delta} \barh^{\rm ret/S}_{\nu\gamma}(x).
	\label{eq:self_force_retarded_singular_definition}
\eeq
There is a subtlety here. Initially, the operator $k^{\mu\nu\gamma\delta} := k^{\mu\nu\gamma\delta}(x_{p})$,
is only defined on the worldline.  If one is to define the kinematic tensor away from the wordline, as in 
\eqn{self_force_retarded_singular_definition}, then one can extend the defintion as done so in Refs.~\cite{Barack:2001gx, 
Barack:2010tm}, where $k^{\mu\nu\gamma\delta} := k^{\mu\nu\gamma\delta}(x, x_{p})$ is defined at the field point $x$.
Within this definition, $k^{\mu\nu\gamma\delta}$ has the same expression as \eqn{projection_operator}, with 
$g^{\mu\nu}$ evaluated at $x$ and $u^{\mu}$ evaluated at $x_{p}$.

In terms of a practical computation, one must tame this divergence of the retarded and singular metric perturbation
in such a manner as to only deal with finite contributions.  
At the level of the full metric perturbation in \eqn{self_force_retarded_singular_split} this is not possible.  
Therefore, one must reformulate the GSF problem in a rigorous manner to do this subtraction.

One such method to overcome this subtraction problem is to do this subtraction at the level of the field equations
themselves, thereby solving directly for the regular piece of the metric perturbation.  This is known as the
\emph{effective-source} approach and is essential for doing GSF computations up to second-order \cite{Miller:2023ers}.
To do this, \eqn{lorenz_gauge_eqn} is rewritten in terms of the regular and singular metric perturbation through
the split given in \eqn{regular_field_definition}.  However, the local singularity that appears through the singular 
field is not accessible, and thus one must replace the singular field with a \emph{puncture} field, a local expansion
of the singular field that is truncated at some finite order.  This puncture field will only have compact support in 
the vicinity of the worldline so it is usually translated to zero at some arbitrary, but finite distance from the
worldline.  One can ensure this translation back to the full metric perturbation by solving within a \emph{worldtube method}
as in Ref.~\cite{Miller:2023ers} or via an appropiate \emph{window function} as in Ref.~\cite{Wardell:2015ada}.
By definition, the puncture field will only be equal to the singular field up to some finite expansion order,
and therefore the right-hand-side of the field equations will not be zero as is the case for the regular field.
Hence, one will be left with a set of inhomogeneous field equations with an \emph{effective-source}.
This approach has been demonstrated to work in the scalar case with hyperboloidal slicing and spectral methods
in Paper I and Ref.~\cite{PanossoMacedo:2024pox}.  

Here we shall forgo this method and instead adopt the first-order specific scheme of \emph{mode-sum regularisation}.
This is the traditional method pioneered by Barack and Ori in \cite{Barack:1999wf}, following studies of the singular
structure of the metric perturbation by Mino \emph{et al.} in \cite{Mino:1996nk}.
The central premise of the mode-sum method is the observation that whilst the retarded and singular metric 
perturbations are themselves divergent, their multipole modes are finite.  If one individually decomposes the
multipole modes of the singular and regular metric perturbations, the subtraction we see in \eqn{self_force_retarded_singular_split}
can be done on a mode-by-mode basis, leaving an individual regularized mode.  
These finite modes are then summed to obtain a physical self-force.

At this juncture, a crucial question arises: which multipole basis should be used for decomposing the fields? 
While it might initially seem natural to select a basis of tensorial harmonics, given the decomposition 
presented in \secref{lorenz_gauge}, it is important to note that, prior to Ref.~\cite{Wardell:2015ada}, most 
existing calculations relied on a mode-sum prescription utilizing a scalar spherical harmonic basis.
This choice, more natural when tackling the scalar self-force problem, in the gravitational case requires
a projection of the tensor harmonic modes of the retarded field, $F^{{\ell}, {\rm ret}}_{\mu}$, 
onto a basis of scalar spherical harmonic modes, $F^{\hat{\ell}, {\rm ret}}_{\mu}$.  Here, we have
used the notation of Ref.~\cite{Wardell:2015ada}, denoting $\hat{\ell}$ super/subscript as a scalar-harmonic
multipole contribution (summed over $m$), so as to not be confused with tensor harmonic multipoles
that are denoted with $\ell$.  

This projection results in troubling mode-coupling between the harmonics that 
leads to substantial computational inefficiency.  
To compute the self-force, for example, up to a given spherical harmonic mode, $\hat{\ell}$,
one must compute $\hat{\ell} + 3$ tensor harmonic modes \cite{Wardell:2015ada}.
Therefore, for a given retarded field scalar $\hat{\ell}$-modes that one wishes to regularise to obtain
a physical quantity, one must compute a greater number of tensorial modes, $\ell$, which are subsequently
lost during the projection to scalar spherical harmonics.  Furthermore, this projection involves 
mode-coupling formulas that are irksome to deal with since they are quite lengthy and need to be derived
independently for each component of the self-force.

We therefore follow the approach in Sec.VI of Ref.~\cite{Wardell:2015ada}, and pursue a mode-sum 
regularisation approach directly with tensor-harmonic modes, circumventing the unnecessary step of
projecting onto scalar spherical harmonics.  Let us first consider the mode-sum expression for the
regular field,
\beq
	h^{\rm R}_{\mu\nu} = \sum^{\infty}_{\ell = 0} \left[ h^{\rm ret,\, \ell}_{\mu\nu} 
	- h^{\rm S,\, \ell}_{\mu\nu} \right],
	\label{eq:regular_field_mode_sum}
\eeq
where $h^{\rm ret,\, \ell}_{\mu\nu}$ are the retarded field modes evaluated at the particle.
The tensor harmonic decomposition of the singular field has the form 
$h^{\rm S,\, \ell}_{\mu\nu} = h^{[0]}_{\mu\nu} + {\cal O}(\ell^{-2})$ for the non-spinning case\footnote{If 
one was to consider the more complex case of spinning bodies, as in Ref.~\cite{Mathews:2021rod}, then the leading term
would be one order higher in $\ell$, 
i.e. $h^{\rm S,\, \ell}_{\mu\nu} = (2\ell + 1) h^{[-1]}_{\mu\nu} + h^{[0]}_{\mu\nu} + {\cal O}(\ell^{-2})$},
hence the metric perturbation and its derivative have the schematic form
\begin{align}
	h^{\rm R}_{\mu\nu} &= \sum^{\infty}_{\ell = 0} \left[ h^{\rm ret,\, \ell}_{\mu\nu}
	- h^{[0]}_{\mu\nu} \right], \label{eq:regular_metric}\\
	h^{\rm R}_{\mu\nu, \gamma} &= \sum^{\infty}_{\ell = 0} \left[ h^{\rm ret,\, \ell}_{\mu\nu, \gamma}
	- (2\ell + 1) h^{[-1]}_{\mu\nu, \gamma} - h^{[0]}_{\mu\nu, \gamma} \right] \label{eq:regular_metric_derivative}
\end{align}
where $h^{[-1]}_{\mu\nu}$ and $h^{[0]}_{\mu\nu}$ are the leading regularisation parameters.

The self-force can similarly expressed as a mode-sum such that
\beq
	F_{\mu}^{\rm self} = \sum^{\infty}_{\ell = 0} \left[ F_{\mu, \ell \pm}^{\rm ret}
	- (2\ell + 1) F_{\mu, \pm}^{[-1]} - F_{\mu}^{[0]} \right] - D_{\mu},
	\label{eq:self_force_mode_sum}
\eeq
where the $\ell$-modes of the ``retarded force" can be computed from the metric perturbation,
\beq
	F_{\mu, \ell \pm}^{\rm ret} = \mu \lim_{r \rightarrow \rp^{\pm}} \sum^{\ell}_{m = -\ell}
	k^{\mu\nu\gamma\delta}\nabla_{\delta} \barh^{\rm ret}_{\nu\gamma}(x) 
	\bigg|_{\substack{\theta = \pi/2, \\ \varphi = 0}}.
\eeq
Here $F_{\mu, \pm}^{[-1]}$, $F_{\mu}^{[0]}$, and $D^{\mu}$ are the regularisation parameters for the
self-force.  The $\pm$ notation indicates that the tensorial mode contributions to the retarded force are 
finite and $C^{-1}$-differentiable at the particle’s position. Consequently, their one-sided limits, 
as $r \rightarrow \pm \rp$, differ depending on the direction of approach.

The regularisation parameters are constants in $\ell$ that are derived from a local expansion of the 
Detweiler-Whiting singular field.
This procedure is intricate, and for the case under consideration, it has already been carried 
out for circular orbits in Schwarzschild spacetime by Wardell \emph{et al.} in 
Ref.~\cite{Wardell:2015ada}. For a more detailed explanation, we direct the reader to that reference.
We shall merely use their results that we will summarise here.

At the level of the metric perturbation, the regularisation parameters from Eqs.~(\ref{eq:regular_metric}) and 
(\ref{eq:regular_metric_derivative}) are given explicitly in Appendix~\ref{sec:metric_perturbation_regularisation_parameters}.
These expressions can also be used to compute the regularisation parameters for the self-force in \eqn{self_force_mode_sum},
\begin{subequations}
	\begin{eqnarray}
		F^{r\pm}_{[-1]} &=& \mp \frac{\mu^2}{2 \rp^2}\left(1-\frac{3M}{\rp}\right)^{1/2}
		\pm \bigg[\frac{2 \mu^2 M (2 M-\rp)}{\rp^{5/2} (\rp-3 M)^{3/2}}\bigg]_{\ell<1} \nn \\
		&& \qquad \pm \bigg[\frac{\mu^2 M^2}{2 \rp^{5/2} (\rp-3 M)^{3/2}}\bigg]_{\ell<2},
		\\
		F^r_{[0]} &=& \frac{\mu^2 \rp {\cal E}^2}{\pi({\cal L}^2 + \rp^2)^{3/2}}\left[E - 2K\right] \nn \\
		&& - \frac{1}{(2\ell-1)(2\ell+3)} \times \nn \\
		&& \qquad \frac{2 \mu^2 M (\rp-2M)^{1/2} [(6\rp-17M) E + 2 M K]}{\pi \rp^3 (\rp-3M)^{3/2}} \nn \\
		&& + \frac{1}{(2\ell-3)(2\ell-1)(2\ell+3)(2\ell+5)} \times \nn \\
		&& \qquad \frac{105 \mu^2 M^2 (\rp-2M)^{1/2} (E+2K)}{\pi \rp^3 (\rp-3M)^{3/2}}.
		\label{eq:self_force_regularisation_parameters}
	\end{eqnarray}
\end{subequations}
Here $K := \int^{\pi/2}_{0} \left( 1 - \frac{M}{\rp - 2M} \sin^{2}x \right)^{-1/2} dx$ and 
$E := \int^{\pi/2}_{0} \left( 1 - \frac{M}{\rp - 2M} \sin^{2}x \right)^{1/2} dx$ are the complete elliptic 
integrals of the first and second kind, respectively.  
We have also been careful to omit a factor of the small mass $\mu$ in the expressions for reasons of clarity.
In some of the regularisation parameters, we have included the subscript notation $[\ell \geq n]$ or $[\ell < n]$ 
to indicate where certain terms are only non-zero for $\ell \geq n$ or $\ell < n$ respectively.  We should also note that
in the Lorenz gauge, $D_{\mu}$ vanishes if one derives these regularisation parameters from the Detweiler-Whiting singular field
\cite{Barack:2005nr, Barack:2007tm}.

Furthermore, only the radial component of the self-force requires regularisation parameters becuase this is the only
component with singular contributions.  The temporal and angular components of the self-force 
converge exponentially in the mode-sum, whereas the radial component converges more slowly, 
at a rate of $\ell^{-2}$ \cite{Wardell:2015ada}. Consequently, one important consideration here is that truncating our 
formally infinite sums at a particular multipole mode, $\ellmax$, will introduce an error in our radial self-force 
calculations. This error can be mitigated by introducing a ``tail correction", as described in 
Refs.~\cite{Barack:2007tm,Akcay:2010dx,Akcay:2013wfa,Warburton:2021kwk,Shah:2012gu,Thompson:2018lgb}, which accounts for the 
missing higher-order modes by numerically fitting the $\ell$-falloff. We define this tail correction for the 
radial component as
\beq
	F^{r}_{\rm self} = F^{r}_{\ell \leq \ellmax} + F^{r}_{\ell > \ellmax}
	\label{eq:radial_self_force_definition}
\eeq
where we have defined,
\beq
	F^{r}_{\ell \leq \ellmax} := \sum^{\ellmax}_{\ell = 0}F^{r, \ell}_{\rm reg}, \quad\quad
	F^{r}_{\ell > \ellmax} := \sum^{\infty}_{\ellmax + 1}F^{r, \ell}_{\rm reg},
	\label{eq:radial_self_force_truncation}	
\eeq
and the regularized radial component of the self-force is given by
\beq
	F^{r, \ell}_{\rm reg} := F^{r, \ell \pm}_{\rm ret} - (2\ell + 1) F^{r, \pm}_{[-1]} - F^{r}_{[0]}.
	\label{eq:radial_self_force_regularized}
\eeq
$F^{r}_{\ell \leq \ellmax}$ is computed numerically with the mode-sum method, while $F^{r}_{\ell > \ellmax}$ is
found by numerically fitting the high $\ell$-falloff of $F^{r, \ell}_{\rm reg}$ using the ansatz
\beq
	F^{r}_{\ell > \ellmax} \simeq \sum^{k_{\rm max}}_{k = 1} \frac{F^{r}_{[2k]}}{P_{2k}(\ell)}.
	\label{eq:fitting_ansatz}
\eeq
Here the summand begins at $k = 1$ to match the $\ell$-falloff the regularised data, 
$\{F^{r}_{[2k]}\}^{k_{\rm max}}_{k = 1}$, are constant fitting parameters, and $P_{2k}(\ell)$ is a polynomial
of order $\ell^{2k}$.  The polynomial is choosen such that
\beq
	\sum_{\ell = 0}^{\infty} \frac{1}{P_{2k}(\ell)} = 0,
	\label{eq:polynomial_condition}
\eeq
meaning that each term in \eqn{fitting_ansatz} will vanish when summed over all $\ell$.
Following the approach in Refs.~\cite{Barack:2007tm,Shah:2012gu,Thompson:2018lgb}, we use the polynomial
\beq
	P_{2k}(\ell) = \prod_{k^{\prime} = 0}^{k} (2\ell - 2k^{\prime} - 1)(2\ell + 2k^{\prime} + 3),
	\label{eq:regularisation_polynomial}
\eeq
which is a logical choice, as its terms appear within the higher-order regularisation parameter in 
\eqn{self_force_regularisation_parameters} when $k^{\prime} = 0$.  Bringing this all together, our
final expression for the radial self-force with accelerated convergence is given by
\begin{align}
	F^{r}_{\rm self} &= \sum^{\ellmax}_{\ell = 0} 
	\left[ F^{r, \ell \pm}_{\rm ret} - (2\ell + 1) F^{r, \pm}_{[-1]} - F^{r}_{[0]} \right] \\
	&+ \sum_{\ell_{\rm max} + 1}^{\infty} \sum_{k = 1}^{k_{\rm max}} \frac{F^{r}_{[2k]}}{P_{2k}(\ell)}
	+ {\cal O}(\ell^{-(2k_{\rm max} + 1)}).
	\label{eq:radial_self_force_accelerated}
\end{align}
This is the similar expression to what is used in \cite{Thompson:2018lgb} and \cite{Shah:2012gu}.
The value of the cut-off, $k_{\rm max}$, and the number of modes used in the numerical fitting is somewhat open
to experimentation, and at some point as we shall see will depend on the numerical precision one can calculate
the multipole modes to.  We shall discuss this in more detail in \secref{results}.

Finally, we discuss the effects of the self-force on the motion of the secondary body.  
In this work, our secondary body can be assumed to be a point-particle within the background spacetime,
obeying the forced geodesic equation,
\beq
	\mu u^{\mu}\nabla_{\mu}u^{\nu} = F^{\nu}_{\rm self}.
	\label{eq:forced_geodesic_equation}
\eeq
The spherical symmetry of our systems allows one to immediately see that $F^{\theta}_{\rm self} = 0$ will vanish.
The remaining components of the self-force can be further constrained by ensuring the four-velocity
remains normalised, $u_{\mu}u^{\mu} = -1$, along the worldline.  This will yield a orthogonality condition
on the self-force such that $u_{\mu}F^{\mu}_{\rm self} = 0$.  The orthogonality condition can be used to see
the interdependency of the self-force components, as for the circular orbits we are considering one can 
expand this relation to find: $u_{t}F^{t}_{\rm self} + u_{\varphi}F^{\varphi}_{\rm self} = 0$.
Thus, the temporal and azimuthal components of the self-force can be directly related through linear order in
the mass-ratio \cite{Barack:2001ph} as $F^{\varphi}_{\rm self} = ({\cal E} / {\cal L}) F^{t}_{\rm self}$.
This relation simplifies the self-force calculation, as only the temporal component needs to be computed to derive the 
azimuthal component. Additionally, it serves as an internal consistency check for our calculation.

At first-order in the mass-ratio, one can consider the motion of the secondary body to be instantaneously geodesic, but
over time the self-force will cause the orbit to deviate from the geodesic path.  
The long secular drift in the value of the constants of motion, energy $({\cal E})$ and angular momentum $(\cal L)$,
is the dissipative effect of the self-force.  This effect is described by considering the temporal and azimuthal components
of \eqn{forced_geodesic_equation},
\beq
	\diff{{\cal E}}{t} = - (\mu u^{t})^{-1} F_{t}^{\rm self}, \quad\quad
	\diff{{\cal L}}{t} = (\mu u^{t})^{-1} F_{\varphi}^{\rm self},
	\label{eq:constants_of_motion_evolution}
\eeq
where we have expressed the derivative in terms of proper time along the worldline in terms of the 
Schwarzschild coordinate time, $t$.  If we take the \emph{adiabatic approximation}, whereby $T_{\rm rad} \gg T_{\rm orb}$ 
such that the effect of radiation reaction is considered to be small over the timescale of the orbit, then $d{\cal E}/{dt}$
and $d{\cal L}/{dt}$ can be considered the average rate of change of energy and angular momentum
over an orbital period.  Owing to energy conservation, the loss of energy and angular momentum must be precisely 
balanced by the radiation of these quantities in gravitational wave emission to null-infinity and the primary 
black hole horizon.  This means one can write down a balance law from \eqn{constants_of_motion_evolution} and
Eqs.~(\ref{eq:energy_flux_null_{i}nfinity})-(\ref{eq:angular_flux_horizon}),
\begin{align}
	\dot{E} &:= \dot{E}^{\scri^{+}} + \dot{E}^{\cal H} = \mu \dot{\cal E} = - \frac{F_{t}^{\rm self}}{u^{t}}, 
	\label{eq:energy_balance_law} \\
	\dot{J} &:= \dot{J}^{\scri^{+}} + \dot{J}^{\cal H} = \mu \dot{\cal L} = \frac{F_{\varphi}^{\rm self}}{u^{t}}.
	\label{eq:angular_balance_law}
\end{align}
These quantities provide an excellent consistency check for our calculations in two key ways. 
First, the full local self-force is gauge-dependent, and the balance laws provide a gauge-invariant check of our
calculation.
Second, while the local self-force is evaluated along the worldline and relies 
on the metric reconstruction procedure, the fluxes are computed at the boundaries of our compactified domain 
directly from the RWZ fields. Verifying the balance laws in \eqn{energy_balance_law} and \eqn{angular_balance_law} 
serves as a robust qualitative check on the accuracy of our calculation.
\section{Detweiler Redshift}
\label{sec:detweiler_redshift}
The self-force, while central to the calculations presented in this work, is inherently complicated by 
its manifest gauge dependence \cite{Barack:2001ph}. Consequently, the focus is often on gauge-invariant 
characterisations of the self-force effects. In the context of quasi-circular orbits in 
Schwarzschild spacetime, the primary quantity of interest is the 
\emph{Detweiler-Redshift invariant} \cite{Detweiler:2008ft}.

One can understand this invariant by a short, heuristic argument.  The orbit of the secondary, whilst influenced
by the self-force, will be accelerated away from its original geodesic path in the background spacetime.  However,
in the \emph{effective metric} in the smooth, physical perturbed spacetime, the motion will be geodesic.
The four-velocity with respect to proper time along this effective geodesic can be written as
$\tilde{u}^{\mu} = \{\tilde{u}^{t}, 0, 0, \tilde{u}^{\varphi}\}$, where the individual components can be shown
to be gauge invariant under transformations that preserve the helical symmetry of the effective metric and have
the form $\xi^{\mu} = {\cal O}(\epsilon)$ \cite{Barack:2001ph, Akcay:2012ea}.
One can then define a gauge invariant quantity as the ratio of proper time in the effective metric to
coordinate time (on the accelerated worldline),
\beq
	z := (\tilde{u}^{t})^{-1} = \diff{\tilde{\tau}}{t}.
	\label{eq:detweiler_redshift_definition}
\eeq
By virtue of the gauge invariance of the other quantities entering into this expression, then $z$ 
is also gauge invariant. 
In this work, we shall call the quantity, $z$, the Detweiler-Redshift invariant, rather than its 
inverse, as has sometimes been done elsewhere in the literature.  A more in depth discussion of the redshift 
invariant, and its gauge invariance up-to-and-including second-order has been explored by Pound in 
Ref.~\cite{Pound:2014koa}. 

The redshift invariant, $z$, can be expressed as an expansion in the small-mass ratio, such that
$z = z_{0} + \epsilon \Delta z + {\cal O}(\epsilon^{2})$.  The zeroth-order piece, $z_{0}$, is the geodesic
limit, $z_{0} = (1 - 3M/\rp)^{-1/2}$, and hence most numerical comparisons are made with the first-order correction,
which for the case of quasi-circular orbits in Schwarzschild spacetime can be written as
\beq
	\Delta z = \left(1 - \frac{3M}{\rp}\right)^{-1/2} \left[  
	\frac{1}{2}\tilde{u}^{\mu}\tilde{u}^{\nu} h^{\rm R}_{\mu\nu} \right].
	\label{eq:detweiler_redshift_regge_wheeler}
\eeq
The double contraction of the metric perturbation with the four-velocity is often denoted 
$h^{\rm R}_{uu} := h^{\rm R}_{\mu\nu}\tilde{u}^{\mu}\tilde{u}^{\nu}$, and we shall use that shorthand here.
We should also note that $h^{\rm R}_{uu}$ is invariant for circular orbits \cite{Akcay:2015pza}.

The Detweiler redshift, however, remains gauge invariant only within a suitable class of gauges that 
preserve the periodic nature of the orbit (and, in this case, helical symmetry) 
\emph{and} whose monopole contributions to the metric are asymptotically flat.
There presents an additional subtlety when computing this invariant in Lorenz gauge.  
First discussed in \cite{Sago:2008id}, a gauge transformation from Regge-Wheeler
gauge to Lorenz gauge \emph{does not} leave the redshift gauge invariant since the Lorenz gauge metric
perturbation is not asymptotically flat i.e., it does not vanish in the limit to spatial infinity.
In particular, the $h^{\rm L}_{tt}$ component of the Lorenz gauge metric perturbation is non-zero in the limit
$r \rightarrow \infty$, whilst the other components remain regular.  This irregular behaviour of the $h^{\rm L}_{tt}$
originates from the static piece of the monopole perturbation and, as a result, is only dependent on the 
orbital parameters and has no angular dependence \cite{Sago:2008id, Akcay:2015pza}.  

We discuss the monopole in more detail in Appendix.~\ref{sec:monopole_static}, and show how this artifact can
be removed by a suitable gauge transformation.  Here we shall simply state the final result.
Introducing this transformation, and evaluating everything in terms of the gauge invariant radius,
$x := (M\Omega_{\varphi})^{2/3}$, one finds
\beq
	\Delta z (x) = \sqrt{1 - 3x} \left[\frac{1}{2} h^{\rm R, L}_{uu} 
	- \frac{x(1 - 2x)}{(1 - 3x)^{3/2}} \right].
	\label{eq:detweiler_redshift_gauge_invariant}
\eeq
Note that the $h^{\rm R, L}_{uu}$ is evaluated from GSF calculations on an unperturbed, bound geodesic,
but to leading order in the mass-ratio, one finds $r^{-1} = x + {\cal O}(\epsilon)$, hence in 
\eqn{detweiler_redshift_gauge_invariant} one can replace 
$h^{\rm R, L}_{uu}(r) \rightarrow h^{\rm R, L}_{uu}(x)$ \cite{Akcay:2012ea}.

The final piece of the computation of the redshift is the construction of appropiate regular field
contraction, $h^{\rm R, L}_{uu}$.  We shall do this in the same manner as the self-force in the
preceeding section, by constructing the regular field from the mode-sum expression in \eqn{regular_metric},
then accelerating this convergence through use of the tail correction in we used
in  \eqn{radial_self_force_accelerated}.  This procedure follows from the discussion in \secref{self_force}
so we will not repeat it here, and present the final expression:
\begin{align}
	h^{\rm R}_{\mu\nu} &= \sum^{\ellmax}_{\ell = 0} 
	\left[ h^{\rm ret,\, \ell}_{\mu\nu} - h^{[0]}_{\mu\nu} \right] \\
	&+ \sum_{\ell_{\rm max} + 1}^{\infty} \sum_{k = 2}^{k_{\rm max}} \frac{h_{\mu\nu}^{[2k]}}{P_{2k}(\ell)}
	+ {\cal O}(\ell^{-(2k_{\rm max} + 1)}).
	\label{eq:metric_redshift_accelerated}
\end{align}
Here we choose to regularize the individual components of the metric perturbation, $h_{\mu\nu}$, and then
perform the contraction with the four-velocity.  An equally valid method would be to combine the regularisation
parameters in Eqs.~(\ref{eq:h_0_tt})-(\ref{eq:h_0_thetatheta}) into a single parameter for $h_{uu}$, as is
done in Ref.~\cite{Wardell:2015ada} and Ref.~\cite{Thompson:2018lgb}.
\section{Numerical Methods}
\label{sec:numerical_methods}
In this section, we describe the numerical methods used to compute our conformal fields, that are
then used to reconstruct the metric perturbation and compute the self-force and Detweiler-Redshift invariant.
Our methods follow those outlined in Paper I, which itself is based on the work of Ansorg \emph{et al.} in 
Refs.~\cite{Ansorg:2003br,Ansorg:2006gd,Ansorg:2016ztf,Ansorg13}, and the methods outlined in 
Refs.~\cite{Boyd,Grandclement:2007sb,Meinel:2008kpy}.  Since the release of Paper I, another multi-domain
spectral method for self-force calculations using a similar scheme, but with an $m$-mode decomposition
has been developed by Macedo \emph{et al.} in Ref.~\cite{PanossoMacedo:2024pox}.

\subsection{Multi-domain spectral method}
\label{sec:multi_domain_spectral_method}
To solve for our hyperboloidal fields on the compact domain $\sigma \in [0, 1]$, we use a \emph{collocation-point
spectral method}, primarily based on the method developed in Paper I.  In this method, we shall divide this domain
into two subdomains, $\sigma \in {\bf D}_{1} \cup {\bf D}_{2}$, such that $\{ {\bf D}_{1} \} = [0, \sigma_{p}]$ and
$\{ {\bf D}_{1} \} = [\sigma_{p}, 1]$.  
One could implement more subdomains here, as has been done previously in the scalar self-force case, but we find this
unecessary for the configurations being considered here.  If one attempted to tackle the problem of regularisation with
an effective source, one would require a scheme with at least three subdomains as in Paper I.
Within each subdomain, the radial coordinate is further mapped into the new
coordinate $x \in [-1, 1]$ via the relations,
\begin{align}
	\sigma &= \frac{1}{2} \left[ \sigma_{i_{d}}(1 + x) + \sigma_{i_{d - 1}}(1 - x) \right], \\
	x &= \frac{2\sigma - (\sigma_{i_{d}} + \sigma_{i_{d - 1}})}{\sigma_{i_{d}} - \sigma_{i_{d - 1}}}.
\end{align}
Here $\sigma_{i_{d}}$ and $\sigma_{i_{d - 1}}$ label the boundaries of the $i_{d}$-th subdomain, which will run
from $\sigma_{i_{d}} = 0, 1$.  This remapping precisely allows discretisation onto a Chebyshev spectral grid within
each domain.

Let us consider a real-valued function that is defined on the domain, ${\bf D}_{d}$, denoted $f^{(i_{d}, i_{f})}(x)$.
The index $i_{f}$ indicates the particular field that we are considering. Each hyperboloidal RWZ field is complex
valued within each domain.  Thus one requires two fields for each domain corresponding to the real and imaginary 
component of the full conformal field.
The main difference between our new gravitational case, and the SSF case in Paper I, is that we now have to deal with
up to four fields in each domain for the even parity case, see Table.~\ref{tbl:rwz_fields}.
In general, the label $i_{f} = 0,\ldots, N_{f}$, corresponds to one of the total, $n_{f} = N_{f} + 1$ fields in the
same way the label $i_{d} = 0,\ldots, N_{d}$, corresponds to one of the total $n_{d} = N_{d} + 1$ domains.

Our numerical scheme approximates the field $f^{(i_{d}, i_{f})}(x)$ by a truncated Chebyshev expansion,
\beq
	f^{(i_{d}, i_{f})}_{N_{i_{d}}}(x) = \sum^{N_{i_{d}}}_{k = 0}c_{k}^{(i_{d}, i_{f})}T_{k}(x),
	\label{eq:chebyshev_expansion}
\eeq
where $T_{k}(x) = \cos[k\arccos(x)]$ are the Chebyshev polynomials of the first kind, 
$c_{k}^{(i_{d}, i_{f})}$ are the Chebyshev coefficients, and $N_{i_{d}}$ is the truncation order of the expansion.

We shall introduce a discrete grid for each domain, $\{x_{i_{d}}\}_{i_{d}}^{N_{i_{d}}}$, which are each 
\emph{Chebyshev-Lobatto} grids, such that the grid points are given by
\beq
	x_{i} = \cos\left( \frac{\pi i}{N_{i_{d}}} \right), \quad i = 0, 1, \ldots, N_{i_{d}}.
	\label{eq:chebyshev_lobatto_grid}
\eeq
From a computational standpoint, the Chebyshev grids are optimal for this problem with transition conditions at
domain interfaces as well as regularity conditions at the boundaries since these grids include the entire interval.
In addition, as is desired, the resulting conformal result will be known throughout the spacetime including the 
horizon and null infinity.

As in Paper I, we shall fix the Chebyshev coefficients with a collocation method by imposing that our 
approximation to the field that appears in \eqn{chebyshev_expansion} is equal to the exact value of the field
at each point on a discrete, collocation grid.  This implies one would need to invert
$f^{(i_{d}, i_{f})}_{N_{i_{d}}}(x_{i}) = f^{(i_{d}, i_{f})}(x_{i})$ to obtain the values of 
$c_{k}^{(i_{d}, i_{f})}$.  This inversion cannot be perfomed directly since we are solving an ODE problem 
rather than simply an interpolation, and therefore one must recast our differential equation into a form that
can be inverted.  This is done by introducing a set of Chebyshev-Lobatto differentiation matrices to
apply spectral derivatives to our approximate functions on the discrete grids,
\beq
	\pdiff{f^{(i_{d}, i_{f})}_{N_{i_{d}}}(x_{i})}{x} = \sum^{N_{i_{d}}}_{j = 0} {\rm D}_{i, j}
	f^{(i_{d}, i_{f})}_{N_{i_{d}}}(x_{i}),
	\label{eq:spectral_derivative}
\eeq
where ${\rm D}_{i, j}$ are the Chebyshev-Lobatto differentiation matrices, given by
where the matrix $D_{ij}$ is given by \cite{Boyd}
\beq
    {\rm D}_{i,j} = 
    \displaystyle
		\begin{cases}
		\displaystyle
				\frac{\kappa_i (-1)^{i-j}}{\kappa_j (x_i-x_j)} & i \neq j\\[1.25em]
		\displaystyle
				-\frac{x_j}{2 \left[1-\left(x_j\right)^2 \right] } &0< i = j < N\\[1.75em]
		\displaystyle
				\frac{2 N^2 + 1}{6} & i = j = 0\\[1.25em]
		\displaystyle
				-\frac{2 N^2 + 1}{6} & i = j = N
    \end{cases},
    \label{eq:chebyshev_derivative_matrices}
\eeq
where
\beq
    k_{i} = 
    \begin{cases}
        2 & i=0,N, \\
        1 & i \neq 0,N.
    \end{cases}
	\label{eq:chebyshev_derivative_matrix_k}
\eeq
For our differential equations we require the second-order derivative, which is given by applying the 
same differentiation matrix again.

Collecting the discretised equations, along with the transition equations at the domain interface at the
worldline given in \secref{hyperboloidal_rwz_equation} and \secref{gauge_fields}, then one can collect 
the values into an algebraic system of linear equations.  More concretely, the values of the function on
the discrete Chebyshev grids into a vector such that
\beq
	\vec{X} = \left( f^{(i_{d}, i_{f})}_{N_{i_{d}}}(x_{i}) \right)
	\quad {\rm for}
	\begin{cases}
		i_{d} = 0, 1 \\
		i_{f} = 0, \ldots , N_{f} \\
		i = 0, \ldots, N_{i_{d}}.
	\end{cases}
\eeq
We note that this vector will have a size of
\beq
	n_{\rm total} = N_{\rm total} + 1 = n_{f}\sum_{i_{d} = 0}^{n_{d}}(N_{i_{d}} + 1).
	\label{eq:total_vector_size}
\eeq
This formalism allows us to write the discretised system as a simplex matrix equation,
\beq
	\vec{F}(\vec{X}) = \hat{J} \cdot \vec{X} - \vec{S} = 0,
	\label{eq:spectral_matrix_equation}
\eeq
where $\hat{J}$ is the Jacobian matrix, that encodes the discretisation of the differential operators 
and the related transition conditions at the domain interfaces, and $\vec{S}$ is the source term.
This representation of the right-hand-sides of our differential equations contains only non-zero elements
for the gauge field, $\tilde{\chi}$, due to the unbounded validity of the source, whereas for the 
fields with distributional sources, these elements will be zero.

Our problem, therefore, has been reduced to solving this linear system, $\vec{F}(\vec{X})$, for the vector
$\vec{X}$, which is done directly using a Lower-Upper (LU) decomposition algorithm as in Paper I.  
Alternative methods, such as iterative methods like \emph{Bi-Conjugate Stabilized method} (Bi-CGStab)
\cite{Bi-CGSTAB} could be used here as in Ref.~\cite{PanossoMacedo:2024pox}, but we find have not 
implemented this here as we find the LU decomposition to be effecient enough for our purposes.
Within each domain, we expect the functions that are being approximated to be smooth, hence we can expect
the numerical solution to converge exponentially with the truncation order, $N_{i_{d}}$.  
Hence, as the LU decomposition algorithm scales as $\sim n_{\rm total}^{3}$ for each domain, 
the algorithm is still sufficiently fast.

\subsection{Analytical Mesh Refinement}
\label{sec:analytical_mesh_refinement}
Following Paper I, we also implement an analytic mesh refinement (AnMR) scheme to counteract the strong 
gradients that are present in the conformal fields near the worldline for certain parts of the parameter 
space.  This technique is based on the ingenious function in Paper I,
\begin{align}
	x_{i} &= x_{B} \left( 1 - 2\cosh(2\kappa)\sinh[\kappa(1 - x_{B}\chi_{i})] \right), \nn \\
	\chi_{i} &= \arccos\left(\frac{\pi i}{N_{i_{d}}}\right),\quad i = 0, \ldots, N_{i_{d}},
	\label{eq:anmr_mapping}
\end{align}
that defines a mapping from the uniform Chebyshev-Lobatto grid, $x_{i} \in [-1, 1]$, to a non-uniform grid
that concentrates points near a specific boundary of the domain.  Here, $x_{B}$ is a parameter that defines
which boundary the points are concentrated towards.  If $x_{B} = -1$, the grid will be concentrated around
the left boundary, whilst if $x_{B} = 1$, the grid will be concentrated around the right boundary.  
The parameter $\kappa$ is the mesh refinement parameter, which controls the rate at which the mesh is 
concentrated around the boundary of choice.  This concentration allows for higher resolution in regions 
of the computational domain where the solution may exhibit rapid variations, 
while using fewer points in regions where the solution is smoother.

In our numerical computations, we will leverage the findings from Paper I to determine the optimal mesh 
refinement parameter, $\kappa^{*}$. As demonstrated in Paper I, for large-radius orbits, the 
compactification of the domain leads to strong gradients in the retarded fields near the worldline, 
potentially causing a stagnation in the rate of exponential convergence. If left unaddressed, 
this stagnation can significantly affect the accuracy of the conformal field solutions for 
large-radius calculations. However, the AnMR technique was shown to resolve this issue by concentrating 
the mesh points near the worldline, thereby improving the rate of convergence and, consequently, 
the accuracy of the solution for a given resolution. Specifically, we found that the optimal value 
of the mesh refinement parameter, $\kappa^{*}$, for ${\bf D_{2}}$ with $x_{B} = -1$, exhibited 
a logarithmic dependence on the radius of the orbit $\rp$ such that
\beq
	\kappa^{*}(\rp) \sim 0.5\ln \left(\frac{\rp}{M}\right),
	\label{eq:optimal_kappa}
\eeq
while the angular dependence of each of the $(\ell, m)$ modes was found to merely act as a 
constant shift to this function. We will utilize this functional form in our numerical computations 
to enhance their accuracy and efficiency.
\section{Results}
\label{sec:results}
In the following section we present several numerical results that demonstrate the accuracy and
effectiveness of our approach.  We first present the results for the conformal RWZ fields and
gauge fields, which form the building blocks of our metric reconstruction and self-force 
calculations.  In doing so, we aim to demonstrate that computing Lorenz gauge metric perturbations
and gauge invariants can be solved to high accuracy with these alternative methods, hence motivating
this approach for future, more generic calculations.

\subsection{Regge-Wheeler-Zerilli}
\begin{figure}[h!]
	\centering
	\includegraphics[width=\columnwidth]{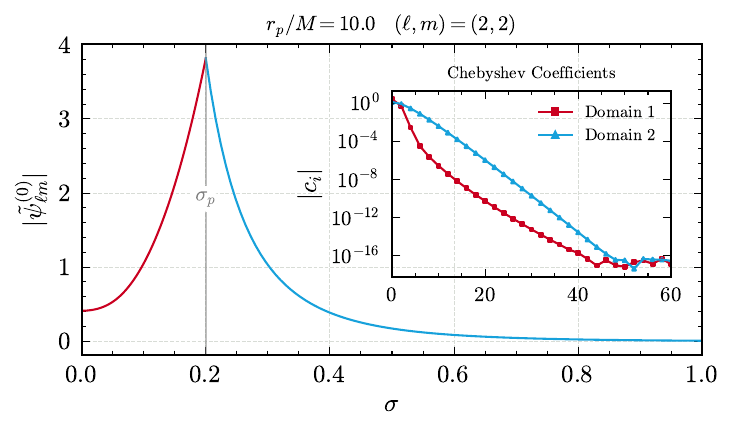}
	\caption{
		A plot of the conformal RWZ field mode, $\tilde{\psi}^{(0)}_{\lm}$, for $(\ell, m) = (2, 2)$.
		In this configuration, the particle is located at $\sigma_{p} = 0.2$ (or $\rp/M = 10$).
		The field in domain 1, ${\bf D_{1}}$, is given in red, whilst the field in domain 2, ${\bf D_{2}}$
		is given in blue.
		The continuous, but finite differentiability of the field at the worldline is evident, as well as
		the spectral convergence shown by the exponential decay of the Chebyshev coefficients in the inset.
		We observe a machine precision floor has been reached by $N_{i_{d}} \sim 60$ within each domain.
		}
	\label{fig:rwz_s0_ell2_m2}
\end{figure}

\begin{figure}[h!]
	\centering
	\includegraphics[width=\columnwidth]{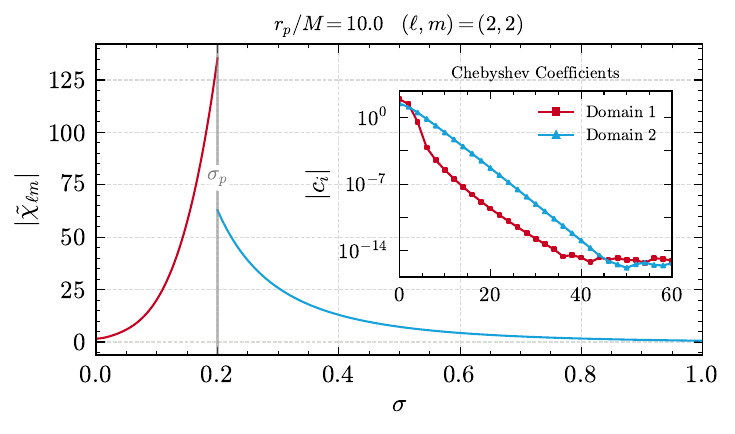}
	\caption{
		A plot of the conformal RWZ field mode, $\tilde{\chi}_{\lm}$, for $(\ell, m) = (2, 2)$ and
		the same orbtial configuration as in \fig{rwz_s0_ell2_m2}.  
		Unlike \fig{rwz_s0_ell2_m2}, however, this field is discontinuous at the worldline due to the
		distributional nature of the source.  Yet, as desired, we still see spectral convergence in each
		domain.
		We observe this conformal field, despite the re-scaling, is significantly larger in magnitude 
		than the RWZ fields.  Hence care must be taken when considering the addition of fields during 
		metric reconstruction.
		}
	\label{fig:chi_ell2_m2}
\end{figure}

Firstly we show the computation of the even parity conformal RWZ field, $\tilde{\psi}^{(0)}_{\lm}$, 
and the gauge field $\tilde{\chi}_{\lm}$ in Figs.~\ref{fig:rwz_s0_ell2_m2} and \ref{fig:chi_ell2_m2}
respectively.  In these examples, we have used $N_{i_{d}} = N_{1} = N_{2} \leq 60$ for the number of
Chebyshev nodes within each subdomain.
One observes that within both subdomains the solutions are smooth and are spectrally convergent as
shown in the insets of each Figure.

The conformal fields satisfy the expected behaviour at the worldline, with the 
$\tilde{\psi}^{(0)}_{\lm}$ field being continuous and $C^{1}$-differentiable across the worldline,
whilst $\tilde{\chi}_{\lm}$ is discontinuous due to the distributional nature of the source. We also
highlight that the fields are defined throughout the entire spacetime and remain
regular at the horizon and null infinity, as anticipated.  This is a stringent test of our numerical
code since the fields are solved independently in each domain, with the transition conditions at the
boundary providing the only connection between the two.  
We only present a sample of the fields here for the even-parity case, since the other RWZ fields, both 
in the even- and odd-parity case, have broadly similar properties to the field shown in 
\fig{rwz_s0_ell2_m2}, except with varying degrees of smoothness at the worldline.
We should emphasise though, that we have verified our fields are identical
to those produced by the BHPT \texttt{ReggeWheeler} package \cite{BHPToolkit} to machine precision, 
when transforming back to the original, non-conformal geometry.

One can see in \fig{chi_ell2_m2} that the gauge field, $\tilde{\chi}_{\lm}$, is significantly larger
in magnitude than the RWZ field, $\tilde{\psi}^{(0)}_{\lm}$, which is true for the other RWZ fields.
This is also evident from the results in \cite{Durkan:2022fvm}, where the gauge field $M_{2af}$ and
the auxilarily distributional field $\psi_{0b}$ were calculated seperately. This served as the motivation
for combining both quantities into a single function, reducing the potential for round-off error due to 
cancellations between the two fields.  The additional conformal factor we introduce in transforming to 
the conformal field in \eqn{regularised_chi} also helps to reduce the magnitude between this field and
the RWZ fields.

However, one must be cautious, since as we observed in Paper I, the accuracy of such a field can be
hindered by the presence of large gradients towards the worldline.  Therefore, when we proceed to metric
reconstruction and self-force calculations, one must be careful to ensure this field, as well as the RWZ fields
at sufficiently for sufficiently large orbital radius, are appropiately resolved.  For this reason, we use the AnMR
technique described in \secref{analytical_mesh_refinement} to ensure the accuracy of our results.

\subsection{Metric Reconstruction}
\label{sec:metric_reconstruction}
\begin{figure*}[hbt!]
	\centering
	\includegraphics[width=\textwidth]{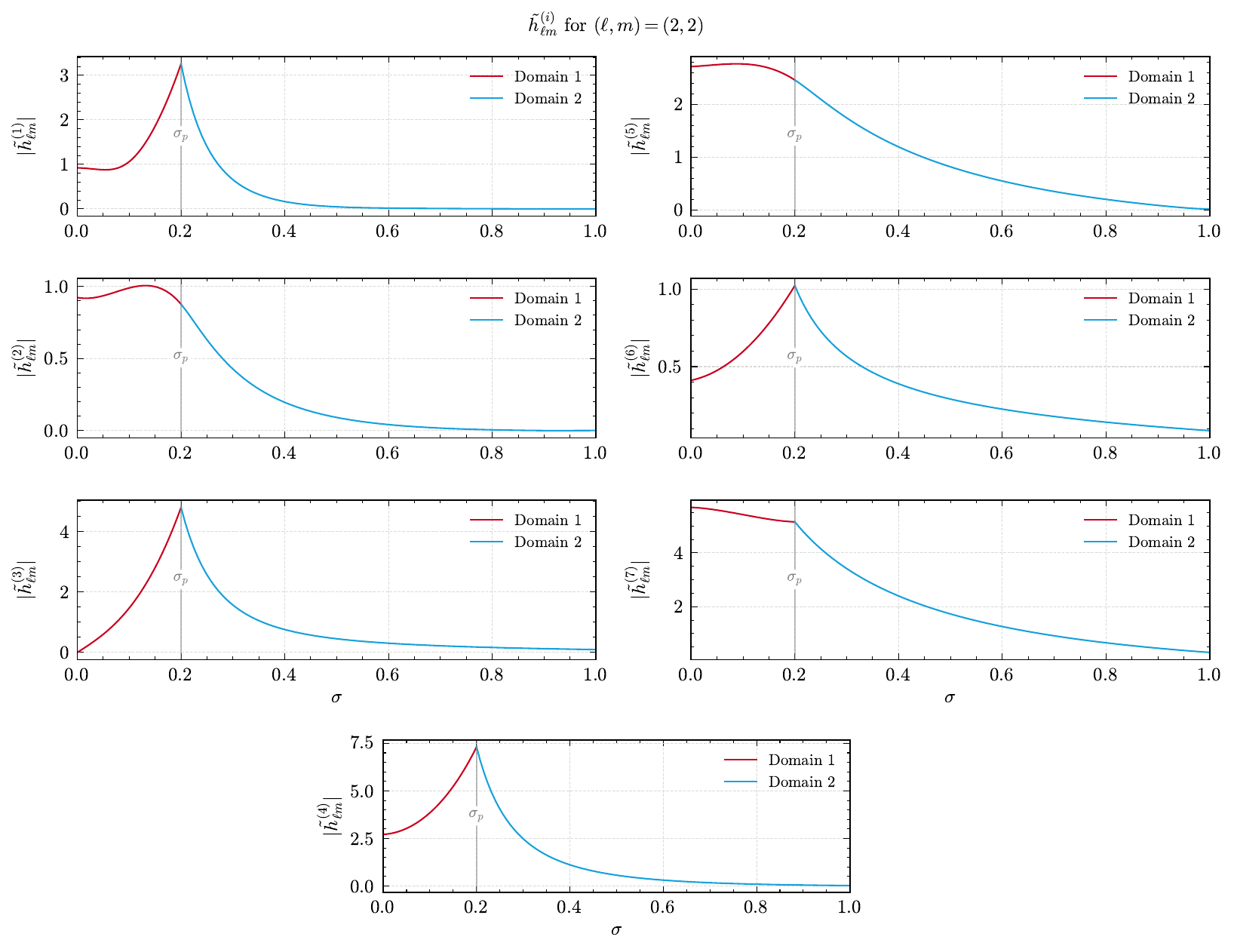}
	\caption{
		A plot of the nonzero (conformal) $\tilde{h}^{(i)}_{\lm}$ BLS modes for $\ell = 2$ and $m = 2$ 
		throughout the spacetime that extends from the null-infinity ($\sigma = 0$) to the horizon ($\sigma = 1$).
		In this orbital configuration, the particle is located at $\sigma_{p} = 0.2$.
		The field in Domain 1, $\{ {\bf D}_{1} \} = [0, \sigma_{p}]$, is shown by the red curves, while the fields
		in Domain 2, $\{ {\bf D}_{2} \} = [\sigma_{p}, 1]$, are shown by the blue curves.  
		We note that whilst all the modes are continuous and regular across the domain, particular modes 
		($i = 2$ and $5$) are at least $C^{2}$-differentiable at the worldline.  This is a stringent test of our
		numerical method since the fields in each domain are solved independently in on the Chebyshev-Lobatto grids
		in each domain, with the transition condition at the boundaries being the only connection between the two.
		}
	\label{fig:h_BLS_ell2_m2}
\end{figure*}

Since we now have all the prerequisite building blocks, we proceed with the metric reconstruction and the
calculation of the gravitational self-force.  We shall first present the results for the even-parity metric
perturbation in terms of the conformal BLS components in \fig{h_BLS_ell2_m2}, which are directly computed from 
the results in Figs.~\ref{fig:rwz_s0_ell2_m2} and \ref{fig:chi_ell2_m2}.

A key distinction of our computation is that this approach enables us to determine the metric perturbation across 
the entire spacetime, encompassing both the horizon and null infinity.  Another important observation,
is that despite calculating the component RWZ fields within each domain, the appropiate differentiability of
the full metric perturbation is achieved.  This is particularly presient for the $i = 2$ and $i = 5$ BLS fields,
since these fields are at least $C^{2}$-differentiable at the worldline.  The fact we recover this behaviour
in our results shows we have reached a sufficient level of accuracy in our calculations.

We have also verified our numerical results with previous data for the BLS modes of the Lorenz gauge metric
perturbation used in the first second-order gravitational self-force calculations in 
Refs.~\cite{Pound:2019lzj,Warburton:2021kwk,Wardell:2021fyy}.  We find that our results are in excellent agreement
when transforming back to the non-conformal geometry, which is another strong validation of our calculation here.
\begin{figure*}[hbt!]
	\centering
	\includegraphics[width=\textwidth]{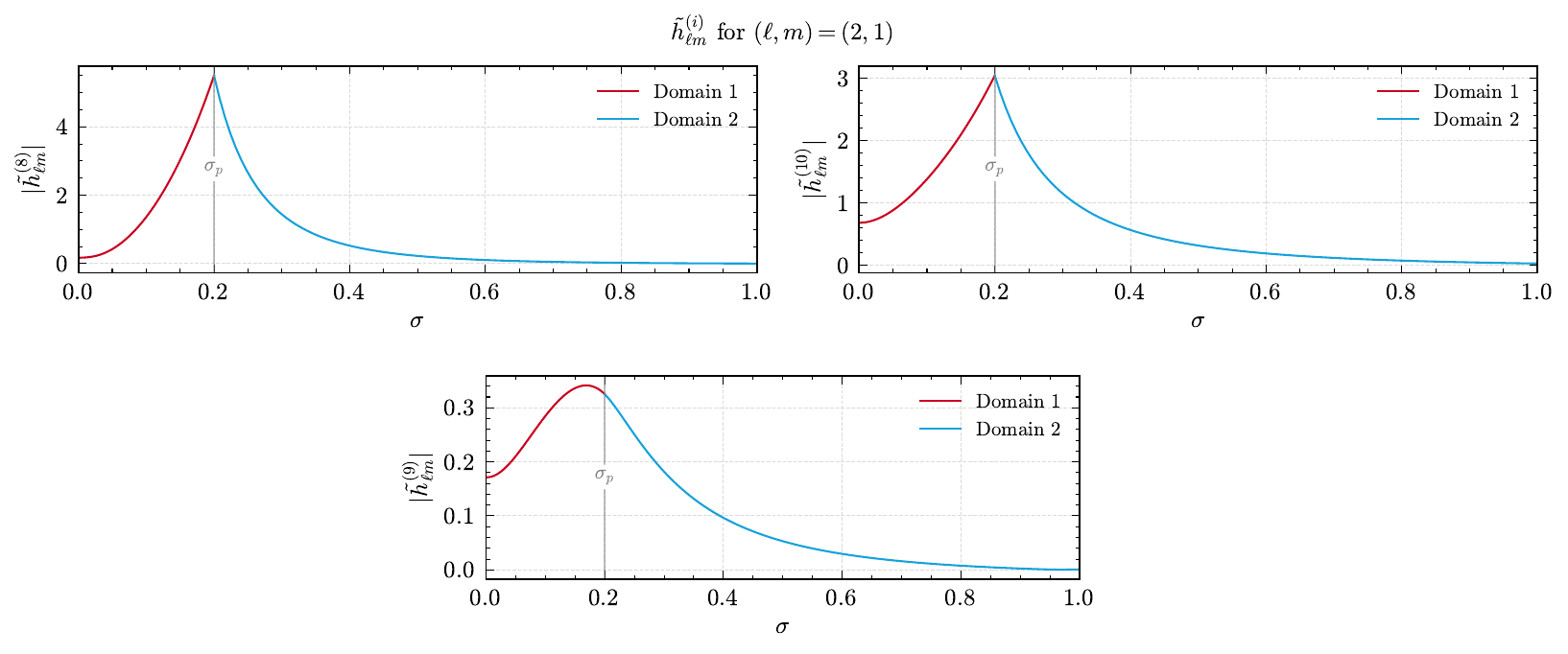}
	\caption{
		A plot of the nonzero (conformal) $\tilde{h}^{(i)}_{\lm}$ BLS modes for $\ell = 2$ and $m = 1$ 
		throughout the spacetime that extends from the null-infinity ($\sigma = 0$) to the horizon ($\sigma = 1$).
		This plot is for the same orbital configuration as in Fig.~\ref{fig:h_BLS_ell2_m2}.
		The modes here are continuous and regular across the domain, whilst the $i = 9$ mode in particular is
		at least $C^{2}$-differentiable at the worldline.  
		}
	\label{fig:h_BLS_ell2_m1}
\end{figure*}

We also present, in \fig{h_BLS_ell2_m1}, the BLS modes for the odd-parity $(\ell, m) = (2, 1)$ case for the same
orbital configuration.  
Note that for the odd-parity case, there are far fewer non-zero BLS modes than in the even-parity
case, whilst the $i = 9$ mode is the only mode that is at least $C^{2}$-differentiable at the worldline.
Furthermore, the gauge field $\tilde{\chi}_{\lm}$ is zero for the odd-parity case.  We observe similar agreement
as the even-parity case with the aforementioned Lorenz-gauge metric perturbation data.

Thus far, we have only presented results for the retarded metric perturbation, which is not the complete picture. 
For full self-force calculations, whether at first- or second-order, the regularised metric perturbation 
is required.  We have outlined our tensor mode-sum regularisation procedure in \secref{self_force}, which 
we have used to compute the regular metric perturbation.  It should be noted, as in Paper I, one could also
implement the alternative \emph{effective-source regularisation} method with our numerical scheme.  The only
difference would be changing to a four-domain code, as in Paper I, and changing the source and transition 
conditions accordingly.  We shall present two results that verify the accuracy of our method, the calculation of
the full gravitational self-force and the Detweiler-Redshift invariant.

\subsection{Self-force}
\label{sec:self_force}
\begin{figure}[h!]
	\centering
	\includegraphics[width=\columnwidth]{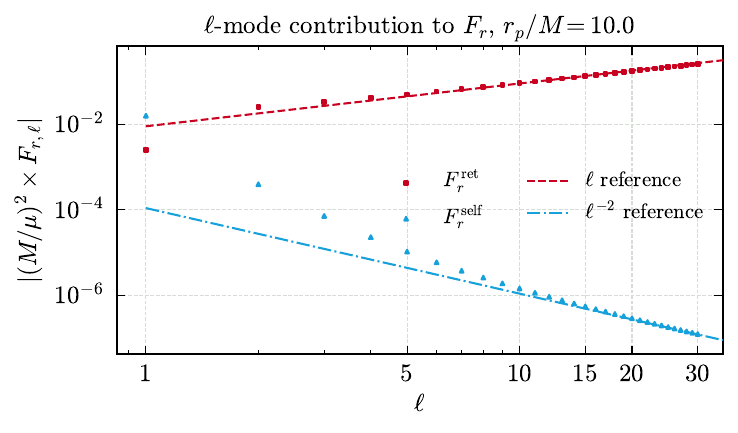}
	\caption{
		A plot of the $\ell$-mode contribution to the radial self-force, $F_{r, \ell}$ for the differing 
		degrees of regularisation we have discussed in this work for a secondary on an orbit of $\rp/M = 10$.
		We show the results for the unregularised (red squares) and the mode-sum regularised (blue triangles)
		radial self-force.  
		The ``retarded" modes of 
		self-force are the quantity calculated from the unregularised metric perturbation and diverge for
		large-$\ell$ as $\sim \ell$.  The regularised modes, however, have a finite sum and converge as
		$\sim \ell^{-2}$ for large $\ell$, as expected.  However, this convergence can be accelerated 
		further by fitting for the tail contribution for large-$\ell$ to compensate for the trunacation
		of the $\ell$-sum.
		}
	\label{fig:Fr_ell}
\end{figure}

\begin{figure}[h!]
	\centering
	\includegraphics[width=\columnwidth]{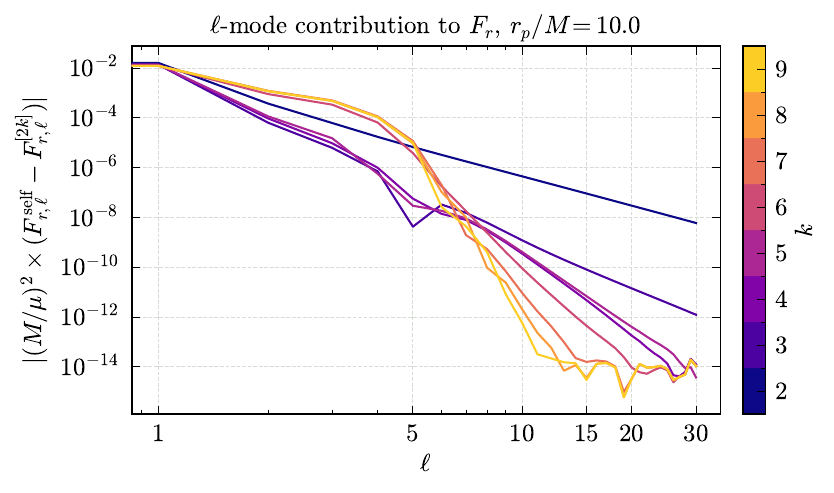}
	\caption{
		A plot of the $\ell$-mode contribution to the radial self-force, $F_{r, \ell}$, using the high-order
		regularisation parameters found by fitting for the tail contribution.  
		Here we show the residuals for subtracting the higher-order regularisation parameters 
		$F^{[2k]}_{r}$ from the modes regularised with the analytic regularisation parameters 
		$F^{[-1]}_{r}$ and $F^{[0]}_{r}$, i.e.,
		$F_{r}^{\,\rm self} - F_{r}^{[2k]}$.  We find that the optimal
		regularisation parameters up to and including $F^{[2k]}_{r}$, with $k = 9$, yield the best convergence
		down to our precision floor.  We utilise these regularisation parameters in our final self-force calculation
		shown in \tblref{F_r}.
		}
	\label{fig:Fr_ell_tail}
\end{figure}

We present results for the gravitational self-force, which is computed from the regular metric perturbation
and its derivatives.  In \fig{Fr_ell}, we show the $\ell$-mode contribution to the radial self-force, $F_{r, \ell}$,
for the different regularisation schemes we have discussed in this work for a secondary on a circular orbit of radius $\rp/M = 10$.  
We show the results for the unregularised modes of the self-force, which diverge for large-$\ell$, as well as the 
mode-sum regularised modes.
The regularised modes converge as $\sim \ell^{-2}$ for large $\ell$, as expected, but we also see that this 
asymptotic convergence for large $\ell$ would require a large number of $\ell$ modes to achieve machine level precision.
We choose to accelerate this convergence further through the tail correction method we have outlined in 
\secref{self_force} in \fig{Fr_ell_tail}.  
Fitting for this tail contribution allows us to compensate for the truncation of the $\ell$-sum
and accelerate the convergence to something nearly exponential in $\ell$ until our machine precision floor is reached.

\begin{figure}[h!]
	\centering
	\includegraphics[width=\columnwidth]{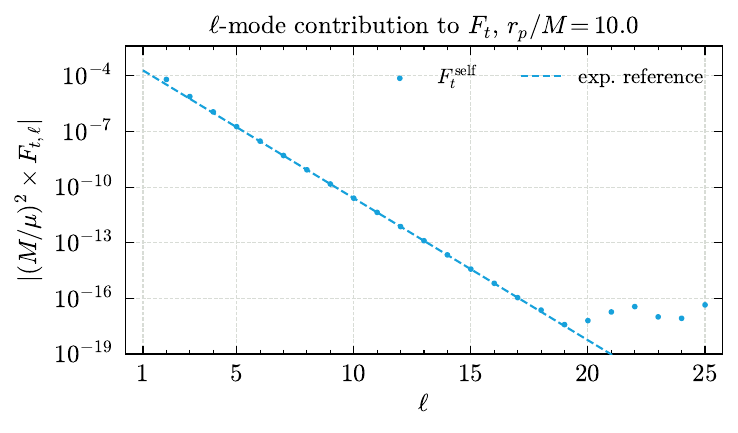}
	\caption{
		A plot of the $\ell$-mode contribution to the temporal component of self-force, $F_{t, \ell}$ 
		for a secondary on an orbit of $\rp/M = 10$.
		The $t$-component of the self-force in this configuration does not require any regularisation
		since it is completely dissipative and converges exponentially with $\ell$, as clearly seen
		in the plot.  This is in contrast to the radial component of the self-force in \fig{Fr_ell}, 
		which requires regulatisation in order to converge.  The rapid convergence of $F_{t, \ell}$ 
		means we encounter a precision floor of our numerical calculations around $\ell \sim 20$.
		}
	\label{fig:Ft_ell}
\end{figure}

We also show the $\ell$-mode contribution to the temporal component of the self-force, $F_{t, \ell}$, in \fig{Ft_ell}.
This is a purely dissipative component of the self-force within this configuration, hence this component does not
require any regularisation to achieve exponential convergence with $\ell$.  Such rapid convergence with $\ell$ means
the numerical precision floor is reached around $\ell \sim 20$ in this configuration.  Similar convergence is also
observed for the azimuthal component of the self-force, $F_{\varphi, \ell}$, which we do not show here.

For the calculation of the components of the self-force, we present sample results of the radial component, $F^{r}$, 
in \tblref{F_r} and find good agreement with previous results in the literature.  
We present two alternative results for the radial self-force. 
The first is derived using the Lorenz gauge monopole solution from 
Refs.~\cite{Barack:2007tm, Akcay:2010dx, Akcay:2013wfa, Wardell:2015ada}, which exhibits irregular behavior 
at null infinity. The second is based on the solution by Berndtson \cite{Berndtson:2007gsc}, which remains 
regular across the entire domain.  Further discussion on the calculation and behaviour of the Lorenz gauge
monopoles used in this work can be found in Appendix~\ref{sec:monopole_static}.
{\renewcommand{\arraystretch}{1.2}
\begin{table*}[hbt!]
\begin{tabular*}{\textwidth}{ @{\extracolsep{\fill}} c S[table-format=3.3] S[table-format=3.3] S[table-format=3.3]  
S[table-format=3.3]}
\toprule
$r_{p} / M$	
& \multicolumn{1}{c}{$F^{r}_{\rm Reg} \times (M/\mu)^{2}$}
& \multicolumn{1}{c}{$F^{r}_{\rm Irreg} \times (M/\mu)^{2}$}
& \multicolumn{1}{c}{$1 - | F^{r}_{\rm Reg} / F_{\rm Bern}^{r} |$}
& \multicolumn{1}{c}{$1 - | F^{r}_{\rm Irreg} / F_{\rm Ward}^{r} |$} \\
\toprule
$5$	
&		\num{7.7252d-2}		
&		\num{2.3974d-2}
&		\num{-6.05d-12}
&		\num{-1.88d-12} \\
$6$	
&		\num{4.96857d-2}		
&		\num{2.44665d-2}
&		\num{7.89d-8}
&		\num{3.88d-8} \\
$7$
&		\num{3.56247d-2}
&		\num{2.14991d-2}
&		\num{6.20d-12}
&		\num{3.73d-12} \\
$8$
&		\num{2.71128d-2}
&		\num{1.83578d-2}
&		\num{-5.13d-9}
&		\num{-3.48d-9} \\
$9$
&		\num{2.14527d-2}
&		\num{1.56371d-2}
&		\num{-3.39d-11}
&		\num{-2.47d-11} \\
$10$
&		\num{1.74546d-2}
&		\num{1.33895d-2}
&		\num{-2.34d-10}
&		\num{-1.80d-10} \\
$11$
&		\num{1.45072d-2}
&		\num{1.15517d-2}
&		\num{7.26d-11}
&		\num{5.78d-11} \\
$12$
&		\num{1.22634d-2}
&		\num{1.00462d-2}
&		\num{-4.59d-11}
&		\num{-3.76d-11} \\
$13$
&		\num{1.05112d-2}
&		\num{8.80489d-2}
&		\num{-4.40d-9}
&		\num{-3.69d-9} \\
$14$
&		\num{9.11467d-2}
&		\num{7.77306d-2}
&		\num{-2.12d-9}
&		\num{-1.81d-9} \\
$15$
&		\num{7.98222d-2}
&		\num{6.90815d-2}
&		\num{-2.76d-10}
&		\num{-2.39d-10} \\
$20$
&		\num{4.5873d-3}
&		\num{4.15706d-3}
&		\num{3.76d-10}
&		\num{-3.86d-10} \\
$30$
&		\num{2.09124d-3}
&		\num{1.96982d-3}
&		\num{-1.87d-9}
&		\num{-1.76d-9} \\
$40$
&		\num{1.19293d-3}
&		\num{1.14288d-3}
&		\num{-4.39d-9}
&		\num{-4.20d-10} \\
$50$
&		\num{7.70228d-4}
&		\num{7.44949d-4}
&		\num{-1.85d-9}
&		\num{-1.79d-9} \\
$60$
&		\num{5.38113d-4}
&		\num{5.23614d-4}
&		\num{-1.21d-10}
&		\num{-1.74d-10} \\
$70$
&		\num{3.97083d-4}
&		\num{3.8801d-4}
&		\num{8.42d-10}
&		\num{8.23d-10} \\
\botrule
\end{tabular*}
\caption{Sample numerical results for the radial component of the self-force, $F^{r}$, for a range of values of the
orbital radius of the secondary, $\rp$.  We show the results for the self-force constructed from a metric perturbation
with an \emph{asymptotically regular} Berndtson monopole solution, $F^{r}_{\rm Reg}$, and the 
\emph{asymptotically irregular} monopole solution, $F^{r}_{\rm Irreg}$.  $F^{r}_{\rm Irreg}$ corresponds to
the solutions used in Refs.~\cite{Barack:2007tm, Akcay:2010dx, Akcay:2013wfa, Wardell:2015ada}, whilst $F^{r}_{\rm Reg}$
corresponds to the solutions used in Refs.~\cite{Berndtson:2007gsc} and the second-order calculations presented in
\cite{Pound:2019lzj, Warburton:2021kwk, Wardell:2021fyy}.  In the fourth and fifth columns we compute the relative 
error between the solutions $F^{r}_{\rm Irreg}$ and $F^{r}_{\rm Reg}$ with Lorenz gauge data calculated with 
a high-precision frequency domain code used to compute the first-order metric perturbation for second-order calculations 
based on Ref.~\cite{Wardell:2015ada}, labelled $F_{\rm Ward}^{r}$.}.
\label{tbl:F_r}
\end{table*}
}

Furthermore, we also present numerical results for the $t$-component of the self-force, $F^{t}$, in \tblref{F_t}.
We find that the results for the temporal component of the self-force are also in excellent agreement with pre-existing
results.  In addition, one can perform a further internal consistency check of our approach by numerically verifying the
balance laws in Eqs.~(\ref{eq:energy_balance_law}) and (\ref{eq:angular_balance_law}) between the energy and angular momentum 
fluxes and the components of the self-force at first-order in the mass-ratio.  These results clearly demonstrate the
internal consistency and the high accuracy of our numerical approach.  Encouraged by these results, we will now proceed to 
calculate the gauge-invariant Detweiler redshift over the full range of the parameter space.
{\renewcommand{\arraystretch}{1.2}
\begin{table*}[hbt!]
\begin{tabular*}{\textwidth}{ @{\extracolsep{\fill}} c S[table-format=3.3] S[table-format=3.3] S[table-format=3.3]  
S[table-format=3.3] S[table-format=3.3]}
\toprule
$r_{p} / M$	
& \multicolumn{1}{c}{$F^{t} \times (M/\mu)^{2}$}
& \multicolumn{1}{c}{$F^{\varphi} \times (M/\mu)^{2}$}
& \multicolumn{1}{c}{$1 - | F^{t} / F_{\rm Ward}^{t} |$}
& \multicolumn{1}{c}{$1 - | F_{t} / u^{t} \dot{E} |$}
& \multicolumn{1}{c}{$1 - | F_{\varphi} / u^{t} \dot{J} |$} \\
\toprule
$5$	
&	\num{7.7252d-2}		
&	\num{2.3974d-2}
&	\num{-1.88d-12}
&	\num{-1.88d-12}
&	\num{-1.07d-12} \\
$6$	
&	\num{-1.99476d-3}		
&	\num{-5.42905d-4}
&	\num{-2.77d-13}
&	\num{-2.77d-13}
&	\num{-2.76d-13} \\
$7$
&	\num{-7.41113d-4}
&	\num{-2.00082d-4}
&	\num{-6.45d-13}
&	\num{-6.45d-13}
&	\num{-6.45d-13} \\
$8$
&	\num{-3.30740d-4}
&	\num{-8.77006d-5}
&	\num{1.42d-13}
&	\num{1.42d-13}
&	\num{1.42d-13} \\
$9$
&	\num{-1.66810d-4}
&	\num{-4.32471d-5}
&	\num{-1.11d-12}
&	\num{-1.11d-12}
&	\num{-1.11d-12} \\
$10$
&	\num{-9.19076d-5}
&	\num{-2.32510d-5}
&	\num{-1.37d-12}
&	\num{-1.37d-12}
&	\num{-1.37d-12} \\
$11$
&	\num{-5.41623d-5}
&	\num{-1.33614d-5}
&	\num{-3.40d-13}
&	\num{-3.41d-13}
&	\num{-3.40d-13} \\
$12$
&	\num{-3.36596d-5}
&	\num{-8.09723d-6}
&	\num{-5.04d-12}
&	\num{-5.04d-12}
&	\num{-5.04d-12} \\
$13$
&	\num{-2.18392d-5}
&	\num{-5.12525d-6}
&	\num{4.73d-12}
&	\num{4.73d-12}
&	\num{4.73d-12} \\
$14$
&	\num{-1.46854d-5}
&	\num{-3.36415d-6}
&	\num{-1.43d-11}
&	\num{-1.43d-11}
&	\num{-1.43d-11} \\
$15$
&	\num{-1.01771d-5}
&	\num{-2.27736d-6}
&	\num{-1.33d-11}
&	\num{-1.33d-11}
&	\num{-1.33d-11} \\
$20$
&	\num{-2.25544d-6}
&	\num{-4.53898d-7}
&	\num{-3.03d-11}
&	\num{-3.03d-11}
&	\num{-3.03d-11} \\
$30$
&	\num{-2.80819d-7}
&	\num{-4.78522d-8}
&	\num{-2.83d-11}
&	\num{-2.83d-11}
&	\num{-2.83d-11} \\
$40$
&	\num{-6.51229d-8}
&	\num{-9.78199d-9}
&	\num{-5.50d-11}
&	\num{-5.50d-11}
&	\num{-5.50d-11} \\
$50$
&	\num{-2.10846d-8}
&	\num{-2.86254d-9}
&	\num{-1.43d-10}
&	\num{-1.43d-10}
&	\num{-1.43d-10} \\
$60$
&	\num{-8.41279d-9}
&	\num{-1.04988d-9}
&	\num{-3.77d-10}
&	\num{-3.77d-10}
&	\num{-3.77d-10} \\
$70$
&	\num{-3.87410d-9}
&	\num{-4.49813d-10}
&	\num{-1.26d-9}
&	\num{-1.26d-9}
&	\num{-1.26d-9} \\
\botrule
\end{tabular*}
\caption{Sample numerical results for the temporal and azimuthal components of the self-force, $F^{t}$, and, $F^{\varphi}$, respectively, 
for a range of values of the orbital radius of the secondary, $\rp$.  In the third column we calculate the relative error of $F^{t}$ with
Lorenz gauge data calculated with a high-precision frequency domain code used to compute the first-order metric perturbation for 
second-order calculations based on Ref.~\cite{Wardell:2015ada}.  This code was also used for the comparisons in \tblref{F_r}.
In the fourth and fifth columns we compute the relative 
error between the left-hand-side and right-hand-side of the balance laws in in Eqs.~(\ref{eq:energy_balance_law}) 
and (\ref{eq:angular_balance_law}) using our own numerical results.  We find that the balance laws are satisfied to a high degree of
accuracy up to our numerical precision floor.}
\label{tbl:F_t}
\end{table*}
}

\subsection{Detweiler Redshift}
\label{sec:detweiler_redshift}
Finally, we shall present results for the Detweiler redshift invariant, $\Delta z$.  First, we present sample numerical results
of the Detweiler redshift invariant for a range of values of the orbital radius of the secondary, $\rp$, in \tblref{redshift}.
We find good agreement with previous results in the literature, specifically comparing to previous results from Dolan \emph{et al.}
\cite{Dolan:2014pja}, with our relative errors at a similar level of numerical precision as our previous self-force calculations.
{\renewcommand{\arraystretch}{1.2}
\begin{table}[hbt!]
\begin{tabular*}{\columnwidth}{ @{\extracolsep{\fill}} c S[table-format=3.3] S[table-format=3.3] }
\toprule
$r_{p} / M$	
& \multicolumn{1}{c}{$\Delta z \times (M/\mu)$}
& \multicolumn{1}{c}{$1 - |\Delta z / \Delta z_{\rm Dolan}|$} \\
\toprule
$7$	
&	\num{-2.20848d-1}		
&	\num{3.03d-10} \\
$8$
&	\num{-1.77720d-1}
&	\num{3.51d-10} \\
$9$
&	\num{-1.49361d-1}
&	\num{1.75d-10} \\
$10$
&	\num{-1.29122d-1}
&	\num{8.22d-11} \\
$12$
&	\num{-1.01936d-1}
&	\num{2.30d-11} \\
$14$
&	\num{-8.43820d-2}
&	\num{8.49d-12} \\
$16$
&	\num{-7.20551d-2}
&	\num{3.85d-12} \\
$18$
&	\num{-6.29019d-2}
&	\num{5.50d-12} \\
$20$
&	\num{-5.58277d-2}
&	\num{3.07d-12} \\
$30$
&	\num{-3.57783d-2}
&	\num{2.60d-12} \\
$40$
&	\num{-2.63397d-2}
&	\num{3.48d-12} \\
$50$
&	\num{-2.08447d-2}
&	\num{1.229d-13} \\
$60$
&	\num{-1.72476d-2}
&	\num{-5.70d-12} \\
$70$
&	\num{-1.47096d-2}
&	\num{7.98d-12} \\
$80$
&	\num{-1.28230d-2}
&	\num{2.53d-12} \\
$90$
&	\num{-1.13653d-2}
&	\num{-9.78d-13} \\
$100$
&	\num{-1.02053d-2}
&	\num{-6.24d-13} \\
\botrule
\end{tabular*}
\caption{Sample numerical results for the Detweiler redshift invariant, $\Delta z$ for a range of values of the orbital radius of the
secondary $\rp$.  In the third column we calculate the relative error of $\Delta z$ with results from Dolan \emph{et al.} in
Ref.~\cite{Dolan:2014pja}.  We find excellent agreement with the results from Ref.~\cite{Dolan:2014pja} down to a similar numerical precision
floor we have seen elsewhere in our calculations.}
\label{tbl:redshift}
\end{table}
}

To demonstrate the effectiveness of our hyperboloidal spectral method in calculating the Lorenz gauge metric perturbation, 
we will compute the Detweiler redshift invariant across a wide range of orbits, extending to very large orbital radii.
This allows us to compare our results to a post-Newtonian (PN) expansion of the Detweiler redshift invariant in the weak field.
Kavanagh \emph{et al.} derived a 21.5PN expression for the Detweiler redshift invariant, $\Delta z^{(21.5 \rm{PN})}$, 
in Ref.~\cite{Kavanagh:2015lva}, avaliable through the BHPToolkit $\texttt{PostNewtonianSelfForce}$ \cite{niels_warburton_2023_8112975} 
package, which we will use for comparison.
The full expression for $\Delta z^{(21.5 \rm{PN})}$ is incredibly lengthy and cumbersome, so we will not repeat it here.
\begin{figure*}[hbt!]
	\centering
	\includegraphics[width=\textwidth]{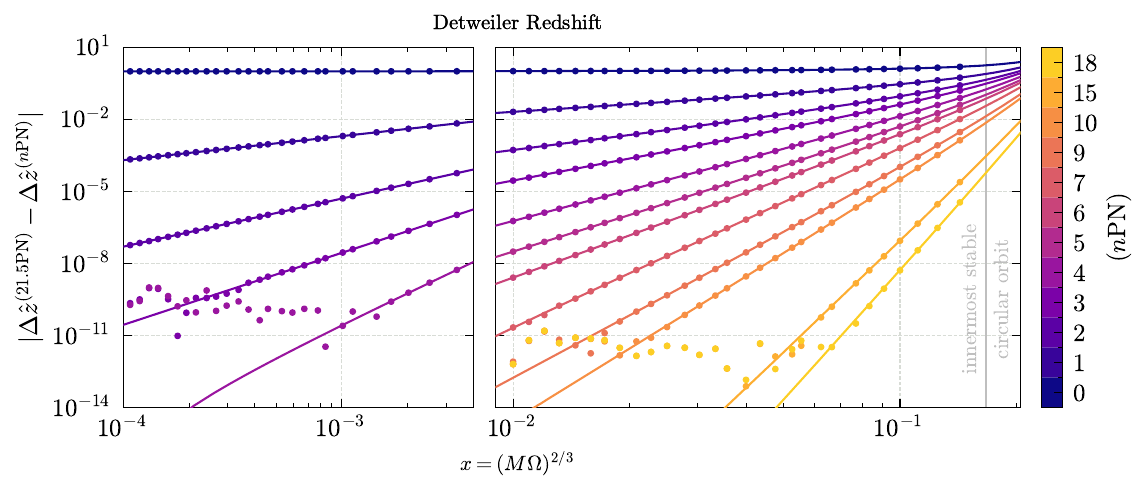}
	\caption{
		A comparison of the Newtonian-normalised Detweiler redshift, $\Delta \hat{z}$, with the 21.5PN expression
		from Kavanagh \emph{et al.} from Ref.~\cite{Kavanagh:2015lva}.  The redshift is plotted as a function 
		of the gauge invariant radius, $x = (M\Omega_{\varphi})^{2/3}$, hence the rightmost boundary of the 
		plot is towards the central black hole horizon. 
		We have labelled the innermost stable circular orbit
		(ISCO) at $x = 1/6$ to help illustrate this.  
		Our (Newtonian-normalised) redshift data is shown by the top (blue) dots and the full 21.5PN expression
		is plotted as the solid (blue) curve.
		We subtract successive PN terms from the leading-order
		normalised redshift and compare these residuals to the residuals of the successive PN series,
		indicated by the colour bar on the right-hand side of the plot. In this plot we subtract the first 18PN terms
		from the full 21.5PN series.
		In the weak field regime, we see excellent agreement between the residuals and the PN expressions
		down to a numerical floor due to the precision of our calculations.  This shows we can acheieve a high
		level of accuracy for calculating the metric perturbation and associated gauge invariants 
		with our numerical for even very weak field orbits.
		}
	\label{fig:detweiler_redshift}
\end{figure*}

We compare the results of our method with the 21.5PN series in \fig{detweiler_redshift}. 
To ensure a valid comparison with this PN expression, recall from \secref{detweiler_redshift} that the Detweiler redshift 
invariant must be calculated in an asymptotically flat gauge. This is achieved by using the appropriate monopole solution 
for the metric perturbation, which has been applied in the comparison shown in \tblref{F_r}. We plot the Newtonian-normalized 
redshift invariant (i.e., normalized by the leading term in the PN series) as a function of the gauge-invariant radius, $x$. 
We denote this normalized quantity with an overhat, $\Delta \hat{z}$.
As higher-order PN series terms are subtracted from our numerical results, the residuals behave as expected in the weak-field 
regime: the residuals of our numerical data closely match those of the corresponding PN series, down to a numerical floor set 
by the precision of our calculations. For instance, after subtracting the leading PN order term, the residual of our numerical 
data follows the ${\cal O}(x)$ scaling of the $|\Delta \hat{z}^{(21.5 \rm{PN})} - \Delta \hat{z}^{(1 \rm{PN})}|$ 
residual PN series. The next residual follows the anticipated ${\cal O}(x^{2})$ scaling, and this pattern continues for higher-order residuals.
Thus, in general, we find: ${\cal O}(|\Delta \hat{z} - \Delta \hat{z}^{(n \rm{PN})}|)
\sim {\cal O}(|\Delta \hat{z}^{(21.5 \rm{PN})} - \Delta \hat{z}^{(n \rm{PN})}|) \sim x^{n}$
until we reach the limit of our numerical precision. This agreement with the PN series demonstrates that our numerical 
scheme achieves a high level of accuracy for calculating the metric perturbation and associated gauge invariants, 
even for very weak-field orbits as large as $\rp/M \sim 10^{4}$.
\section{Conclusion}
\label{sec:conclusion}
In this work, we presented a comprehensive framework for the calculation of the gravitational self-force in the Lorenz gauge
for a secondary particle in a circular orbit around a Schwarzschild black hole using hyperboloidal slicing and spectral methods.
Building on the toy model introduced in Paper I, we extended these methods to solve for gravitational perturbations in the
Lorenz gauge, marking a significant step forward in the application of hyperboloidal and spectral techniques to self-force
calculations.  By adopting hyperboloidal slicing, we circumvent the challenges inherent to the traditional methods of variation
of parameters, which suffer from numerical error due to large integration domains and degenerate boundary conditions, especially
at low frequencies.

Our numerical results demonstrated the power of the hyperboloidal-spectral approach. We performed GSF calculations for a range of 
circular orbits, finding excellent agreement with known analytical results for energy and angular momentum fluxes at null infinity
and the black hole horizon. 
The use of hyperboloidal slicing facilitated accurate evaluation of these fluxes directly at the boundaries, avoiding the numerical 
difficulties associated with truncating the domain at finite radii. The spectral method provided rapid convergence across the 
entire domain, even for high spherical harmonic modes, which are particularly challenging for standard Fourier domain approaches. 
Furthermore, the accuracy of the method was preserved for both small- and large-radius orbits, confirming its robustness in diverse regimes. 

In particular, our results for gauge invariant quantities like the Detweiler redshift for orbits in the weak-field show that the
challenges that were surmounted in Paper I for the scalar-field case are equally applicable for the gravitational case.
Compactification, coupled with the analytical mesh refinement procedure we introduced in Paper I, allowed for detailed 
comparison with post-Newtonian expressions and we observed agreement to a high level of precision between our results.
One would expect such a methodology will continue to provide a valuable link between gravitational self-force calculations
and post-Newtonian theory.

We also observe, in line with the findings of Paper I, that the hyperboloidal-spectral method offers a notable increase in 
efficiency compared to Fourier-domain approaches. Hyperboloidal slicing alleviates the need to resolve the highly oscillatory 
behaviour of the fields near null infinity and the black hole horizon, thereby reducing the resolution requirements and 
associated computational cost. This leads to significant reductions in runtime without any loss of precision. 
The radial profiles of the conformal fields remain smooth across the numerical domain, and the spherical harmonic modes exhibit 
more homogeneous structure. As a result, the spectral method achieves uniform and rapid convergence across all modes. 
Furthermore, since each mode is solved independently, the approach is inherently parallelisable and scales efficiently on 
modern computational architectures.

Looking forward, this work opens several avenues for further research. The success of hyperboloidal slicing in a 
Schwarzschild background motivates the extension of these methods to Kerr spacetime, where the lack of separability in the 
field equations presents a formidable challenge.  The adaptability of the hyperboloidal approach, for which the foliation in
Kerr spacetime is already presented in \cite{PanossoMacedo:2019npm}, offers a promising direction for overcoming the difficulties 
associated with mode coupling in the Kerr geometry via the methods presented by \emph{Dolan et al.} in 
Refs.~\cite{Dolan:2023enf, Wardell:2024yoi}.  
This approach to metric reconstruction, which has parallels to the Berndtson method utilised in this work, could produce
highly accurate Lorenz gauge metric perturbations in Kerr spacetime.

This is not only valuable at first-perturbative order,
where there are preciously little calculations avaliable \cite{Dolan:2023enf, Isoyama:2014mja}, but could an important 
input for future calculations at second-order.
Second-order GSF calculations rely at some point on metric perturbations calculated in the Lorenz gauge and the ability
of the methods presented here to produce accurate and regular perturbations throughout the entire spacetime could be 
beneficial to extending the fidelity of second-order calculations.  The weak field regime, and comparisons with
post-Newtonian theory, have so far proved difficult for existing second-order GSF implementations \cite{Warburton:2024xnr}.

Finally, the methods presented here and in Paper I, demonstrate the accuracy and efficacy for GSF calculations for circular
orbits.  But one of the key challenges within the parameter space of EMRIs is the extension to eccentricity.  EMRIs formed
thorough formation channels like a dynamical capture \cite{Babak:2017tow}, are expected to retain a siginifcant level of 
eccentricity in the LISA band.  
Whilst there are existing eccentric Fourier-domain GSF calculations that leverage methods based on variation 
of parameters, our spectral method presented here would need considerable modifications to be used for secondaries on
eccentric geodesics. This is because particle in our framework is no longer confined to a solitary value of
$\rp$ for a given geodesic, but instead librates within a minimum and maximum radius.  For a distributional or 
effective-source in the Fourier domain, this presents longstanding issues with Gibbs phenomenon.  One could envisage
a solution based on the methods presented in Ref.~\cite{Leather:2023dzj}, but we shall leave this to future work.

\acknowledgments
The author would like to thank Niels Warburton, Barry Wardell, Sarp Akcay, Leanne Durkan and Adrian Ottewill for 
helpful discussions.
Special thanks are also due to Sarp Akcay, Zach Nasipak, Niels Warburton and Rodrigo Panosso Macedo for 
their insightful feedback and  comments on an earlier draft of this manuscript.
We would also like to thank Barry Wardell and Adrian Ottewill for sharing their even-parity analytic solutions to the
static modes of the Lorenz gauge metric perturbation.
Some of the calculations in this work were performed on the Hypatia cluster at the 
Max Planck Institute for Gravitational Physics.
This work makes use of the Black Hole Perturbation Toolkit \cite{BHPToolkit}. 

\appendix

\section{Stress-energy tensor}
\label{sec:stress_energy_tensor}
The stress-energy tensor in \eqn{point_particle_stress_energy_tensor} can be decomposed in the same manner as the 
metric peturbation into tensor spherical harmonics and then seperating the even- and odd-parity components, 
$T^{\lm}_{\mu\nu} = T^{e, \lm}_{\mu\nu} + T^{o, \lm}_{\mu\nu}$.
The even-parity components can then be written in matrix form as,
\begin{widetext}
	\beq
		T_{\mu\nu}^{e,\lm} = 
		\begin{pmatrix}
			\vspace{0.2cm} T^{e, \lm}_{00} Y_{\lm} 
			& T^{e, \lm}_{01} Y_{\lm}  
			& T^{e, \lm}_{02} \dfrac{\partial Y_{\lm}}{\partial \theta} 
			& T^{e, \lm}_{02} \dfrac{\partial Y_{\lm}}{\partial \varphi} \\
			\vspace{0.2cm} * & T^{e, \lm}_{11} Y_{\lm} 
			& T^{e, \lm}_{12} \dfrac{\partial Y_{\lm}}{\partial \theta} 
			& T^{e, \lm}_{12} \dfrac{\partial Y_{\lm}}{\partial \varphi} \\
			\vspace{0.2cm} * & * 
			& U^{e, \lm}_{22} Y_{\lm} + T^{e, \lm}_{22} W_{\lm}
			& \sin\theta\, T^{e, \lm}_{22} X_{\lm} \\
			\vspace{0.2cm} * & * & * 
			& \sin^2\theta \left( U^{e, \lm}_{22} Y_{\lm} + T^{e, \lm}_{22} W_{\lm} \right)
		\end{pmatrix},
	\eeq
where the components of the stress-energy tensor are given in the Fourier domain for circular, equatorial orbits 
as \cite{Berndtson:2007gsc, Durkan:2022fvm}
\begin{align}
	T^{e, \lm}_{00} &= \mu u^{t} \frac{f(r)^{2}}{r^{2}} \delta(r - \rp) Y^{*}_{\lm}\left(\frac{\pi}{2}, 0\right),\\
	T^{e, \lm}_{01} &= 0, \\
	T^{e, \lm}_{11} &= 0, \\
	T^{e, \lm}_{02} &= i \omega_{m} \mu u^{t} \frac{f(r)}{\ell(\ell + 1)}
	\delta(r - \rp) Y^{*}_{\lm}\left(\frac{\pi}{2}, 0\right), \\
	T^{e, \lm}_{12} &= 0, \\
	T^{e, \lm}_{22} &= \mu u^{t} \frac{(r^{2}\Omega_{\varphi})^{2} \left( \ell(\ell + 1) - 2 m^{2} \right)}
	{2\ell(\ell - 1)(\ell + 1)(\ell + 2)} \delta(r - \rp) Y^{*}_{\lm}\left(\frac{\pi}{2}, 0\right), \\
	U^{e, \lm}_{22} &= \frac{1}{2} \mu u^{t} (r\, \Omega_{\varphi})^{2} 
	\delta(r - \rp) Y^{*}_{\lm}\left(\frac{\pi}{2}, 0\right).
\end{align}
The odd-parity components of the stress-energy tensor can also be written in matrix form as,
\beq
	T_{\mu\nu}^{o,\lm} = 
	\begin{pmatrix}
		\vspace{0.2cm} 0
		& 0
		& T^{o, \lm}_{02} \csc\theta \dfrac{\partial Y_{\lm}}{\partial \theta} 
		& -T^{o, \lm}_{02} \sin\theta  \dfrac{\partial Y_{\lm}}{\partial \theta} \\
		\vspace{0.2cm} * & 0 
		& T^{o, \lm}_{12} \csc\theta \dfrac{\partial Y_{\lm}}{\partial \varphi} 
		& -T^{o, \lm}_{12} \sin\theta \dfrac{\partial Y_{\lm}}{\partial \theta} \\
		\vspace{0.2cm} * & * 
		& -T^{o, \lm}_{22} X_{\lm}
		& T^{o, \lm}_{22}\sin\theta W_{\lm} \\
		\vspace{0.2cm} * & * & * 
		& T^{o, \lm}_{22} \sin^2\theta  X_{\lm}
	\end{pmatrix},
\eeq
where the odd-sector stress-energy components for the same orbital configuration are given by,
\begin{align}
	T^{e, \lm}_{02} &= \mu u^{t} \frac{f(r)\Omega_{\varphi}}{\ell(\ell + 1)} \delta(r - \rp)
	\dfrac{\partial Y^{*}_{\lm}}{\partial \theta}\bigg|_{\substack{\theta = \pi/2, \\ \varphi = 0}},\\
	T^{e, \lm}_{12} &= 0, \\
	T^{e, \lm}_{22} &= -2im\mu u^{t} \frac{(r\Omega_{\varphi})^{2}}{2\ell(\ell - 1)(\ell + 1)(\ell + 2)} 
	\delta(r - \rp) \dfrac{\partial Y^{*}_{\lm}}{\partial \theta}
	\bigg|_{\substack{\theta = \pi/2, \\ \varphi = 0}}.
\end{align}
\end{widetext}

\section{Regge-Wheeler and Regge-Wheeler-Zerilli source terms}
\label{sec:RWZ_source_terms}
We present here the source terms for the Regge-Wheeler and Regge-Wheeler-Zerilli equations that are
utilised in our calculations within this work.  The full expressions can be found in 
Refs.~\cite{Berndtson:2007gsc, Durkan:2022fvm}; we reproduce them here for completeness, 
adapted to the notation used in this work.

We begin with the even-parity, RWZ source terms, where the coefficients $\kappa_{s}$ and $\eta_{s}$
of the Dirac delta functions in \eqn{generic_source_term_hyperboloidal} are given by,
\begin{widetext}
\begin{align}
	\kappa^{\lm}_{0} &= \frac{16 \pi \mu u^{t}}{\rp^{3}} \frac{f(r) \left( Mr^{2} - \rp^{3} f(r) \right)}{r} Y^{*}_{\lm}\left(\frac{\pi}{2}, 0\right), \\
	\eta^{\lm}_{0} &= 0, \\
	\kappa^{\lm}_{0b} &= 4 \pi \mu u^{t} \frac{f(r)}{r^{2}} 
	\left[ r^{3}\left( \frac{8 - 2\ell(\ell + 1) + 8\mu_{\ell}}{\ell(\ell + 1)} - \frac{M(7 + 6\mu_{\ell})}{\rp^{3}\,\omega} \right) 
	+ \frac{r(2 + 3\mu_{\ell})}{\omega^{2}} - \frac{M(5 + 6\mu_{\ell})}{\omega^{2}}\right] 
	Y^{*}_{\lm}\left(\frac{\pi}{2}, 0\right), \\
	\eta^{\lm}_{0b} &= -\frac{8 \pi \mu u^{t}}{\ell(\ell + 1)\omega^{2}} (1 + \mu_{\ell}) f(r)^{2} Y^{*}_{\lm}\left(\frac{\pi}{2}, 0\right), \\
	\kappa^{\lm}_{1} &= \frac{8i \pi \mu u^{t}}{3\ell(\ell + 1) \rp^{3}\,\omega} \frac{3\ell(\ell + 1) M r^{3} 
	+ 2 \rp^{3} f(r) \left(8M - \ell (\ell + 1)r\right)}{r^{2}} Y^{*}_{\lm}\left(\frac{\pi}{2}, 0\right), \\
	\eta^{\lm}_{1} &= \frac{32i \pi \mu u^{t} f(r)^{3}}{3\ell(\ell + 1)\omega} Y^{*}_{\lm}\left(\frac{\pi}{2}, 0\right), \\
	\kappa^{\lm}_{2} &= \frac{16\pi \mu u^{t}}{\ell(\ell + 1)} \frac{f(r)^{2} 
	\left( M(3 + \mu_{\ell}) - r\mu_{\ell}(1 + \mu_{\ell}) \right)}{(3M + r \mu_{\ell})^{2}} Y^{*}_{\lm}\left(\frac{\pi}{2}, 0\right), \\
	\eta^{\lm}_{2} &= \frac{16\pi \mu u^{t}}{\ell(\ell + 1)} \frac{r f(r)^{3}}{(3M + r\mu_{\ell})} Y^{*}_{\lm}\left(\frac{\pi}{2}, 0\right).
\end{align}
The odd-parity components, on the other hand, are
\begin{align}
	\kappa^{\lm}_{0} &= 0, \\
	\eta^{\lm}_{0} &= 0, \\
	\kappa^{\lm}_{1} &= \frac{32i m \pi M \mu u^{t}}{3\ell(\ell - 1)(\ell + 1)(\ell + 2) \rp^{3}}
	\frac{\mu_{\ell} f(r)(4M - 3r)}{r} \dfrac{\partial Y^{*}_{\lm}}{\partial \theta}
	\bigg|_{\substack{\theta = \pi/2, \\ \varphi = 0}}, \\
	\eta^{\lm}_{1} &= -\frac{32i m \pi M \mu u^{t}}{\ell(\ell - 1)(\ell + 1)(\ell + 2) \rp^{3}}
	r \mu_{\ell} f(r)^{2} \dfrac{\partial Y^{*}_{\lm}}{\partial \theta}
	\bigg|_{\substack{\theta = \pi/2, \\ \varphi = 0}}, \\
	\kappa^{\lm}_{2} &= -\frac{32i m \pi M \mu u^{t}}{\ell(\ell - 1)(\ell + 1)(\ell + 2) \rp^{3}}
	\left( 6 M^{2} - 5Mr + r^{2} - r^{2} f(r)^{2}\right) \dfrac{\partial Y^{*}_{\lm}}{\partial \theta}
	\bigg|_{\substack{\theta = \pi/2, \\ \varphi = 0}}, \\
	\eta^{\lm}_{2} &= \frac{16i m \pi \mu u^{t}}{\ell(\ell - 1)(\ell + 1)(\ell + 2) \rp^{3}} 
	r f(r)^{2} \dfrac{\partial Y^{*}_{\lm}}{\partial \theta}
	\bigg|_{\substack{\theta = \pi/2, \\ \varphi = 0}}.
\end{align}
\end{widetext}

\section{Hyperboloidal Barack-Lousto-Sago Components}
\label{sec:metric_perturbation_hyperboloidal}
In this appendix, we present expressions for the regular, hyperboloidal BLS components of the metric perturbations 
in the Lorenz gauge.  These expressions are derived from the full radiative components of the metric perturbation 
given in Berndtson's thesis, Ref.~\cite{Berndtson:2007gsc}, and our hyperboloidal transformation given in
\eqn{hyperboloidal_bls_transformation}.  One further simplification we have made for the sake of brevity we
have neglected the stress energy terms that did appear in Eqs.~(\ref{eq:h0L_RW_tranformation})-
(\ref{eq:h2L_RW_tranformation}).  This is because, as we noted in \secref{gauge_transformations}, the 
Lorenz gauge metric perturbation is $C^{0}$-differentiable at the worldline.  One is therefore free to use the
homogeneous versions of Berndtson's expressions off the worldline, but on the worldline the stress-energy terms
will be exactly cancelled by the distributional content emerging from the derivatives of the RWZ functions.
Hence, if the distributional content of the component RWZ functions is correct, i.e. the jumps at the worldline
we enforce in our spectral method, then our Lorenz gauge metric perturbations shall have the correct 
differentiability at the worldline.

\subsection{Even Parity}
We shall first present the expressions for the conformal BLS components of the metric perturbation for the even-parity,
$\ell \geq 2$ with $\omega \neq 0$.  The seven non-zero components for $i \in \{1, \dots, 7\}$ are given by,
\begin{widetext}
\begin{align}
	\tilde{h}^{(1)}_{\lm}(\sigma)  &= \left(-\frac{\zeta^2}{\sigma} + 4 (1 + \mu_{\ell}) (\sigma - 1) \sigma 
	- \zeta (-2 + \sigma) (2 \sigma^2 - 1)\right) \tilde{\chi} 
	+ \left(-3 - \frac{\zeta}{\sigma} + 2 (1 + \zeta) \sigma\right) \tilde{\psi}_{0} \nn \\
	&+ \frac{2 (1 + \mu_{\ell}) (\zeta^2 - 4 (1 + \mu_{\ell}) (\sigma - 1) \sigma^2 
	+ \zeta (\sigma - 2 \sigma^3)) \tilde{\psi}_{1}}{M \zeta} \nn \\
	&+ \frac{2 \mu_{\ell} (1 + \mu_{\ell})}{3 \zeta^2 (2 \mu_{\ell} + 3 \sigma)^2} 
	\bigg[ \zeta^2 (2 \mu_{\ell} + 3 \sigma)^2 
	- 2 (\sigma - 1) \sigma^2 (8 \mu_{\ell}^3 + 18 \mu_{\ell} \sigma^2 + 9 \sigma^3 + 4 \mu_{\ell}^2 (2 + 3 \sigma)) \nn \\
	&- \zeta \sigma (2 \mu_{\ell} + 3 \sigma) (2 \sigma^2 - 1) (3 \sigma^2 
	+ \mu_{\ell} (-4 + 6 \sigma)) \bigg] \tilde{\psi}_{2} 
	- 2 (\sigma - 2) (\sigma - 1) \sigma^2 \diff{\tilde{\chi}}{\sigma} \nn \\
	&+ 2 (\sigma - 1) \sigma \diff{\tilde{\psi}_{0}}{\sigma}
	- \frac{4 (1 + \mu_{\ell}) (\sigma - 1) \sigma^3}{M \zeta} \diff{\tilde{\psi}_{1}}{\sigma}
	- \frac{4 \mu_{\ell} (1 + \mu_{\ell}) (\sigma - 1) \sigma^3 (3 \sigma^2 
	+ \mu_{\ell} (-4 + 6 \sigma))}{3 \zeta^2 (2 \mu_{\ell} + 3 \sigma)} \diff{\tilde{\psi}_{2}}{\sigma}, \\
	\tilde{h}^{(2)}_{\lm}(\sigma)  &= \frac{\zeta (\zeta + \sigma^2 - 2 \zeta \sigma^2) \tilde{\chi}}{\sigma} 
	+ \left(1 + \frac{\zeta}{\sigma} - \frac{2 (3 + 2 \mu_{\ell}) (\sigma - 1) \sigma}{\zeta} - 2 \sigma^2 \right) \tilde{\psi}_{0} 
	+ \frac{2 (1 + \mu_{\ell}) (-\sigma + \zeta (2 \sigma^2 - 1)) \tilde{\psi}_{1}}{M} \nn \\
	&+ \frac{2 \mu_{\ell} (1 + \mu_{\ell}) (\mu_{\ell} (4 - 6 \sigma) \sigma - 3 \sigma^3 
	+ \zeta (2 \mu_{\ell} + 3 \sigma) (2 \sigma^2 - 1)) \tilde{\psi}_{2}}{\zeta (6 \mu_{\ell} + 9 \sigma)} 
	- 2 \zeta (\sigma - 1) \sigma \diff{\tilde{\chi}}{\sigma} \nn \\
	&- \frac{2 (\sigma - 1) \sigma^2}{\zeta} \diff{\tilde{\psi}_{0}}{\sigma}
	+ \frac{4 (1 + \mu_{\ell}) (\sigma - 1) \sigma^2}{M} \diff{\tilde{\psi}_{1}}{\sigma}
	+ \frac{4 \mu_{\ell} (1 + \mu_{\ell}) (\sigma - 1) \sigma^2}{3 \zeta} \diff{\tilde{\psi}_{2}}{\sigma}, \\
	\tilde{h}^{(3)}_{\lm}(\sigma)  &= 2 (\zeta + 2 (1 + \mu_{\ell}) \sigma - 2 \zeta \sigma^2) \tilde{\chi} + 2 \tilde{\psi}_{0} 
	- \frac{2 (1 + \mu_{\ell}) \sigma (\zeta + 4 (1 + \mu_{\ell}) \sigma - 2 \zeta \sigma^2) \tilde{\psi}_{1}}{M \zeta} \nn \\
	&- \frac{4 \mu_{\ell} (1 + \mu_{\ell}) \sigma (4 \mu_{\ell} (1 + \mu_{\ell}) \sigma 
	+ 6 \mu_{\ell} \sigma^2 + 6 \sigma^3 
	+ \zeta (2 \mu_{\ell} + 3 \sigma) (2 \sigma^2 - 1)) \tilde{\psi}_{2}}{3 \zeta^2 (2 \mu_{\ell} + 3 \sigma)} 
	- 4 (\sigma - 1) \sigma^2 \diff{\tilde{\chi}}{\sigma} \nn \\
	&+ \frac{4 (1 + \mu_{\ell}) (\sigma - 1) \sigma^3}{M \zeta} \diff{\tilde{\psi}_{1}}{\sigma}
	- \frac{8 \mu_{\ell} (1 + \mu_{\ell}) (\sigma - 1) \sigma^3}{3 \zeta^2} \diff{\tilde{\psi}_{2}}{\sigma}, \\
	\tilde{h}^{(4)}_{\lm}(\sigma)  &= \ell (\ell + 1) \bigg[ -2 \zeta \tilde{\chi} + \frac{(2 (1 + \zeta) \sigma^2 
	-\zeta - 2 \sigma) \tilde{\psi}_{0}}{\zeta} + \frac{(8 (1 + \mu_{\ell}) \sigma 
	+ \zeta (2 \sigma^2 - 1)) \tilde{\psi}_{1}}{2 M} \nn \\
	&+ \frac{2 \mu_{\ell} (4 \mu_{\ell} (1 + \mu_{\ell}) \sigma + 6 \mu_{\ell} \sigma^2 
	+ 6 \sigma^3 + \zeta (2 \mu_{\ell} + 3 \sigma) (2 \sigma^2 - 1)) \tilde{\psi}_{2}}{\zeta (6 \mu_{\ell} + 9 \sigma)} 
	+ \frac{2(\sigma - 1) \sigma^2}{\zeta} \diff{\tilde{\psi}_{0}}{\sigma} \nn \\
	&+ \frac{(\sigma - 1) \sigma^2}{M} \diff{\tilde{\psi}_{1}}{\sigma}
	+ \frac{4\mu_{\ell} (\sigma - 1) \sigma^2}{3 \zeta} \diff{\tilde{\psi}_{2}}{\sigma} \bigg], \\
	\tilde{h}^{(5)}_{\lm}(\sigma)  &= \ell (\ell + 1) \bigg[ -2 (-\zeta - 2 \sigma + 2 (1 + \zeta) \sigma^2) \tilde{\chi} 
	+ \tilde{\psi}_{0} \nn \\
	&+ \frac{ (\zeta^2 + 16 (1 + \mu_{\ell}) (\sigma - 1) \sigma^2 
	+ 4 \zeta \sigma (1 + 2 \mu_{\ell} + \sigma) (2 \sigma^2 - 1)) \tilde{\psi}_{1}}{2 M \zeta} \nn \\
	&+ \frac{2  \mu_{\ell}}{3 \zeta^2 (2 \mu_{\ell} + 3 \sigma)^2} 
	\Big(\zeta^2 (2 \mu_{\ell} + 3 \sigma)^2 
	+ 2 \zeta \sigma (2 \mu_{\ell} + 3 \sigma) (2 \mu_{\ell} (2 + \mu_{\ell}) 
	+ (3 + \mu_{\ell}) \sigma) (2 \sigma^2 - 1) \nn \\ 
	&- 8 \mu_{\ell} (\sigma - 1) \sigma^2 (4 \mu_{\ell}^2 
	+ 3 \sigma (1 + \sigma) + \mu_{\ell} (4 + 6 \sigma))\Big) \tilde{\psi}_{2}
	- 4 (\sigma - 1) \sigma^2 \diff{\tilde{\chi}}{\sigma} \nn \\
	&+ \frac{4 (\sigma - 1) \sigma^3 (1 + 2 \mu_{\ell} + \sigma)}{M \zeta}\diff{\tilde{\psi}_{1}}{\sigma}
	+ \frac{8 \mu_{\ell} (\sigma - 1) \sigma^3 (2 \mu_{\ell} 
	(2 + \mu_{\ell}) + (3 + \mu_{\ell}) \sigma)}{3 \zeta^2 (2 \mu_{\ell} + 3 \sigma)} \diff{\tilde{\psi}_{2}}{\sigma} \bigg],
\end{align}
\end{widetext}
\begin{widetext}
\begin{align}
	\tilde{h}^{(6)}_{\lm}(\sigma)  &= \left(4 (1 + \mu_{\ell}) \sigma + \zeta (2 - 4 \sigma^2)\right) \tilde{\chi} + \tilde{\psi}_{0} 
	- \frac{2 (1 + \mu_{\ell}) \sigma (\zeta + 4 (1 + \mu_{\ell}) \sigma - 2 \zeta \sigma^2) \tilde{\psi}_{1}}{M \zeta} \nn \\
	&- \frac{4 \mu_{\ell} (1 + \mu_{\ell}) \sigma (4 \mu_{\ell} (1 + \mu_{\ell}) \sigma + 6 \mu_{\ell} \sigma^2 
	+ 6 \sigma^3 + \zeta (2 \mu_{\ell} + 3 \sigma) (2 \sigma^2 - 1)) \tilde{\psi}_{2}}{3 \zeta^2 (2 \mu_{\ell} + 3 \sigma)} 
	- 4 (\sigma - 1) \sigma^2 \diff{\tilde{\chi}}{\sigma} \nn \\
	&+ \frac{4 (1 + \mu_{\ell}) (\sigma - 1) \sigma^3}{M \zeta} \diff{\tilde{\psi}_{1}}{\sigma}
	- \frac{8 \mu_{\ell} (1 + \mu_{\ell}) (\sigma - 1) \sigma^3}{3 \zeta^2} \diff{\tilde{\psi}_{2}}{\sigma}, \\
	\tilde{h}^{(7)}_{\lm}(\sigma)  &= \frac{(\ell - 1) \ell (\ell + 1) (\ell + 2) \, \sigma}{3} \bigg[ -6 \, \tilde{\chi} 
	+ \frac{3 (4 (1 + \mu_{\ell}) \sigma + \zeta (2 \sigma^2 - 1)) \tilde{\psi}_{1}}{M \zeta} \nn \\
	&+ \left(\frac{3}{\sigma} + \frac{4 \mu_{\ell} (3 + 2 \mu_{\ell}) (2 \sigma^2 - 1)}{\zeta (2 \mu_{\ell} + 3 \sigma)} 
	+ \frac{4 \mu_{\ell} \sigma (4 \mu_{\ell} (1 + \mu_{\ell}) (3 + \mu_{\ell}) + 6 \mu_{\ell} \sigma 
	+ 3 (3 + \mu_{\ell}) \sigma^2)}{\zeta^2 (2 \mu_{\ell} + 3 \sigma)^2}\right) \tilde{\psi}_{2} \nn \\
	&+ \frac{6 (\sigma - 1) \sigma^2}{M \zeta} \diff{\tilde{\psi}_{1}}{\sigma}
	+ \frac{8 \mu_{\ell} (3 + 2 \mu_{\ell}) (\sigma - 1) \sigma^2}{\zeta^2 (2 \mu_{\ell} + 3 \sigma)} 
	\diff{\tilde{\psi}_{2}}{\sigma}\bigg]
\end{align}
\end{widetext}
\subsection{Odd Parity}
The non-zero, odd-parity, conformal BLS components for $\ell \geq 2$ with $\omega \neq 0$ for $i \in \{8, 9, 10\}$
given by,
\begin{widetext}
\begin{align}
	\tilde{h}^{(8)}_{\lm}(\sigma)  &= -\frac{8M\ell(\ell + 1)}{\zeta} 
	\left( \tilde{\psi}_{1} + \frac{2\mu_{\ell}}{3} \tilde{\psi}_{2} \right), \\
	\tilde{h}^{(9)}_{\lm}(\sigma)  &= -\frac{8M\ell(\ell + 1)}{3\zeta^{2}}
	\bigg[ 3(\zeta + 4\sigma - 2(2 + \zeta)\sigma^{2})\tilde{\psi}_{1}
	+ 2\mu_{\ell}(\zeta - 2\sigma - 2(\zeta - 1)\sigma^{2})\tilde{\psi}_{2} \nn \\
	&- 2(\sigma - 1)\sigma^{2} \bigg(3\diff{\tilde{\psi}_{1}}{\sigma} 
	+ 2\mu_{\ell}\diff{\tilde{\psi}_{2}}{\sigma} \bigg) \bigg], \\
	\tilde{h}^{(10)}_{\lm}(\sigma)  &= \frac{8M(\ell - 1)\ell(\ell + 1)(\ell + 2)}{3\zeta^{2}}
	\bigg[ 3\zeta \tilde{\psi}_{2} - 2\sigma\bigg( 3\tilde{\psi}_{1} + (3 + 2\mu_{\ell}
	+ 3(\zeta - 1)\sigma)\tilde{\psi}_{2}
	- 3(\sigma - 1)\sigma \diff{\tilde{\psi}_{2}}{\sigma} \bigg) \bigg].
\end{align}
\end{widetext}

\section{Metric Perturbation Regularisation Parameters}
\label{sec:metric_perturbation_regularisation_parameters}
In this appendix, we present the metric perturbation regularisation parameters used in our self-force calculations.
These expressions are given in terms of the complete elliptic integrals of the first and second kind, $E$ and $K$, respectively,
and the auxilarily quantities $\Lambda_{1} := \frac{\ell(\ell + 1)}{(2\ell + 1)(2\ell + 3)}$ and 
$\Lambda_{2} := \frac{(\ell - 1)\ell(\ell + 1)(\ell + 2)}{(2\ell - 3)(2\ell - 1)(2\ell + 3)(2\ell + 5)}$.
The explicit expressions for each component are given by,
\begin{subequations}
	\allowdisplaybreaks
	\begin{eqnarray}
	  h^{[0]}_{tt} &=& \frac{4 (\rp - M) K}{\pi\, \rp^2}\sqrt{\frac{\rp - 2M}{\rp - 3M}}, 
	  \label{eq:h_0_tt}
	  \\
	  h^{[0]}_{t\varphi} &=& - \frac{32 M^{1/2} K}{\pi \, \rp^{1/2}} \sqrt{\frac{\rp - 2M}
	  {\rp - 3M}} \Lambda_1,\quad 
	  \\
	  h^{[0]}_{rr} &=& \frac{4 K}{\pi}\frac{(\rp - 3M)^{1/2}}{(\rp - 2M)^{3/2}}, 
	  \\
	  h^{[0]}_{\theta\theta} &=& \frac{4 \rp K}{\pi} \sqrt{\frac{\rp - 2M}{\rp - 3M}} \nn 
	  \\
	  && \quad- \frac{64 M \rp K}{\pi (\rp - 2M)^{1/2}(\rp - 3M)^{1/2}} \Lambda_2, 
	  \label{eq:h_0_thetatheta}
	  \\
	  h^{[0]}_{\varphi\varphi} &=& \frac{4 \rp K}{\pi} \sqrt{\frac{\rp - 2M}{\rp - 3M}} \nn 
	  \\
	  && \quad+ \frac{64 M \rp K}{\pi (\rp - 2M)^{1/2}(\rp - 3M)^{1/2}} \Lambda_2,
	  \\
	  h^{[-1]}_{tt,r} &=& \mp\frac{(\rp - M)}{\rp^{5/2} (\rp - 3 M)^{1/2}}, 
	  \\
	  h^{[0]}_{tt,r} &=& \frac{2 (\rp - M) [(\rp - 2 M) E - 2 (\rp - 4 M) K]}
	  {\pi \rp^3 (\rp - 3 M)^{1/2} (\rp - 2 M)^{1/2}},
	  \\
	  h^{[-1]}_{rr,r} &=& \mp\frac{(\rp - 3 M)^{1/2}}{\rp^{1/2} (\rp - 2 M)^2}, 
	  \\
	  h^{[0]}_{rr,r} &=& \frac{2 (\rp - 3 M)^{1/2} [(\rp - 2 M) E - 2 \rp K]}
	  {\pi \rp (\rp - 2 M)^{5/2}}, 
	  \\
	  h^{[-1]}_{t\varphi,r} &=& \pm \bigg[\frac{2 M^{1/2}}{\rp (\rp - 3 M)^{1/2}}\bigg]_{\ell\ge1}, 
	  \\
	  h^{[0]}_{t\varphi,r} &=& -\frac{16 M ^{1/2} [(\rp - 2 M) E + 2 M K]}
	  {\pi \rp^{3/2} (\rp - 3 M)^{1/2} (\rp - 2 M)^{1/2}} \Lambda_1, 
	  \\
	  h^{[-1]}_{\varphi\varphi,r} &=& \mp \sqrt{\frac{\rp}{\rp-3M}} \nn \\
	  && \quad \mp \bigg[\frac{M \rp^{1/2}}{(\rp-2M)(\rp-3M)^{1/2}}\bigg]_{\ell\ge2}, \\
	  \\
	  h^{[0]}_{\varphi\varphi,r} &=& \frac{2 (E + 2 K)}{\pi} \sqrt{\frac{\rp - 2 M}{\rp - 3 M}} \nn \\
	  && \quad + \frac{32 M (E + 2 K)}{\pi (\rp - 3 M)^{1/2} (\rp - 2 M)^{1/2}} \Lambda_2,
	  \\
	  h^{[-1]}_{\theta\theta,r} &=& \mp \sqrt{\frac{\rp}{\rp-3M}} \pm 
	  \bigg[ \frac{M \rp^{1/2}}{(\rp-2M)(\rp-3M)^{1/2}} \bigg]_{\ell\ge2}, \nn \\
	  \\ 
	  h^{[0]}_{\theta\theta,r} &=& \frac{2 (E + 2 K)}{\pi} \sqrt{\frac{\rp - 2 M}{\rp - 3 M}} \\
	  && \quad- \frac{32 M (E + 2 K)}{\pi (\rp - 3 M)^{1/2} (\rp - 2 M)^{1/2}} \Lambda_2, 
	  \\
	  h^{[0]}_{tr,\varphi} &=& - \frac{32((\rp - 2 M) E - (\rp - 3 M) K)}
	  {\pi M^{1/2} \rp^{3/2}} \sqrt{\frac{\rp - 2 M}{\rp - 3 M}} \Lambda_1, \nn \\
	  \\ 
	  h^{[0]}_{r\varphi,\varphi} &=& \frac{16[(\rp - 2 M) E - (\rp - 3 M) K]}
	  {\pi (\rp - 2 M)^{1/2} (\rp - 3 M)^{1/2}} (\Lambda_1 + 4 \Lambda_2).
	\end{eqnarray}
\end{subequations}

\section{Static Modes}
\label{sec:static_modes}
In this section, we shall present the explicit analytical solutions for the static modes of the Lorenz gauge metric perturbation
that have been derived previously in the literature.  This section will emphasise the important subtleties of the static modes,
but it should be considered as a review of the existing literature rather than a novel contribution.  We therefore 
refer the reader to references therein for further details.

\subsection{Even parity $(\ell \geq 2)$ modes}
\label{sec:even_parity_static}
For the even parity static modes with $\ell \geq 2$, we shall use the approach from an upcoming publication 
by Ottewill and Wardell \cite{OttewillStatic}, which obtains analytic solutions for the even-parity Lorenz gauge static modes 
using a similar set of expressions for reconstructing the metric as the even-parity non-zero frequency case.  
We stop short of deriving these expressions, as this is presented at length in Sec 3.2.1 of Ref.~\cite{Berndtson:2007gsc}, 
but repeat the expressions here for completeness.  
Our strategy will be to calculate scaled homogeneous solutions to the BLS metric perturbations in Lorenz gauge,
then impose jump conditions at the worldline to determine the full inhomogeneous metric perturbation.
This is a similar strategy to the one used in the non-zero frequency case in Ref.~\cite{Durkan:2022fvm}.

The seven homogeneous non-zero metric perturbation functions are given in
terms of the zero-frequency solutions of the generalised RWZ equation, and the gauge function $M_{2a}$ 
in the nomeclature of Berndtson, as
\begin{widetext}
\begin{align}
	{\mathbb H}_{0} &= \frac{(3 + 2\mu_{\ell})M^2 - 2\mu_{\ell}(1 + \mu_{\ell})Mr + \mu_{\ell}(1 + \mu_{\ell})r^2}{(1 + \mu_{\ell})r(3M + \mu_{\ell} r)} \diff{\psi_{2}}{r}
	+ \frac{2M M_{2a}}{r^4} - \frac{M\psi_0}{2(1 + \mu_{\ell})r^2} \nn \\
	&+ \frac{(3(3 + 4\mu_{\ell})M^3 + 15\mu_{\ell}(1 + \mu_{\ell})M^2r + 4\mu_{\ell}^2(1 + \mu_{\ell})Mr^2 
	+ \mu_{\ell}^2(1 + \mu_{\ell})^2r^3)\psi_{2}}{(1 + \mu_{\ell})r^2(3M + \mu_{\ell} r)^2}
	+ \frac{2M M_{2a}}{r^3} - \frac{M}{2r + 2\mu_{\ell} r} \diff{\psi_{0}}{r} \label{eq:H0_even_static}\\
	{\mathbb H}_{1} &= -\frac{\psi_{1}}{r^{2}} 
	+ \frac{M}{(1 + \mu_{\ell})r^{2}} \diff{\psi_{1}}{r}, \\
	{\mathbb H}_{2} &= \frac{-(3 + 2\mu_{\ell})M^2 + 2(3 + \mu_{\ell} - \mu_{\ell}^2)Mr + \mu_{\ell}(1 + \mu_{\ell})r^2}{(1 + \mu_{\ell})r(3M + \mu_{\ell} r)} \diff{\psi_{2}}{r} 
	+ \frac{(6M - 4(2 + \mu_{\ell}))M_{2a}}{r^4} + \frac{(2(2 + \mu_{\ell})r - 3M)}{2(1 + \mu_{\ell})r^2} \diff{\psi_{0}}{r}  \nn \\
	&+ \frac{9(3 + 4\mu_{\ell})M^3}{(1 + \mu_{\ell})^2(3M + \mu_{\ell} r)^2} + 3\mu_{\ell}(11 + 13\mu_{\ell})M^2r + 6\mu_{\ell}(-1 + \mu_{\ell} + \mu_{\ell}^2)M r^2 
	+ \mu_{\ell}^2(1 + \mu_{\ell}^2)r^3 \psi_{2} \nn \\
	&- \frac{(6M + 4\mu_{\ell} r)}{r^3} \diff{M_{2a}}{r}
	+ \frac{(3M - 2r)}{2r + 2\mu_{\ell} r}\diff{\psi_{0}}{r},
\end{align}
\end{widetext}
\begin{widetext}
\begin{align}
	{\mathbb K} &= \frac{(3 + 2\mu_{\ell})M(2M - r)}{(1 + \mu_{\ell})r(3M + \mu_{\ell} r)} \diff{\psi_{2}}{r}
	+ \frac{(-4M + 2(2 + \mu_{\ell})r)M_{2a}}{r^4} - \frac{(2M + r)\psi_0}{2(1 + \mu_{\ell})r^2} \nn \\
	&+ \frac{6(3 + 4\mu_{\ell})M^3 + 3\mu_{\ell}(8 + 9\mu_{\ell})M^2r + \mu_{\ell}(-3 + 5\mu_{\ell})Mr^2}{(1 + \mu_{\ell})^2 r^2(3M + \mu_{\ell} r)^2}
	+ \frac{(4M - 2r)M_{2a}}{r^3}
	+ \frac{(2M + r)}{4 + 4\mu_{\ell}}\diff{\psi_{0}}{r}, \\
	{\mathbb G} &= -\frac{M_{2a}}{r^3}, \\
	{\rm h}_{0} &= 
	- \frac{(r - 2M)}{2r + 2\mu_{\ell} r} \diff{\psi_{1}}{r}, \\
	{\rm h}_{1} &= - \frac{r((6 + 5\mu_{\ell})M + \mu_{\ell}(1 + \mu_{\ell})r)}{2(1 + \mu_{\ell})(3M + \mu_{\ell} r)} \diff{\psi_{2}}{r}
	+ \frac{4M_{2a}}{r^2} - \frac{\psi_0}{4 + 4\mu_{\ell}}
	+ \frac{\mu_{\ell}(3M^2 + 6(1 + \mu_{\ell})Mr + \mu_{\ell}(1 + \mu_{\ell})r^2)}{2(1 + \mu_{\ell})(3M + \mu_{\ell} r)^2} \diff{\psi_{2}}{r} \nn \\
	&- \frac{2}{r}\diff{M_{2a}}{r} + \frac{r}{4 + 4\mu_{\ell}}\diff{\psi_{0}}{r}. \label{eq:h1_even_static}
\end{align}
\end{widetext}
With these expressions in hand, one requires solutions for the generalised RWZ equation and $M_{2a}$ in order
to compute the full metric perturbation.

Let us first consider the zero frequency, homoegeneous solutions of the generalised RWZ equation.  
The generalised homogeneous RW equation with zero frequency can be written in terms of the compactified 
hyperboloidal coordinate $\sigma$ as
\begin{multline}
	\sigma^{2}(1 - \sigma) \sdiff{\psi_{s}}{\sigma} + 2\sigma(1 - 3\sigma^{2})\diff{\psi_{s}}{\sigma} \\
	+ \left( \ell (\ell + 1) + \sigma(1 - s^{2}) \right) \psi_{s} = 0.
	\label{eq:RWZ_eqn_zero_freq}
\end{multline}
\eqn{RWZ_eqn_zero_freq} omits two independent homogeneous solutions, $\psi^{+}_{s}$ and $\psi^{-}_{s}$, which are given by
\begin{align}
	\psi^{+}_{s}(\sigma) &= \sigma^{\ell}\, _{2}F_{1}\left(1 + \ell - s, \ell + s + 1, 2(\ell + 1); \sigma \right) \\
	\psi^{-}_{s}(\sigma) &= \sigma^{-\ell-1}\, _{2}F_{1}\left(- \ell - s, s - 1, -2\ell; \sigma\right),
\end{align}
where $_{2}F_{1}(c_{1}, c_{2}, c_{3}; x)$ is the Gauss hypergeometric function.  The solutions $\psi^{+}_{s}$ and $\psi^{-}_{s}$ 
are bounded near the horizon and null-infinity, respectively.  Note that since \eqn{RWZ_eqn_zero_freq} is obtained using the standard
RW-potential, then homogeneous solutions in this case are for odd-parity modes.  The even-parity homogeneous solutions can then be obtained 
by considering an intertwinning operator \cite{Field:2011rna, Anderson:1991kx},
\beq
	D^{\pm} := \sigma^{2}(1  - \sigma) \diff{}{\sigma} \pm \left[ \frac{2}{3}\mu_{\ell}(\mu_{\ell} + 1) - \frac{3\sigma^{2}(\sigma - 1)}{3\mu_{\ell} + 2\sigma} \right],
	\label{eq:intertwinning_operator}
\eeq
which will transform the solutions such that $\psi^{\pm}_{s,\, \rm{RWZ}}(\sigma) = D^{\pm} \psi^{\pm}_{s,\, \rm{RW}}(\sigma)$.  In practice, this intertwinning
operator is only required to find the RWZ spin-weight $s = 2$ solutions.

The solutions, $\psi_{s}$, can then be appropiately rescaled with appropiate constants, whereby in Eqs.~(\ref{eq:H0_even_static})-(\ref{eq:h1_even_static}),
$\psi_{s} \rightarrow c^{\pm}_{s}\psi^{\pm}_{s}$, to give rise to two differing solutions for each of the seven metric perturbation functions, 
$({\mathbb H}^{\pm}_{0}, {\mathbb H}^{\pm}_{1}, {\mathbb H}^{\pm}_{2}, {\mathbb K}^{\pm}, {\mathbb G}^{\pm}, 
{\rm h}^{\pm}_{0}, {\rm h}^{\pm}_{1})$ that are regular at the horizon and null-infinity.  
The constants, $c^{\pm}_{s}$, will be constrained by the jump conditions at the worldline, which will be discussed shortly.

The last function required for the even-parity static modes is the gauge function homogeneous solutions, $M^{\pm}_{2a}$.  
Unlike $M_{2af}$ in Berndtson's original treatment, the different spin-weight RWZ solutions do not decouple in the static sector in the same manner as the 
radiative sector.  However, the function $M_{2a}$ can be found be as a generalised $s = 0$ RWZ solution, which is given by
\cite{Berndtson:2007gsc}
\begin{widetext}
\begin{align}
	{\cal L}_{0} M_{2a} &= \frac{(r - 2M)^{2}}{2(1 + \mu_{\ell})r} \diff{\psi_{0}}{r} 
	- \frac{(r - 2M)^{2}((6 + 5\mu_{\ell})M + \mu_{\ell}(1 + \mu_{\ell})r)}{r(1 + \mu_{\ell})(3M + \mu_{\ell} r)} \diff{\psi_{2}}{r}
	- \frac{(r - 2M)^{2}}{2(1 + \mu_{\ell})r^{2}}\psi_{0} \nn \\
	&- \frac{\mu_{\ell}(r - 2M)^{2}(3M^{2} + 6(1 + \mu_{\ell})Mr + \mu_{\ell}(1 + \mu_{\ell})r^{2})}{r^{2}(1 + \mu_{\ell})(3M + \mu_{\ell} r)^{2}}\psi_{2}.
	\label{eq:M_2a}
\end{align}
Here, one can use the following ans\"atze \cite{OttewillStatic},
	\begin{align}
		M^{-}_{2a}(r) &= a^{-}_{\ell}(r)r^{2} f(r) + r M \ln\left(2M / r\right)P_{\ell}\left(x\right)
		\times \frac{c^{-}_{0} - \frac{3}{4}(\ell - 1)^{3}\ell^{2}(\ell + 1)^{2}(\ell + 2)^{3} c^{-}_{2}}{(2\ell - 1)(2\ell + 1)(2\ell + 3)}
		+ c^{-}_{3}\psi^{-}_{0}, \\
		M^{+}_{2a}(r) &= a_{\ell}^{+}(r) + r M\, {\rm Li}_{2}\left(2M / r\right)P_{\ell}\left(x\right) 
		\times \frac{c^{+}_{0} - \frac{3}{4}(\ell - 1)^{3}\ell^{2}(\ell + 1)^{2}(\ell + 2)^{3} c^{+}_{2}}{(2\ell - 1)(2\ell + 1)(2\ell + 3)} \nn\\
		&- r M \ln\left(2M / r\right)Q_{\ell}\left(x\right) 
		\times \frac{c^{+}_{0} - \frac{3}{4}(\ell - 1)^{3}\ell^{2}(\ell + 1)^{2}(\ell + 2)^{3} c^{+}_{2}}{(2\ell - 1)(2\ell + 1)(2\ell + 3)}
		+ c^{+}_{3}\psi^{+}_{0},
	\end{align}
\end{widetext}
where $x:= r/2M - 1$ and ${\rm Li}_{2}$ is the dilogarithm function, defined as the infinite series ${\rm Li}_{2}(z) := \sum^{\infty}_{k = 1} \frac{z^{k}}{k^{2}}$,
and $c^{\pm}_{3}$ are further constants to be determined.
The functions, $a^{\pm}_{\ell}(r)$, meanwhile, are simple power series in $r$,
\begin{align}
	a_{\ell}^{+}(r) &= \ln f(r) \sum^{\ell+3}_{n = 0} a^{+,\ln}_{\ell, n} r^{n} + \sum^{\ell+2}_{n = 0} a^{+}_{\ell, n} r^{n}, \\
	a_{\ell}^{-}(r) &= \sum^{\ell + 1}_{n = 0} a^{-}_{n} r^{n},
\end{align}
The cofficients $a^{+,\ln}_{\ell, n}$ and $a^{+}_{\ell, n}$ are determined for each $\ell$ by substitution into the differential equation for $M_{2a}$, \eqn{M_2a},
and comparing coefficients.

The BLS components of the homogeneous metric perturbation can now be constructed from the expressions in \eqn{even_bls_berndtson} through the replacement
$(\psi_{s}, M_{2a}) \rightarrow (c^{\pm}_{s}\psi^{\pm}_{s}, M^{\pm}_{2a})$ in Eqs.~(\ref{eq:H0_even_static})-(\ref{eq:h1_even_static}), such that
$\barh^{(1), \pm} = r f ({\mathbb H}_{0}^{\pm} + {\mathbb H}_{2}^{\pm})$, ect.  
The construction of homogeneous BLS mode solutions to the metric perturbation now enables the determination of the inhomogeneous solution, 
as the jump conditions at the worldline can be analytically derived from the source terms.
Put simply, the jump conditions at the worldline fully constrain our inhomogeneous solution either side of the worldline.
To do this, one can reformulate the problem in cannonical matrix form, where the metric perturbations are decomposed into
the following matrix, 
\beq
	\renewcommand{\arraystretch}{2} 
	\bmat{E} = 
	\left(
	\begin{array}{c|c}
		-\barh^{(i), -}_{[j]} & \barh^{(i), +}_{[j]}  \\
		\hline
		-\partial_{r}\barh^{(i), -}_{[j]} & \partial_{r}\barh^{(i), +}_{[j]}
	\end{array}
	\right)_{r = \rp},
	\label{eq:even_coefficient_matrix}
\eeq
where the ``basis vectors", $\barh^{(i), -}_{[j]}$, are defined as the cofficients of the yet undetermined constants, such that
\beq
	\barh^{(i), \pm} =  \sum^{3}_{j} c_{j}\barh^{(i), \pm}_{[j]}.
	\label{eq:even_basis_vectors}
\eeq
In reality, $i$, need not run-over the full set of BLS harmonic modes, since the only non-zero fields are $i \in [1, 3, 5, 6, 7]$.
Since the $i = 6, 7$ modes can be obtained from the equations eminating from the gauge conditions, we shall utilise the $i \in [1, 3, 5]$ modes,
hence the matrix in \eqn{even_coefficient_matrix} will be $6 \times 6$ in size.  
Assembling a coefficient vector, $\bmat{C}$, of the constants, $c^{\pm}_{j}$, 
\beq
	\bmat{C} = \left( c^{-}_{j} \,\big|\, c^{+}_{j} \right)^{T},
	\label{eq:even_coefficient_vector}
\eeq
we can write our matrix problem as
\beq
	\bmat{E} \cdot \bmat{C} = \bmat{J},
	\label{eq:even_matrix_problem}
\eeq
where $\bmat{J} = (\bm{0}_{3}, {\cal J}^{(i)})^{T}$, with ${\cal J}^{(i)}$ $(i \in [1, 3, 5])$ being the source terms for the $i$-th BLS mode.  
Explicitly,
\beq
	{\cal J}^{(i)} = -\frac{16\pi\alpha_{i} {\cal E} }{\rp} Y^{*}_{\ell 0}(\pi/2, 0),
\eeq
where $\alpha_{i}$ is given by
\beq
	\alpha_{1} = 1, \quad \alpha_{3} = 1 / f_{p},\quad \alpha_{5} = 0.
\eeq
$\bmat{C}$ is therefore fully determined analytically by matrix inversion of \eqn{even_matrix_problem}, leaving us with analytic solutions for
metric perturbation functions in the Lorenz gauge for the even-parity static modes.

\subsection{Odd parity $(m = 0)$ modes}
\label{sec:odd_parity_static}
The odd-parity, $m = 0$, static modes of the metric perturbation also admit a fully analytic solution.
For the metric perturbation in the BLS formalism, the $\ell \geq 1$, $m = 0$ modes associated with $i = 9, 10$ are devoid of source terms. 
These modes remain uncoupled from all other BLS modes, leading to the historic conclusion (e.g. see Refs.~\cite{Barack:2005nr,Akcay:2013wfa})
that the solutions for these modes can trivially be zero: $\bar{h}^{(9)}_{\ell 0} = \bar{h}^{(10)}_{\ell 0} = 0$.
However, this choice in fact leads to a gauge singularity at the horizon.  Whilst this is not a problem when considering the computation of
the first-order self-force and related quantities, it is a problem when considering the problem of second-order computations 
\cite{Miller:2020bft} as the gauge singularity will manifest as divergent behaviour at the horizon.  
We shall utilise the horizon regular solution, which is tantamount to the previous solution first obtained by Zerilli in 
Ref.~\cite{PhysRevD.2.2141}, but with an additional purely gauge piece by having a non-zero, homogeneous solution 
for $\bar{h}^{(9)}_{\ell 0}$.

Detweiler and Poisson in Ref.~\cite{PhysRevD.69.084019} were the first to highlight the analytical solution of Zerilli 
is also a Lorenz gauge solution.  The solution was then later rederived by Barack and Lousto in Ref.~\cite{Barack:2005nr}, who
showed how to construct the analytic solutions for $\bar{h}^{(8)}_{\ell 0}$ and we refer the reader to this reference for a more 
detailed derivation. 
For the case in question the Lorenz gauge field equations in \eqn{lorenz_radial_field_equations} reduce to a single ODE \cite{Barack:2005nr},
\begin{multline}
	\sdiff{\bar{h}^{(8)}_{\ell 0}}{r} - \frac{1}{f(r)}\left( \frac{\ell(\ell + 1)}{r^{2}} - \frac{4M}{r^{3}} \right)\bar{h}^{(8)}_{\ell 0}(r) 
	\\ = \beta_{\ell} \times \delta(r - \rp),
	\label{eq:odd_parity_static_ode}
\end{multline}
where
\beq
	\beta_{\ell} = \frac{16(-1)^{(\ell - 1)/2} \sqrt{\pi(2\ell + 1)}\ell!!}{(\ell - 1)!!} \frac{f^{2}_{p}\, \cal{L}}{\rp^{2}}.
	\label{eq:odd_parity_static_beta}
\eeq

As in the previous section, we shall construct an inhomogeneous solution through two independent homogeneous solutions, each bounded
either side of the particle, and then match these solutions at the worldline to construct a regular solution throughout the spacetime.
For $\ell \geq 2$, \eqn{odd_parity_static_ode} omits two independent homogeneous solutions
\begin{align}
	\bar{h}^{(8), -}_{\ell 0} &= \frac{x}{x + 1} \sum^{\ell + 1}_{n = 0} a^{-}_{\ell, n}x^{n},\\
	\bar{h}^{(8), +}_{\ell 0} &= \bar{h}^{(8), -}_{\ell 0} \ln f + \frac{1}{x + 1} \sum^{\ell + 1}_{n = 0} b^{+}_{\ell, n}x^{n},\\
\end{align}
where the coefficients have a closed form,
\begin{align}
	a^{\ell}_{n} &= \frac{\ell(\ell + 1)(\ell + n - 1)!}{(\ell - n + 1)!(n + 1)!n!}, \\
	b^{\ell}_{n} &= \sum^{\ell - n + 1}_{k = 0} (-1)^{k} \frac{a^{\ell}_{n + k}}{k + 1}.
	\label{eq:odd_parity_static_coefficients}
\end{align}
The unique regular solution for the odd-parity static modes for $\ell \geq 2$ is then given by
\begin{multline}
	\bar{h}^{(8)}_{\ell 0}(r) = -2\lambda\mu_{\ell}\beta_{\ell}
	\Big( \bar{h}^{(8), -}_{\ell 0} \bar{h}^{(8), +}_{\ell 0}(\rp)\Theta[r - \rp] \\
	+ \bar{h}^{(8), +}_{\ell 0} \bar{h}^{(8), -}_{\ell 0}(\rp)\Theta[\rp - r] \Big), 
	\label{eq:odd_parity_{i}_8_ell_m_0_solution}
\end{multline}
where $\Theta$ is the heaviside theta function.

In the specific case of the dipole mode, $\ell = 1$, $\bar{h}^{(8), -}_{\ell 0}$ remains a homogeneous solution 
to \eqn{odd_parity_static_ode}, whereas $\bar{h}^{(8), +}_{\ell 0}$ does not. Instead, we adopt the more general 
homogeneous solutions,
\begin{align}
	\bar{h}^{(8), -}_{10}(r) &= r^{2}, \\
	\bar{h}^{(8), +}_{10}(r) &= \frac{1}{r},
	\label{eq:odd_parity_ell_1_homogeneous}
\end{align}
which can generate an inhomogeneous solution analogously to the even parity static case for $\ell \geq 2$. 
To determine the solution, we require regularity at the horizon, continuity at the worldline, 
and a jump condition derived from the delta function source term in \eqn{odd_parity_static_ode}. 
This results in the solution given by Barack and Lousto in Ref.~\cite{Barack:2005nr} and later by 
Miller and Pound in Ref.~\cite{Miller:2020bft},
\beq
	\bar{h}^{(8)}_{10}(r) = -16\sqrt{\frac{\pi}{3}}{\cal L}
	\Bigg[ \frac{r^{2}}{\rp^{3}} \Theta[r - \rp]
	+ \frac{1}{r}\Theta[\rp - r] \Bigg].
	\label{eq:odd_parity_{i}_8_ell_1_m_0_solution}
\eeq

We can then ensure, \emph{a posteriori}, the regularity of the metric perturbation at the horizon by 
adding a purely gauge $i = 9$ component to the solution. Specifically, we assume a general homogeneous 
solution for $\bar{h}^{(9)}_{\ell 0} = a_{2} / r^{2}$ and determine the constant $a_{2}$ through the 
constraint $\bar{h}^{(9)}_{\ell 0} = \bar{h}^{(8)}_{\ell 0}$ at the horizon. 
The resulting non-zero solution for $\bar{h}^{(9)}_{\ell 0}$ is
\beq
	\bar{h}^{(9)}_{\ell 0} = \sqrt{\frac{\pi}{3}} \times \frac{256 M^{4} {\cal L}}{\rp^{3} r^{2}}.
	\label{eq:odd_parity_ell_1_{i}_9}
\eeq
This solution produces a regular metric perturbation that describes the shift in angular momentum across the surface $r = \rp$,
\beq
	h^{\ell = 1, m = 0}_{t\phi} = -2{\cal L}\sin^{2}\theta
	\bigg[ \frac{r^{2}}{\rp^{3}} \Theta[r - \rp] 
	+ \frac{1}{r}\Theta[\rp - r] \bigg].
	\label{eq:h_ell_1_m_0_tphi}
\eeq

\subsection{The monopole $(\ell, m) = (0, 0)$ mode}
\label{sec:monopole_static}
The final static component of the metric perturbation to compute is the monopole, or $(\ell, m) = (0, 0)$, piece. 
Up to a gauge piece, the monopole mode physically describes the variation in the mass of the background Schwarzschild 
geometry due to the presence of the secondary.
This part of the perturbation can also be addressed analytically; however, it involves significantly more subtlety 
than the previous static cases. The primary challenge arises from a gauge pathology: in the Lorenz gauge, there is no 
globally regular homogeneous solution with nonzero mass \cite{Miller:2020bft}. 
A comprehensive discussion of this gauge issue is provided in Appendix D of Ref.~\cite{Miller:2020bft}. 
Here, we briefly summarise the key points, outlining the two choices of monopole used in this work, while 
referring the reader to the original reference for an in-depth discussion.

Firstly, Let us set up the problem.  For the monopole mode, the Lorenz gauge field equations in \eqn{lorenz_radial_field_equations}
reduce to three ODEs, for the $i \in [1, 3, 6]$ modes, whilst the other modes vanish.  The degrees of freedom for the
monopole mode are then further reduced by the gauge condition in \eqn{gauge_even2}, which can be used to decouple the
$i = 6$ component of the metric perturbation and express it algebraically in terms of the other modes.  One is left
with two coupled, second-order ODEs for the $i \in [1, 3]$ modes and thus one can describe the monopole mode in terms of a
complete four dimensional basis of homoegeneous solutions.  Following \cite{Akcay:2013wfa,Wardell:2015ada}, one can write the
metric perturbation that emenates from the non-vanishing BLS monopole modes as
\begin{align}
	\bmat{H} &:= \frac{M}{\mu} \left\{ h_{tt}, h_{rr}, \frac{h_{\theta\theta}}{r^{2}} 
	= \frac{h_{\varphi\varphi}}{r^{2} \sin^{2}\theta} \right\} \nn \\
	&= \frac{M}{4\sqrt{\pi}r} \left\{ \bar{h}^{(1)}_{00} + f \barh^{(6)}_{00}, \frac{\barh^{(1)}_{00} - f \barh^{(6)}_{00}}{f^{2}}, \barh^{(3)}_{00} \right\}.
	\label{eq:monopole_metric_perturbation}
\end{align}
Inverting this metric perturbation in terms of the BLS modes, one finds
\begin{align}
	\bar{h}^{(1)}_{00} &= \frac{2\sqrt{\pi}}{\mu} r \left( h_{tt} + f^{2} h_{rr} \right), \label{eq:h1_monopole} \\
	\bar{h}^{(3)}_{00} &= \frac{4\sqrt{\pi}}{\mu} \frac{h_{\theta\theta}}{r}, \label{eq:h3_monopole} \\
	\bar{h}^{(6)}_{00} &= \frac{2\sqrt{\pi}}{\mu} \frac{r \left( h_{tt} - f^{2} h_{rr} \right)}{f} \label{eq:h6_monopole}. \\
\end{align}
The complete basis of homoegeneous field equations for the $i \in [1, 3]$ modes is then given by \cite{Akcay:2013wfa},
\begin{align}
	\bmat{H}_{A} &= \left\{ -f, f^{-1}, 1 \right\}, \label{eq:monopole_H_A} \\
	\bmat{H}_{B} &= \left\{ -\frac{f M}{r^{3}} P(r), \frac{1}{r^{3}f} Q(r), \frac{f}{r^{2}} P(r) \right\}, \label{eq:monopole_H_B} \\
	\bmat{H}_{C} &= \left\{ -\frac{M^{4}}{r^{4}}, \frac{M^{3}(3M - 2r)}{r^{4}f}, \frac{M^{3}}{r^{3}} \right\}, \label{eq:monopole_H_C} \\
	\bmat{H}_{D} &= \left\{ -\frac{M}{r^{4}} \left[ W(r) + r P(r) f \ln f - 8M^{3} \ln \left(r/M\right)\right] \right., \nn \\
	&\frac{1}{r^{4}f^{2}} \left[ K(r) - r Q(r) f \ln f - 8M^{3}(2r - 3M) \ln \left(r/M\right)\right], \nn \\
	&\left.\frac{1}{r^{3}} \left[ 3r^{3} - W(r) - r P(r) f \ln f + 8M^{3} \ln \left(r/M\right)\right] \right\}, \label{eq:monopole_H_D}
\end{align}
where $P(r)$, $Q(r)$, $W(r)$, and $K(r)$ are given by
\begin{align}
	P(r) &= r^2 + 2rM + 4M^2, \\
	Q(r) &= r^3 - r^2M - 2rM^2 +12M^3, \\
	W(r) &= 3r^3 - r^2M - 4rM^2 -28M^3/3, \\
	K(r) &= r^3M - 5r^2M^2 - 20rM^3/3 +28M^4.
  \end{align}

The first pair of solutions, $\{ \bmat{H}_{A}, \bmat{H}_{B} \}$ are regular at the future horizon,
but \emph{irregular} towards null infinity since the components decay to a constant value instead
of zero.  In contrast, the other pair of homoegeneous solutions $\{ \bmat{H}_{C}, \bmat{H}_{D} \}$
are singular at the future horizon, yet regular towards null infinity.  We shall use these homogeneous solutions
to construct two seperate inhomogeneous solutions which have been used in the literature, and touch upon the relevant
merits of each.

\subsection{Berndtson Solution}
\label{sec:berndtson_solution}
The first solution we shall discuss is the \emph{Berndtson solution}, which was first derived in Ref.~\cite{Berndtson:2007gsc}.
This solution is arguably the most intutive, given our previous setup, as it is equivalent to taking a weighted sum of the
homogeneous solutions where they are bounded.  Thus inside the (circular) orbit, we utilise $\{ \bmat{H}_{A}, \bmat{H}_{B} \}$
and outside, we utilise $\{ \bmat{H}_{C}, \bmat{H}_{D} \}$ such that
\begin{align}
	\bmat{H}^{-} = C_{A} \bmat{H}_{A} + C_{B} \bmat{H}_{B}, \\
	\bmat{H}^{+} = C_{C} \bmat{H}_{C} + C_{D} \bmat{H}_{D}, \\
\end{align}
where $\{C_{A}, C_{B}, C_{C}, C_{D}\}$ are weighting coefficients, 
with the full solution for the metric perturbation being given by
\beq
	\bmat{H}^{\rm Bern}_{\ell = 0} = \bmat{H}^{-} \Theta[r - \rp] + \bmat{H}^{+} \Theta[\rp - r].
	\label{eq:berndtson_solution_H}
\eeq
To determine the weighting coefficients, one can again use the jump conditions at the worldline, which will fully constrain the solution.
Rewriting our problem, as we did in \secref{even_parity_static}, as a matrix problem, the solution for the monopole is given by
\beq
	\bmat{C} = \bmat{E}^{-1} \cdot \bmat{J},
	\label{eq:berndtson_matrix_solution}
\eeq
where $\bmat{C} = (C_{A}\ C_{B}\ C_{C}\ C_{D})^{T}$ and $\bmat{J} = (\bm{0}_{2}\ {\cal J}^{(1)} {\cal J}^{(3)})^{T}$.
The source terms ${\cal J}^{(1)}$ and ${\cal J}^{(3)}$ are given by the simple expressions for the monopole mode,
\begin{align}
	{\cal J}^{(1)} &= -\frac{8\sqrt{\pi}{\cal E} }{\rp}, \\
	{\cal J}^{(3)} &= \frac{{\cal J}^{(1)}}{f_{p}},
\end{align}
and we have defined the following matrix of homogeneous solutions,
\beq
	\renewcommand{\arraystretch}{1.5} 
	\bmat{E} = 
	\left(
	\begin{array}{c c c c}
		-\barh^{(1)}_{A} & -\barh^{(1)}_{B} & \barh^{(1)}_{C} & \barh^{(1)}_{D}  \\
		-\barh^{(3)}_{A} & -\barh^{(3)}_{B} & \barh^{(3)}_{C} & \barh^{(3)}_{D}  \\
		-\partial_{r}\barh^{(1)}_{A} & -\partial_{r}\barh^{(1)}_{B} & \partial_{r}\barh^{(1)}_{C}  & \partial_{r}\barh^{(1)}_{D} \\
		-\partial_{r}\barh^{(3)}_{A} & -\partial_{r}\barh^{(3)}_{B} & \partial_{r}\barh^{(3)}_{C}  & \partial_{r}\barh^{(3)}_{D}
	\end{array}
	\right)_{r = \rp},
	\label{eq:monopole_homoegeneous_matrix}
\eeq
where for $j \in [A, B, C, D]$, $\barh^{(i)}_{j}$ are constructed through applying Eqs.~(\ref{eq:h1_monopole})-(\ref{eq:h6_monopole})
to the basis of homoegeneous solutions,
\begin{align}
	\bar{h}^{(1)}_{j} &= \frac{2\sqrt{\pi}}{\mu} r \left( \bmat{H}^{j}_{tt} + f^{2} \bmat{H}^{j}_{rr} \right), \label{eq:h1_monopole_{j}} \\
	\bar{h}^{(3)}_{j} &= \frac{4\sqrt{\pi}}{\mu} \frac{\bmat{H}^{j}_{\theta\theta}}{r}, \label{eq:h3_monopole_{j}} \\
	\bar{h}^{(6)}_{j} &= \frac{2\sqrt{\pi}r}{\mu} \frac{r \left( \bmat{H}^{j}_{tt} - f^{2} \bmat{H}^{j}_{rr} \right)}{f} \label{eq:h6_monopole_{j}}. \\
\end{align}
The solutions to the weighting coefficients can then be written in full \cite{Wardell:2015ada},
\begin{align}
	C_{A} &= -\frac{2M }{\sqrt{\rp(\rp-3M)}}, \\
	C_{B} &= \frac{8M + (6M - 2\rp)\ln f}{3\sqrt{\rp(\rp-3M)}}, \\
	C_{C} & = \frac{2}{9M\sqrt{\rp(\rp-3M)}} \nn \\
	&\times \left[8M\rp -3\rp^2 \right. \nn \\
	&\left. - 12M^2+24M(3M-\rp)\ln\left(\rp/M\right)\right], \\
	C_{D} &= \frac{2}{3}\sqrt{1-\frac{3M}{\rp}}.
\end{align}
The monopole metric perturbation arising from this solution we label $h_{\alpha\beta}^{\rm Bern}$.  
This solution is globally regular, and, upon excluding any $t-r$ components from the metric perturbation, 
it emerges as the unique Lorenz gauge solution that maintains asymptotic flatness \cite{Miller:2020bft}.
Despite this regularity, the Berndtson solution was not used widely for Lorenz gauge calculations since
it was seen to not to represent a physical solution \cite{Akcay:2010dx,Dolan:2012jg, Wardell:2015ada}.
The reason the Berndtson solution was considered unphysical was due to the fact that the peturbation eminating
from the monopole mode contains a correction to the primary black hole mass due to the presence of the secondary.
However, as Miller and Pound astutely pointed out in Ref.~\cite{Miller:2020bft}, this is not actually the case.

Quantitatively, the mass within a sphere of radius $r < \rp$ for this monopole solution is given by 
$M_{\rm Bern} = -\mu{\cal E}/\rp f_{p}$. Thus, the perturbation $h_{\alpha\beta}^{\rm Bern}$ introduces an 
additional mass $M_{\rm Bern}$ in the region interior to the secondary, $r < \rp$, effectively altering 
the mass of the primary black hole. Consequently, the background mass $M$ no longer corresponds to the 
primary's physical mass. This type of perturbation represents a physical spacetime configuration where a 
secondary object of mass $\mu$ and energy $\cal E$ orbits a primary black hole with an effective mass 
$M_{\rm BH} = M + \epsilon M_{\rm Bern}$.

The Berndtson solution, owing the to this minutiae, has therefore largely been overlooked for use in
self-force calculations, although this solution is distinctly advantageous for second-order computations
as this solution leads to the most well-behaved Lorenz gauge second-order source \cite{Miller:2020bft,Pound:2019lzj}.
In most computations of the self-force in the Lorenz gauge, an alternative solution is used, which we shall now discuss.

\subsection{Detweiler and Poisson Solution}
\label{sec:detwiler_poisson_solution}
The second solution we shall discuss is the \emph{Detweiler and Poisson solution}, which was first derived in
Ref.~\cite{PhysRevD.69.084019}, based on earlier work by Zerilli \cite{PhysRevD.2.2141}.
Detweiler and Poisson's solution was originally derived analytically by taking a gauge transformation of the
Zerilli monopole solution.  This solution was later rederived by Barack and Lousto in Ref.~\cite{Barack:2005nr}
in terms of components of the metric perturbation.  We choose to present an alternative treatment here, based on
Refs.~\cite{Akcay:2013wfa, Wardell:2015ada}, which is more closely aligned with our discussion.

This solution is constructed from the solution given in the previous section, but with the addition of a
term so as the mass-content of the solution is corrected such that the mass within a sphere of radius $r < \rp$
is equal to the mass of the background spacetime, $M$.  Then for a sphere of radius $r > \rp$, the mass content
will be given by $M + \mu {\cal E}$.  This correction term, $\Delta \bmat{H}^{\delta M}$, will be a linear combination
of the homoegeneous solutions $\bmat{H}_{A}$ and $\bmat{H}_{B}$ since any combination of the other homogeneous solutions
is singular at the horizon.  As highlighed in Refs.~\cite{Dolan:2012jg, Akcay:2013wfa}, the correction term being
considered here must subtract off the contribution to the metric perturbation from $\bmat{H}_{A}$, since this 
is the only solution to contain mass-energy.
The combination can then be further constrained by considering their asymptotic form towards future null infinity,
\begin{align}
	\lim_{r \rightarrow \infty} \bmat{H}_{A} &= \{-1, 1, 1\} \\
	\lim_{r \rightarrow \infty} \bmat{H}_{B} &= \{0, 1, 1\},
	\label{eq:asymptotic_form_monopole_homoegeneous}
\end{align}
which shows that any correction term can only achieve asymptotic flatness in the spatial part of the metric.
Therefore, one finds the correction term is given by
\beq
	\Delta \bmat{H}^{\delta M} = - C_{A}(\bmat{H}_{A} + \bmat{H}_{B}),
	\label{eq:monopole_correction_term}
\eeq
which leaves us with the full solution for the monopole mode,
\beq
	\bmat{H}^{\rm DP}_{\ell = 0} = \Delta \bmat{H}^{\delta M} + \bmat{H}^{\rm Bern}_{\ell = 0}.
	\label{eq:detweiler_poisson_solution_H}
\eeq
As evident from the asymptotic form in \eqn{asymptotic_form_monopole_homoegeneous}, the correction term, 
while enforcing a specific mass content within the spacetime, does not preserve asymptotic flatness.
This is well known in the literature, and this irregularity is manifest is in the $tt$-component of the
metric perturbation, which in the limit as $r \rightarrow \infty$ is given by
\beq
	\lim_{r \rightarrow \infty}  h_{tt, \ell = 0}^{\rm DP} = 2M_{\rm Bern}.
	\label{eq:detweiler_poisson_asymptotic_tt}
\eeq
We previously highlighted this issue with this particular Lorenz gauge monopole in 
our discussion of the redshift in \secref{detweiler_redshift}.
For comparisons, it is preferred to calculate a quantity such as the redshift in a gauge 
that has the same asymptotic reference frame as the background spacetime.  
If the metric perturbation is not asymptotically flat, then this condition will not be satisfied.
Hence, when using this monopole solution, one will need to correct for this gauge artifact. 
This gauge artifact can be removed with by a simple class of gauge transformations,
\beq
	\tilde{\Xi}^{\mu} = \Xi^{\mu} + M_{\rm Bern} t \delta^{\mu}_{t}
	\label{eq:gauge_transformation_redshift}
\eeq
where $\Xi^{\mu}$ is a class of gauge generators that satisfy the helical symmetry of the system, 
$(\delta_{t} + \Omega_{\varphi}\delta_{\varphi})\Xi^{\mu} = 0$.  This transformation is tantamount
to normalising the time coordinate, where $M_{\rm Bern}$ corrects the asymptotic monopole behaviour.  
The contraction of the Lorenz gauge metric perturbation is corrected in the following manner 
under such a transformation,
\begin{align}
	h^{\rm R, DP}_{\mu\nu} &\rightarrow h^{\rm R, DP}_{\mu\nu} + 2M_{\rm Bern}\, g_{tt} (u^{t})^{2} \\
	&= h^{\rm R, DP}_{\mu\nu} + 2M_{\rm Bern}\, {\cal E} u^{t}.
\end{align}
This transformation will lead to the final expression for the redshift seen in 
\eqn{detweiler_redshift_gauge_invariant}.

\bibliography{Bibliography}
\end{document}